\documentclass[%
 reprint,
superscriptaddress,
 amsmath,amssymb,
 aps,
 prx,
]{revtex4-2}

\usepackage{graphicx}
\usepackage{dcolumn}
\usepackage{bm}
\usepackage[colorlinks]{hyperref}
\usepackage{physics}
\usepackage[squaren]{SIunits}
\usepackage{mathtools}
\usepackage{amsthm}
\usepackage{xcolor}
\usepackage{mathrsfs}
\usepackage{times}
\usepackage[normalem]{ulem}
\usepackage{algorithm2e}
\usepackage{comment}
\usepackage{titletoc}

\usepackage{tocloft}

\hypersetup{
	colorlinks=true,
	linkcolor=red,
	filecolor=blue,      
	urlcolor=blue,
	citecolor=blue
}
\newcommand{\rev}[1]{{\textcolor{black}{#1}}}

\newcommand{\Eq}[1]{Eq.\,(\ref{#1})}

\newcommand{\Eqs}[2]{Eqs.\,(\ref{#1}-\ref{#2})}
\newcommand{\Sec}[1]{Section~\ref{#1}}
\newcommand{\Fig}[1]{Fig.\,\ref{#1}}

\newcommand{\NS}{S}
\newcommand{\NQ}{L}

\newcommand{\UI}{\bm{u}}
\newcommand{\EC}{C_T}
\newcommand{\ETC}{\bar{\mathcal{T}}}
\newcommand{\regmat}{\widetilde{\mathbf{F}}_N}
\newcommand{\RM}{\mathbf{F}_N}
\newcommand{\Xs}{\mathcal{X}}

\newcommand{\cu}{\mathbf{\Sigma}}  
\newcommand{\ci}{\mathbf{V}}       
\newcommand{\gr}{\mathbf{G}}       
\newcommand{\gh}{\mathbf{\Lambda}}  
\newcommand{\nsr}{\mathbf{R}}      

\newcommand{\Es}[1]{\mathbb{E}_{\Xs}\!\!\left[#1\right]}
\newcommand{\Covs}{\mathrm{Cov}_{\Xs}}
\newcommand{\Vars}{\mathrm{Var}_{\Xs}}
\newcommand{\EUI}[1]{\mathbb{E}_{\bm{u}}\!\!\left[#1\right]}
\newcommand{\Eu}[1]{\mathbb{E}_{u}\!\!\left[#1\right]}

\begin{document}

\title[]{Tackling Sampling Noise in Physical Systems for Machine Learning Applications: Fundamental Limits and Eigentasks}



\makeatletter

\author{Fangjun Hu}
\thanks{These three authors contributed equally}
\affiliation{Department of Electrical and Computer Engineering, Princeton University, Princeton, NJ 08544, USA}

\author{Gerasimos Angelatos}
\thanks{These three authors contributed equally}
\affiliation{Department of Electrical and Computer Engineering, Princeton University, Princeton, NJ 08544, USA}
\affiliation{Raytheon BBN, Cambridge, MA 02138, USA}

\author{Saeed A. Khan}
\thanks{These three authors contributed equally}
\affiliation{Department of Electrical and Computer Engineering, Princeton University, Princeton, NJ 08544, USA}

\author{Marti Vives}
\affiliation{Department of Electrical and Computer Engineering, Princeton University, Princeton, NJ 08544, USA}
\affiliation{Q-CTRL, Santa Monica, CA 90401, USA}

\author{Esin T\"ureci}
\affiliation{Department of Computer Science, Princeton University, Princeton, NJ 08544, USA}

\author{Leon Bello}
\affiliation{Department of Electrical and Computer Engineering, Princeton University, Princeton, NJ 08544, USA}

\author{Graham E. Rowlands}
\affiliation{Raytheon BBN, Cambridge, MA 02138, USA}

\author{Guilhem J. Ribeill}
\affiliation{Raytheon BBN, Cambridge, MA 02138, USA}

\author{Hakan E. T\"ureci}
\affiliation{Department of Electrical and Computer Engineering, Princeton University, Princeton, NJ 08544, USA}

\date{\today}

\begin{abstract}

The expressive capacity of physical systems employed for learning is limited by the unavoidable presence of noise in their extracted outputs. Though present in physical systems across both the classical and quantum regimes, the precise impact of noise on learning remains poorly understood. Focusing on supervised learning, we present a mathematical framework for evaluating the resolvable expressive capacity (REC) of general physical systems under finite sampling noise, and provide a methodology for extracting its extrema, the eigentasks. Eigentasks are a native set of functions that a given physical system can approximate with minimal error. We show that the REC of a quantum system is limited by the fundamental theory of quantum measurement, and obtain a tight upper bound for the REC of any finitely-sampled physical system. We then provide empirical evidence that extracting low-noise eigentasks can lead to improved performance for machine learning tasks such as classification, displaying robustness to overfitting. We present analyses suggesting that correlations in the measured quantum system enhance learning capacity by reducing noise in eigentasks. The applicability of these results in practice is demonstrated with experiments on superconducting quantum processors. Our findings have broad implications for quantum machine learning and sensing applications.
\end{abstract}

\maketitle





\section{Introduction}

\begin{figure}
    \centering
    \includegraphics[width=\columnwidth]{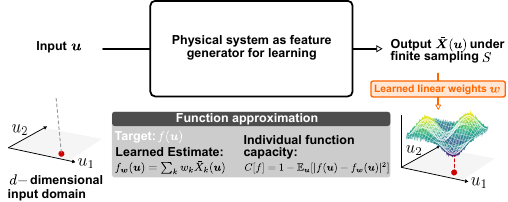}
    \caption{Framework of learning with physical systems we consider in this work: inputs $\UI$ are transformed to a set of outputs $\bar{\bm{X}}(\UI)$ via a parameterized feature generator, implemented using an arbitrary physical system. Outputs are combined with appropriate weights $\bm{w}$ to approximate a desired function $f$. 
    Capacity $C[f]$ quantifies the error in this approximation $f_{\bm{w}}$. We consider normalized functions $\EUI{f^2}=1$, where $\mathbb{E}_{\UI}$ denotes the expectation over the input domain with respect to a chosen measure.
    }
    \label{fig:Gen_Schematic}
\end{figure}

A physical system receiving an input stimulus typically evolves in response to it, such that its degrees of freedom become dependent on said input after a certain period of interaction with it. This everyday observation has a profound implication: 
any dynamical system can be viewed as performing a transformation of its input, realizing an input-output map~\cite{boyd_fading_1985}. This functional map can in principle be optimized, inspiring an emerging approach to learning with analog physical systems, which we will collectively refer to as Physical Neural Networks (PNN) \cite{Wright2022, nakajima_physical_2022, markovic_physics_2020}. 
PNNs employ a wide variety of analog physical systems to compute a trainable transformation on an input signal~\cite{Tanaka2019, mujal_opportunities_2021, cerezo_variational_2021, ortin_unified_2015, lopez-pastor_self-learning_2021, wilson_quantum_2019, garcia-beni_scalable_2022, havlicek_supervised_2019, rowlands_reservoir_2021, canaday_rapid_2018, shen_deep_2017, lin_all-optical_2018, pai_experimentally_2023}.    
More precisely, the role of an idealized (\textit{i.e.}\,completely deterministic, noise-free) physical system in these approaches is that of a high-dimensional feature generator. 
Given inputs $\bm{u}$, the measured degrees of freedom $x_k(\bm{u})$ for $k \in [K]$, generated by the system, act as an input-dependent vector of features. These features are used to approximate a function $f(\bm{u})$ via a learned linear projection with sufficient accuracy, as dictated by a chosen loss function (See Fig.\,\ref{fig:Gen_Schematic}). Different characteristics of the physical system, described by a set of hyperparameters $\bm{\theta}$, may determine its ability to approximate a particular function. Consequently, the relationship between a specific physical system and the classes of functions it can express with high accuracy is a fundamental question in this paradigm of machine learning
~\cite{dambre_information_2012, sheldon_computational_2022, schuld_effect_2021, wu_expressivity_2021, wright_capacity_2019, pai_experimentally_2023}. 


No physical system however exists in isolation, and is therefore necessarily subject to noise. Noise can enter at the input, whereby it evolves under the same dynamical law governing the evolution of the physical system. There may also be variability in this very dynamics of the physical system itself. Finally, there is typically noise associated with the measurement of output features from the physical system. 
As a consequence of these noise sources, the resulting feature map is stochastic: even under identical preparations and inputs $\bm{u}$, the outcome of a measurement $X_k^{(s)}(\bm{u})$ of a feature $k$ can vary between repetitions, each of which is referred to as a ``shot'' $s$. By empirically averaging the outcomes of $\NS$ shots, one can
generally reduce this stochasticity. We will refer to the resulting noise as “sampling noise”. Theoretical analysis and experimental implementations of PNNs have already demonstrated that sampling noise can have a substantial role in the ultimate performance of a physical learning machine~\cite{garcia-beni_scalable_2022, shen_deep_2017, havlicek_supervised_2019}. However, it is also known that this role may be more subtle than a limitation on performance across the board, as evidenced for example in the effective use of noise for regularization to aid generalizability in learning~\cite{bishop_training_1995, neelakantan_adding_2015, noh_regularizing_2017}. 


Often, heuristic descriptions are used to theoretically model such sampling noise and explain its effects on learning~\cite{dambre_information_2012, ortin_unified_2015, Rumyantsev2020, garcia-beni_scalable_2022}. However, when considering physical \textit{quantum} systems for learning, a fundamental microscopic model for sampling noise is provided, and in fact imposed by the quantum theory of measurement. Explicitly, for a quantum system prepared in an initial state density matrix $\hat{\rho}_0$ and evolving under an input-parameterized quantum channel $\mathcal{U}(\bm{\theta},\bm{u})$, the final state is $\hat{\rho}(\UI) = \mathcal{U}(\bm{\theta},\bm{u})\hat{\rho}_0$~\footnote{A few things to note here. 1. The initial state preparation is in practice often realized by an act of measurement as well. Then, the input-evolution-output sequence can be described as the sequence of measurement-evolution-measurement sequence. 2. Some PNN realizations view input as provided through an input state $|\Psi (\UI) \rangle$. Within the framework we adopt, this can be described as a parametric evolution $\mathcal{U}(\UI)$ acting on an initial $\UI$-independent state.}. Sampling noise in measured features from this quantum system is constrained by the choice of measurement projectors $\hat{M}_k$ associated with $\hat{\rho}(\UI)$. Unless the physical transformation defined by the quantum channel is optimized to yield only specific highly localized $\hat{\rho}(\UI)$ in the eigenspace of $\hat{M}_k$ (as in quantum algorithms such as Grover's or Shor's \cite{nielsen2002quantum}) -- a significant design restriction -- or an excessively large number of shots $\NS$ is used -- a significant hardware restriction -- such quantum sampling noise will be an intrinsic component of learning with quantum systems. 



Therefore, a framework is required that can account for sampling noise across generic physical systems, and provide tools for learning when sampling noise is unavoidable. In this paper, we address the following question directly: what is the resolvable function space of an arbitrary physical system when regarded as an input-output machine \textit{in the presence of sampling noise}? This simple objective leads us to a general mathematical framework with important consequences for statistical learning theory, which we now overview.  Our analysis is centered around a specific metric, the Resolvable Expressive Capacity (REC), which is a generalization of the information processing capacity introduced in Ref.\,\cite{dambre_information_2012} (see also the earlier work in Ref.\,\cite{jager2001short}) to account for the presence of sampling noise.  
Specifically, the REC is a quantitative measure of the accuracy with which $K$ system-specific orthogonal functions can be constructed from $K$ stochastic features $\bar{X}_k(\bm{u})$. Remarkably, this accuracy has a tight, calculable $\NS$-dependent upper-bound. 
The special functions, referred to as the eigentasks $y^{(k)}(\bm{u})$ of the physical system, define the maximally-resolvable function space under $\NS$ shots, which sets the stage for the introduction of a learning methodology in the presence of sampling noise. 

Crucially, our framework can be applied to an arbitrary physical system via the solution of a simple matrix eigenproblem. The matrices in question are standard Gram matrix $\gr$ and covariance matrix $\ci$, which can be estimated using stochastic samples from the system as a function of inputs $\bm{u}$ over the domain of interest; the analysis can thus be directly implemented in experimental settings without an internal model of the system. The solution of this linear eigenproblem yields both the eigentasks $\{y^{(k)}(\bm{u})\}$ and associated ``noise-to-signal'' eigenvalues $\{\beta_k^2\}$, which codify the normalized noise power in a construction of $y^{(k)}(\bm{u})$ from finite-$S$ $\bar{X}_k(\bm{u})$. The REC of the system is then only a function of $\{\beta_k^2\}$ and $S$.

In the second part of this paper we develop Eigentask Learning, a means of learning in physical systems where sampling noise dominates, by using the noise-ordered eigentasks to construct a maximally resolvable basis of measured features. Our approach affects a systematic removal of high-noise features during training, which we demonstrate in experiments. These experimental demonstrations provide empirical evidence of robustness to overfitting in supervised learning, enhancing generalizability in the presence of sampling noise. Such a learning scheme built on avoiding features identified as having large noise may in fact be at play in natural physical systems such as biological neural circuits~\cite{montijn_population-level_2016}. 
A well-studied example is that of neural vision: here input visual stimuli drive stochastic dynamics of sensory neurons in the visual cortex, which must together elicit a target response, such as the brain correctly distinguishing two images. Studies have shown that the dynamics of individual neurons under nominally-identical stimuli can exhibit great variability on a shot-by-shot basis~\cite{faisal_noise_2008}; however in spite of the significant noise, the overall driven behavior remains capable of distinguishing visual stimuli with high fidelity. Studies analyzing the robust neural code despite noisy neural activity have found emergent global coding directions in the population activity that evade ``modes" with maximal noise~\cite{montijn_population-level_2016, Rumyantsev2020}. The eigentask construction introduced here can be viewed as a generalization of this idea of noise-ordered modes to function spaces over an arbitrary input domain, and for an arbitrary physical system.

The Eigentask Learning framework is  sensitive not just to the properties of the noise itself (encoded in $\ci$), but also any dependence between it and the noise-free features via the Gram matrix $\gr$.  Such a situation typically prevails when the dominant source of sampling noise is either part of, or evolves under, the same input-output map defined by the physical system, as opposed to a completely uncorrelated noise source downstream. A simple example illustrating this, and one we analyze in detail, is that of a classical optical system, with features measured via photodetection. Here, the sampled features -- the integrated photocurrents -- are subject to shot noise whose variance is related to the mean of the photocurrents themselves.

In the quantum regime of operation of a physical system -- our ultimate focus -- this relation between sampling noise and the state of the physical system emerges in the most general description of quantum measurement as a positive operator-valued measure (POVM). We formulate the computation of REC and eigentasks for arbitrary quantum systems in the presence of this fundamental sampling noise structure; we focus on qubit-based quantum systems (including gate-based circuits and quantum annealers) operated as 
untrained PNNs under static inputs (so-called Extreme Learning Machines (ELMs) \cite{GuangBinHuang, ortin_unified_2015}), but our analysis is applicable to far more general quantum sensing and learning platforms. 
To validate the theoretical findings and emphasize their ready applicability to experimental scenarios, we implemented an ELM through a parameterized quantum circuit encoding on an IBMQ superconducting processor: demonstrating the calculation of REC, the construction of eigentasks, and the application of Eigentask Learning to a classification task. In all cases excellent agreement is seen with numerical simulation, and direct correlation is observed between REC and success at the considered classification task. This invites the exploration of principles to maximize the finite-sampling REC of a quantum system; for the qubit-based systems analyzed here, we show that an increase in measured quantum correlations can aid this goal. 



\begin{figure}
    \centering
    \includegraphics[width=\columnwidth]{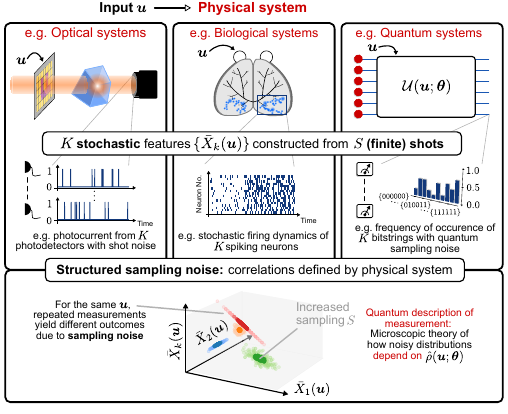}
    \caption{Three distinct examples of physical systems for learning. Extracted information takes the form of $K$ stochastic features $\bar{\bm{X}}$ obtained under $\NS$ shots. For quantum systems, the geometric structure of distributions of these measured features (bottom panel) is fundamentally determined by quantum sampling noise, which depends on the quantum state $\hat{\rho}(\UI;\bm{\theta})$, and hence on the nature of the mapping from input $\UI$ to this state. We show four calculated distributions differing only in the values of inputs $\UI$ to highlight this dependence. }
    \label{fig:NISQRC_Schematic}
\end{figure}


The remainder of this paper is organized as follows. Section \ref{sec:TheoreticalAnalysis} provides the general theory of REC and eigentasks with respect to sampling noise in generic supervised physical learning systems, and presents a calculation for a basic classical optical PNN. Section \ref{sec:QSN} applies REC theory and eigentask construction to machine learning with quantum systems, which is then validated and demonstrated with experiments performed on a 7-qubit IBMQ superconducting processor
in Sec.~\ref{sec:ibmq}. Finally, conclusions are presented in Sec.~\ref{sec:discussion}.
\section{Theoretical Analysis}
\label{sec:TheoreticalAnalysis}

\subsection{Sampling Noise in Learning with Physical Systems}

The most general approach to supervised learning from classical data using a generic physical system is depicted schematically in Fig.\,\ref{fig:Gen_Schematic}. A table with symbols and abbreviations used in the text can be found in Appendix \ref{app:table}. We consider a scheme that begins with ``embedding'' the classical input data $\bm{u}$, sampled from a distribution $p(\UI)$, into the physical system to be used for learning. The form of this embedding is unrestricted beyond the requirement of being physical, and its precise nature will influence the REC and eigentasks; some concrete examples will be provided shortly.

In order to access information from the physical system after its interaction with the input, measurements must be performed on its $K$ accessible  degrees of freedom. For a fixed input $\bm{u}$, a single measurement or ``shot'' $s$ yields \textit{single-shot} random-valued features $\{X_k^{(s)} (\UI)\}$ for each $k \in [K]$. We define the \textit{measured features} $\bar{X}_k$ as $\NS$-shot sample means of $\{X_k^{(s)} \}$:
\begin{align}
    \bar{X}_k(\UI) = \frac{1}{\NS} \sum_{s=1}^{S} X_k^{(s)} (\UI)
    \label{eq:Xsum}
\end{align}
whose expectation (equivalently via the central limit theorem, the $S$-infinite limit) is given by
\begin{align}
    x_k({\UI}) \equiv \Es{ \bar{X}_k(\UI) } = \lim_{S\to\infty} \bar{X}_k({\UI})
    \label{eq:mapping0}
\end{align}
where $\UI$ is regarded as a free variable. To be more precise, the expectation is evaluated over the product distribution of $S$ independent and identically distributed (i.i.d.)~vectors $\Xs(\UI) \equiv \{(X_{0}^{(s)}(\UI), X_{1}^{(s)}(\UI), \cdots, X_{K-1}^{(s)}(\UI))\}_{s \in [S]}$, conditioned on a fixed $\UI$.

With the definition of their expectation in Eq.\,(\ref{eq:mapping0}), the measured features $\bar{\bm X}(\UI) \in \mathbb{R}^K$, a column vector consisting of $\bar{X}_k(\UI)$,
can be conveniently decomposed by extracting its deterministic mean value, together with a zero-mean, input-dependent noise term $\boldsymbol{\zeta}(\UI)$:
\begin{equation}
    \bar{\bm X}(\UI) =  {\bm x}(\UI) +\frac{1}{\sqrt{S}}  \boldsymbol{\zeta}(\UI).
    \label{eq:xbar}
\end{equation}
Here $\boldsymbol{\zeta}$ encodes the statistics of sampling; it generally has nontrivial cumulants of all orders, of which the covariances take the particular $S$-independent form $\cu(\UI) \in \mathbb{R}^{K \times K}$:
\begin{align}
    \cu_{jk}(\UI) \equiv \Covs [{\zeta}_j(\UI), {\zeta}_k(\UI)]
    \label{eq:sigmajku}
\end{align}
and only depend on input $\UI$. We note that Eq.\,(\ref{eq:xbar}) is exact. The factor of $1/\sqrt{\NS}$ is merely extracted for convenience of the analysis to follow, and is \textit{not} meant to suggest an expansion for large $\NS$ at this stage; cumulants of $\boldsymbol{\zeta}$ beyond second-order inherit a complicated $S$-dependence. 

The general input-output relationship $\bm{u} \rightarrow \bar{X}_k(\UI)$ above can be made concrete by considering three example physical systems, depicted in Fig.\,\ref{fig:NISQRC_Schematic}. For an {\it optical system}, the input $\UI$ could for instance be embedded as a collection of pixel values on a spatial light modulator (SLM) in the path of a propagating beam of light. The individual single-shot features $\{X_k^{(s)}\}$ could be generated by integrating the photocurrent from each pixel of a number-resolving CCD camera for a certain hold time. For a {\it biological neural circuit}, the input $\UI$ might be a static-in-time visual stimulus, representing the electromagnetic field intensity incident on photoreceptors in the eye, and $\{X_k^{(s)} (\UI)\}$ can be the action potential of the $k$th neuron integrated over a certain time-period, e.g.\,measured through Ca$^{2+}$ imaging \cite{Grienberger2022}. Finally, for a {\it superconducting quantum processor}, inputs may be embedded via a suitable quantum channel, implemented for example via parameterized quantum gates. The single-shot features are simply the indicator functions of the possible outcome labels after quantum measurement. In all cases, the measured features $\bar{X}_k(\bm{u})$ may be obtained by repeating each experiment $S$ times with the same $\UI$  and constructing the $S$-shot histogram.

The randomness of the measured features derives from the quantum mechanical or the thermodynamical nature of the processes that the physical system is subject to during its evolution, but more importantly in the measurement/detection phase. In the case of neural circuits for instance, even when great care is exercised by presenting identical stimuli, the timing of action potentials of individual neurons can vary significantly over repeated trials on a scale that can be physiologically relevant. 
This noise can be traced to various sources~\cite{faisal_noise_2008} including dynamical changes of internal states of neurons between trials, and random processes neurons are subject to. The source of sampling noise for the optical system discussed in Sec.~\ref{sec:photonicRC} is the shot noise related to the discrete nature of energy exchange between the EM field and the photodetector, an electronic system. For an ideal  quantum computing system, the noise process we consider is due to shot noise in projective measurement, which we refer to as quantum sampling noise. We note that in qubit systems, there are many other potential noise sources, but in modern quantum processors these ought to be sub-leading at least for shallow circuits.  Indeed, in experiments reported in Sec.~\ref{subsec:RECqc} we observe that sampling noise dominates even at the maximum available $S$. Quantum sampling noise will still be the limiting source of noise after the advancement of error-corrected quantum computers.

A last important source of noise is the noise in the input signal to be processed. Visual neural circuits for instance involve the absorption of photons that arrive at the photoreceptors from EM sources that are subject to quantum mechanical or thermodynamical fluctuations. Here we are not concerned with a precise description of the physical nature of the input stimuli, and account for it by assuming an underlying probability distribution $p(\UI)$ from which the inputs are sampled. The most complete treatment of such a process requires a quantum mechanical description of both the signal generating system and its coherent coupling to the physical system that processes it, as has been introduced and analyzed in Ref.\,\cite{khan_quantum_2021}.

\subsection{Resolvable Expressive Capacity and Eigentasks}
\label{sec:defwwast}

Returning to the situation depicted in Fig.\,\ref{fig:Gen_Schematic}, supervised learning in physical systems can generically be cast as encoding data in the system, and then using measurement outputs to approximate a desired function $f(\UI)$ (here assumed to be square-integrable $\EUI{f^2} < \infty$), where the expectation over input data $\mathbb{E}_{\UI}$ is defined with respect to the distribution $p(\UI)$: $\EUI{f} \equiv \int \dd \UI \, p(\UI) f(\UI)$. 
The introduction of the symbol $\mathbb{E}_{\UI}$ for expectation over $\UI$ is necessitated by the use of two types of averages in the analysis of the loss function: over the output samples ($\mathbb{E}_{\Xs}$) and over the input domain ($\mathbb{E}_{\UI}$).

Within the PNN approach considered here, $f(\UI)$ is approximated for finite $S$ as $f_{\bm{W}}(\UI) = \bm{W}^T \bar{\bm{X}}(\UI) = \sum_{k} W_k \bar{X}_k(\UI)$. To quantify the fidelity of this approximation, we introduce a statistical variant of the function capacity~\cite{dambre_information_2012, wright_capacity_2019, martinez-pena_information_2020}, which is the normalized mean-squared accuracy of the estimate $f_{\bm{W}}$,
\begin{align}
    C[f] =  1- \min_{\boldsymbol{W} \in \mathbb{R}^K} \frac{ \EUI{ \mathbb{E}_{\mathcal{X}}[ (f(\UI) - f_{\bm{W}}(\UI) )^2 ]} }{\EUI{f(\UI)^2}}. 
    \label{eq:fcap}
\end{align}
This quantity differs from that introduced in Refs.\,\cite{dambre_information_2012, wright_capacity_2019, martinez-pena_information_2020} in that the squared error term $(f(\UI) - f_{\bm{W}}(\UI) )^2$ is stochastic, and thus both the expectation over the output samples $\mathcal{X}$ and the expectation over the inputs $\bm{u}$ are needed to ensure that Eq.\,(\ref{eq:fcap}) is a deterministic value.
Minimizing the error in the approximation of $f(\UI)$ by $f_{\bm{W}}(\UI)$ over the input domain to determine capacity thus requires finding 
\begin{align}
    \boldsymbol{w} = \underset{\bm{W} \in \mathbb{R}^{K}}{\mathrm{argmin}} \, \EUI{\Es{(f - \bm{W}^T \bar{\bm{X}})^2}}.  \label{eq:optimalw}
\end{align}
This minimization can always be expressed analytically via a pseudoinverse operation (see Appendix \ref{sec:DefCap}). This function capacity is constructed such that $0 \leq C[f]\leq 1$, with the upper limit indicating a perfect approximation. 

The choice of a linear estimator and a mean squared error loss function may appear restrictive at first glance, but the generality of our formalism averts such limitations. The use of a linear estimator applied directly to readout features appears to preclude nonlinear post-processing of measurements; this is intentional and simply meant to ensure the calculated functional capacity is a measure of the ability of the physical system itself, and not of a nonlinear processing layer. Furthermore, the mean squared loss effectively describes the first term in a Taylor expansion of a wide range of arbitrary nonlinear post-processing and non-quadratic loss functions. The most well-known example is that of logistic regression for supervised classification problems, where the sigmoid function $\sigma(\bm{W}^T \bar{\bm{X}})$ (\textit{i.e.,} $\sigma(z)=1/(1+\mathrm{exp}(-z))$) is used for post-processing, while the cross-entropy loss function is used for optimization~(for further details and analysis of non-linear post-processing, see Appendix \ref{app:ComplxNonLin}). 


To extend the notion of capacity to a task-independent metric representing how much classical information about an input can be extracted from a system in the presence of sampling noise, we sum the function capacity over a basis of functions $\{ f_\ell \}_{\ell \in \mathbb{N}}$ which are complete and orthonormal with respect to the input distribution, i.e.\,equipped with the inner product $\langle f_{\ell}, f_{\ell'} \rangle_{p} = \int f_{\ell}(\UI) f_{\ell'}(\UI) p(\UI) \dd \UI = \delta_{\ell \ell'}$. The total \textit{Resolvable Expressive Capacity} (REC) is then $\EC \equiv \sum_{\ell=0}^{\infty} C[ f_{\ell} ]$, which effectively quantifies how many linearly-independent functions can be expressed from a linear combination of $\{\bar{X}_k (\UI) \}$. Our main result -- proven in detail in Appendix \ref{app:EC} -- is that given any $S \in \mathbb{N}^{+}$, the REC for a physical system whose measured features are stochastic variables of the form of Eq.\,(\ref{eq:xbar}) is given by
\begin{align}
    \!\!\EC(\bm{\theta}) 
    = \mathrm{Tr} \left(\! \left( \gr + \frac{1}{\NS} \ci \right)^{\!\! - 1}\!\! \gr \right) = \sum_{k=0}^{K-1} \frac{1}{1 + \beta_k^2(\bm{\theta})/\NS}. 
    \label{eq:EC}
\end{align}
Here we made explicit the dependence on $\bm{\theta}$, the hyperparameters of the input embedding to indicate the important dependence of the $S$-shot REC on the input encoding. 

The first equality, arrived at through straight-forward algebraic manipulation, is written in terms of the expected feature Gram and covariance matrices $\gr \equiv \EUI{\bm{x}\bm{x}^T}$ and $\ci \equiv \EUI{\boldsymbol\Sigma}$ respectively. 
First, we are able to conclude that $\lim_{S\to\infty} C_T = \mathrm{Rank}\{\gr\} \leq K$, recovering the bound of Ref.\,\cite{dambre_information_2012}. Importantly, the rank of the Gram matrix is always equal to the maximal number of linearly-independent functions in the set $\{x_k(\UI)\}$ (see Appendix \ref{app:eigentasks}) In this article, we only consider the case where $\gr$ is full-rank, which is the most interesting case: maximizing the rank of $\gr$ maximizes the highest achievable (i.e.\,infinite-$\NS$) REC for a physical system. Furthermore, this condition is typically met unless the physical system is constrained by special symmetries; in such cases where some features $\{x_k(\UI)\}$ are linearly dependent, the matrix inverse in Eq.\,(\ref{eq:EC}) should be modified to a pseudo-inverse.
We also later demonstrate that both $\gr$ and $\ci$ can be estimated efficiently and accurately in experiment and consequently under finite $\NS$ (see Appendix \ref{app:Spectral_finite_statistics}). 
The second equality in Eq.\,(\ref{eq:EC}) remarkably provides a closed-form expression for $\EC$ at any $S$, which is independent of the specific choice of the generally {\it infinite} set $\{ f_\ell \}_{\ell \in \mathbb{N}}$ (and thus not subject to numerical challenges associated with its evaluation over such a set~\cite{dambre_information_2012}).
Instead, the REC is entirely captured by the function capacity of $K$ distinct functions, and for a given physical system is fully characterized by the spectrum of eigenvalues $\{ \beta^{2}_k \}_{k \in [K]}$ satisfying the generalized eigenvalue problem
\begin{align}
    \ci \bm{r}^{(k)} = \beta_k^2 \gr \bm{r}^{(k)}. \label{eq:eigenprob}
\end{align}
In the above, all quantities depend on $\boldsymbol\theta$ and thus the specific physical system and input embedding via the Gram ($\gr$) and covariance ($\ci$) matrices.
Associated with each $\beta_k^2$ is an eigenvector $\bm{r}^{(k)}$ living in the space of measured features and thus defining a set of $K$ orthogonal functions via the linear transformation
\begin{align}
    y^{(k)} (\UI) = \sum_{j} r_{j}^{(k)} x_{j} (\UI). 
    \label{eq:eigentaskproj}
\end{align}
We refer to $\{y^{(k)}\}$ as {\it eigentasks}, as they form the minimal set of orthonormal functions ($\EUI{y^{(j)} y^{(k)}} = \delta_{jk}$) which saturates the available REC of a physical system and thus the accessible information content present in its measured features. Specifically, the capacity to approximate a given $y^{(k)}$ with $S$ shots is $C[y^{(k)}]=1/(1+\beta^2_k/S)$: the REC in Eq.\,\eqref{eq:EC} is simply a sum of eigentask capacities. This further highlights that a given parameterized system can only approximate a target function to the degree that it can be written as a linear combination of $\{y^{(k)}\}$.  The eigentasks thus serve as a powerful basis for learning, as shall be explored in Sec.\,\ref{sec:eigentasklearning}.

\subsection{Resolvable Expressive Capacity and Eigentasks in practice: measured eigentasks}

Our use of the expectation over distributions of the input, $\EUI{\cdot}$, and finitely-sampled measured features, $\Es{\cdot}$, in principle implies the availability of an infinite number of input and measured samples respectively. Of course, for the practical implementation of any PNN, both these values are finite. However, as we will demonstrate via calculations of the REC and eigentasks using both theoretical and experimental systems, our framework can be applied when these values are constrained to be finite. 

More precisely, we note that in practice only a finite number of values $N$ can be i.i.d.~sampled from the input distribution, namely $\UI^{(n)} \sim p(\UI)$ for any $n \in [N]$. For each discrete input, one set of measured output features constructed from finite $\NS$ is obtained, a single sample from the distribution $\mathcal{X}(\bm{u}^{(n)})$. The collection of both input and output samples constitutes the \textit{complete dataset}, which we denote as $\mathcal{D} \equiv \{(\bm{u}^{(n)}, \mathcal{X}(\bm{u}^{(n)}))\}_{n \in [N]}$. Our calculation of REC and eigentasks will have some dependence on $\mathcal{D}$ via $N$ and $\NS$.

In particular, the practically computed optimal weights in the capacity calculation are not the deterministic weights $\bm{w}$, but $\bm{w}^{\ast}$ computed on a given set of input samples and measured features, and hence depend on the dataset $\mathcal{D}$:
\begin{align}
    \bm{w}^{\ast}(\mathcal{D}) \equiv \underset{\bm{W} \in \mathbb{R}^{K}}{\mathrm{argmin}} \frac{1}{N} \! \sum_{n=1}^{N} \! \left( f(\UI^{(n)}) - \bm{W}^T \bar{\bm{X}}(\UI^{(n)}) \right)^2 \!. 
\end{align}
$\bm{w}^{\ast}(\mathcal{D})$ will vary due to changes in $\mathcal{D}$.
Generally, when $N,S$ are simultaneously finite and $S$ is fixed, a study of the $N$-scaling behavior of the difference between $\bm{w}$ in Eq.\,(\ref{eq:optimalw}) and the average optimized weight $\mathbb{E}_{\mathcal{D}}[\bm{w}^{\ast}(\mathcal{D})]$, as well as the variation of $\bm{w}^{\ast}(\mathcal{D})$ for different $\mathcal{D}$, falls in the realm of training and generalization errors over the input domain, an important area of research in {theoretical} machine learning~\cite{Seung1992, Canatar2021}. We leave this problem for future work; for all calculations and experiments in this paper, we consider the case - always realized in practice (and of particular relevance where \textit{sampling}, and thus the time and resource cost of processing with physical systems, is concerned) - where the dataset consists of a finite number $N$ of input samples.

We do address the important problem of REC and eigentask calculation when this fixed value of $N$ is finite, and using only a given set of measured features constructed under finite sampling $\NS$. As alluded to earlier, in Appendix~\ref{app:Spectral_finite_statistics} we demonstrate how the eigenproblem Eq.\,(\ref{eq:eigenprob}) can be constructed for finite $N$ and $\NS$, and present corrections to the eigenvalues and eigenvectors due to the finiteness of $\NS$. Numerical examples presented in Appendix~\ref{app:Spectral_finite_statistics} demonstrate a favorable match between this correction method and numerical simulations of eigenvalues and eigenvectors (see Fig.\,\ref{fig:NSR_Tilde_Renormalization} and Fig.\,\ref{fig:Tilde_Renormalization}). 

Importantly, we define a set of {\it measured} eigentasks $\bar{y}^{(k)}(\UI) = \sum_j {r}_j^{(k)} {\bar{X}}_j(\UI)$ constructed from a given set of measured features. For these measured eigentasks, we find (see Appendix \ref{sec:noisy_ET}) that $\{\bm{r}^{(k)}\}$ specify a unique linear transformation that simultaneously orthogonalizes not only the signal, but also the associated noise: $\EUI{\Es{\bar{y}^{(j)} \bar{y}^{(k)}} } = \delta_{jk}(1+\beta^2_{k}/S)$. The term $\beta^2_k/S$ is thus the mean squared error, or noise power, associated with the approximation of eigentask $y^{(k)}$; equivalently,  $\bar{y}^{(k)}$ has a signal-to-noise ratio of $S/\beta^2_k$. This leads to a natural interpretation of $\{\beta_k^2\}$ as noise-to-signal (NSR) eigenvalues. The eigentasks, ordered in increasing noise strength $0 \leq \beta^2_0 \leq \beta^2_1 \leq \cdots \leq \beta^2_{K-1} < \infty$, are the orthogonal set of functions maximally robust to sampling noise.

\subsection{Example: Resolvable Expressive Capacity and Eigentasks for a Classical Optical Learning System}
\label{sec:photonicRC}

\begin{figure}[t]
    \centering
    \includegraphics[width=\columnwidth]{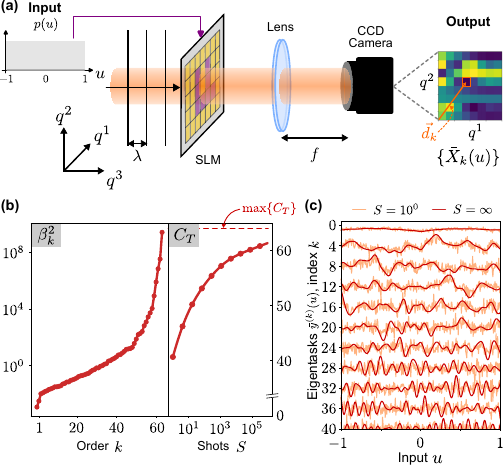}
    \caption{
    (a) 
    A photonic learning system we consider to demonstrate the REC analysis and eigentask construction in a classical optics setup. 
    Inputs $u$ are encoded into the amplitudes and/or phases of propagating light fields of wavelength $\lambda$ via an SLM. The propagating fields are brought to focus via a converging lens of focal length $f$, and the fields are then imaged in the focal plane of the lens using a camera consisting of an array of photodetectors. Output features are given by the integrated stochastic photocurrents measured via these photodetectors. (b) Left panel: Noise-to-signal spectrum $\beta_k^2$ as a function of $k$. Right panel: $\EC$ vs. $\NS$ calculated from the left panel. (c) Eigentasks under infinite sampling ($\NS \to \infty$) and measured eigentasks constructed from features under finite sampling ($\NS=10^0$).
    }
    \label{fig:photonicEC}
\end{figure}

Before presenting more involved examples of physical quantum systems, we discuss an example of the presented framework for noisy \textit{classical} dynamical systems, within the popular PNN platform of photonic ELM \cite{pierangeli_photonic_2021, ortin_unified_2015} and reservoir computing (RC)~\cite{Tanaka2019, dong_optical_2020}.
The specific setup we consider is illustrated in Fig.\,\ref{fig:photonicEC}(a), where computation of inputs $\bm{u}$ is performed via the encoding, propagation, and measurement of propagating electromagnetic (EM) waves in a medium.  Here the entire 3-D space is defined by coordinates $(q^1,q^2,q^3)$, and EM fields of wavelength $\lambda$ propagate in the $q^3$ direction. The electric field distribution is then completely defined by the position vector $\vec{d}$ defined in the plane orthogonal to the propagation direction, so that $\vec{d} = (q^1,q^2)$.~\footnote{We assume the validity of the parabolic approximation here.}

The input embedding of $\bm{u}$ is performed using a spatial light modulator that modulates the amplitude and/or phase of the electric field of the radiation as it passes through. We will restrict this example to 1D inputs $u$ that are uniformly distributed, $p(u)=\mathrm{Unif}[-1,1]$. The scalar $u$ is then mapped to all the pixels of the SLM through a specific mapping discussed in Appendix~\ref{app:photonicRC1}. We consider this rather artificial input encoding for two reasons: for ease of visualization of the computed eigentasks (see \Fig{fig:photonicEC}), and to ensure the distribution is sufficiently sampled. 
In the simulation of the classical optical system we consider here, we choose $N=300$. This is also the number of input samples used in our analysis of qubit-based quantum systems in Sec.\,\ref{sec:ibmq}.

The spatial profile of the electric field $E_0(u;\vec{d})$ following the SLM can be written generally in the form $E_0(u;\vec{d}) \!=\! A_0 \cos (\frac{\varphi_1(u;\vec{d})}{2} ) \exp \! \left\{\! i \!\left(\frac{\varphi_1(u;\vec{d})+2\varphi_2(u;\vec{d})}{2}\right) \!\right\}$, where $A_0$ is the initial electric field amplitude and $\varphi_l(u;\vec{d})$ are input encoding functions~\cite{zhu_arbitrary_2014} (cf.\,\Eqs{eq:inputenc1}{eq:inputenc2}). Following the input encoding, the radiation propagates through free space and then past a thin lens. The electric field in the focal plane of the lens, $E(u;\vec{d})$, can be shown to be related to the initial field $E_0(u;\vec{d})$ via a Fourier transform~\cite{saleh_fundamentals_1991, yariv_photonics_2007}, $E(u;\vec{d}) = \int\!\!\int \dd^2 \vec{d}'~E_0(u;\vec{d}')\exp\left\{\frac{i2\pi}{\lambda f}\left( \vec{d}\cdot\vec{d}' \right) \right\}$, where $f$ is the focal length of the lens. The choice of the optical propagation medium as a lens is again for convenience of analysis, not a limitation. More complex optical systems can be analyzed using the same techniques outlined here. 

Finally, output features are extracted via photodetection (using a CCD camera) in the focal plane of this lens. Modeling this effectively requires us to address the important question of the measurement noise associated with photodetection. First, we consider the camera plane as being comprised of a discrete set of $K=P^2$ photodetectors (here $P=8$), arranged in a $P$-by-$P$ square spatial grid, such that the $k$th photodetector is identified with coordinates $\vec{d}_k = (q^1_k,q^2_k)$, and $k\in[K]$ (See Fig.\,\ref{fig:photonicEC}(a)). This spatial grid ultimately defines the coarse-graining level at which the propagating fields can be probed, and is set by the spatial resolution of the photodetection apparatus, as expected. Then the differential, stochastic photocurrent generated in a given photodetector in a single measurement -- namely the increment in photodetector counts in the time window $[t,t+\dd t]$, which we denote as $\dd N(\vec{d}_k,t)$, follows a Poisson point process (commonly referred to as shot noise)~\cite{wiseman_quantum_2009}. This instantaneous photocurrent $\dd N(\vec{d}_k,t)$ is often integrated over a finite time $T_{\mathrm{int}}$ in so-called integrate-and-dump photodetectors. This defines the single-shot features of this RC scheme, 
\begin{align}
    X^{(s)}_k = \int_{0}^{T_{\mathrm{int}}}~\dd N^{(s)}(\vec{d}_k,t) \in \mathbb{N}.
    \label{eq:XkphotonicRC}
\end{align}
Note that $X^{(s)}_k$ are stochastic, integer quantities, simply counting the total number of photo-generated carriers in a time window $T_{\mathrm{int}}$ of a single measurement.  


We now make the important physical connection between the measured photocurrents and the propagating fields reaching the photodetector. The power incident on the $k$th photodetector is simply set by the Poynting flux of the propagating fields, and is proportional to the electric field intensity, $\alpha|E(u;\vec{d}_k)|^2$, where $\alpha$ is a dimensionful constant that depends for example on the speed of light through the medium of traversal. Then, the expected value of the photocurrent in a time interval $\dd t$ is simply proportional to the incident power, up to a factor $\eta$ that encapsulates the efficiency of photodetection, $\mathbb{E}[\dd N(\vec{d}_k,t)] = \eta \alpha|E(u;\vec{d}_k)|^2~\dd t$.

The complete input-output map defined above fits within our very general framework. In particular, we define measured features $\bar{X}_k(u)$ as $\NS$-shot sample means of $\{X_k^{(s)}\}$, as in Eq.\,(\ref{eq:Xsum}).
Note that in most classical PNN schemes $\NS=1$; here we consider the shot number $\NS \geq 1$. 
We then express the measured features in terms of the decomposition in Eq.\,(\ref{eq:xbar}). First, using Eq.\,(\ref{eq:XkphotonicRC}) and the definition of $x_k(u)$ in Eq.\,(\ref{eq:mapping0}), we find:
\begin{align}
    x_k(u) = \eta\alpha|E(u;\vec{d}_k)|^2 T_{\mathrm{int}}. 
    \label{eq:xCML}
\end{align}
Then, the remaining term in Eq.\,(\ref{eq:xbar}), $\frac{1}{\sqrt{\NS}}\zeta_k(u)$, is a stochastic process with zero mean, and its second-order moment encodes the variance of the Poisson point process in one shot of the experiment, namely that its variance is equal to its mean~(see Appendix~\ref{app:photonicRC} for details),
\begin{align}
    \cu_{jk}(u) = \delta_{jk} x_k(u).
    \label{eq:sigmaCML}
\end{align}
The form of the covariance matrix here is specific to the Poisson nature of the noise process inherited from the classical nature of the source (e.g.\,a coherent light source such as a laser) generating the beam of light. Other types of noise processes will yield distinct covariance matrices, as we will see in examples of quantum systems.

This is an appropriate place to remark that for such a classical description of a physical system for learning, the stochastic photocurrent $\dd N$ for any shot is determined by the \textit{deterministic} electric field incident on the camera plane. In a fully quantum description, on the other hand, the power incident on the photodetectors in a given shot will be determined by the expectation value of the field excitation number operator $\hat{\Psi}^{\dagger}\hat{\Psi}$ (in second-quantized notation) with respect to the \textit{conditional} density matrix $\hat{\rho}^{(c)}$, describing the measurement-conditioned state of the propagating radiation field for that shot. We are compelled to consider a description of this sort when describing physical \textit{quantum} systems in Sec.\,\ref{sec:QSN}. 


Eqs.\,(\ref{eq:xCML}) and (\ref{eq:sigmaCML}) are sufficient to calculate the feature Gram and covariance matrices $\gr$ and $\mathbf{V}$ respectively, as per the discussion following Eq.\,(\ref{eq:EC}). We are thus set up to solve the eigenproblem of Eq.\,(\ref{eq:eigenprob}) and obtain the NSR spectrum and eigentasks for this toy model of a classical optical PNN. We first present the spectrum of NSR eigenvalues $\beta_k^2$ in Fig.\,\ref{fig:photonicEC}(b). The NSR spectrum allows calculation of the REC as a function of $\NS$ using Eq.\,(\ref{eq:EC}); this is shown in Fig.\,\ref{fig:photonicEC}(c). At finite $\NS$, we clearly observe that the REC remains below its upper bound of $K=64$, only approaching it when $\NS$ is increased, reducing the impact of sampling noise on measured features.

Finally, we discuss eigentask construction, also obtained by solving the eigenproblem of Eq.\,(\ref{eq:eigenprob}).  
In Fig.\,\ref{fig:photonicEC}(c) we visualize as a function of $u$ a selection of both the $\NS\to\infty$ eigentasks ${y}^{(k)}(u)$ defined in Eq.\,(\ref{eq:eigentaskproj}), and the measured eigentasks $\bar{y}^{(k)}(u)$ obtained from $\NS = 1$ sampled features (i.e.~single-shot).
We note that the sampled features are obtained by numerically integrating the stochastic differential equation defining independent measurements of the stochastic photocurrents in Eq.~(\ref{eq:XkphotonicRC}). These measured eigentasks exhibit sampling noise, which is evident when compared against the infinite shot eigentasks. 
We clearly see that eigentasks which are higher-order in $k$ are increasingly noisy across the input domain, as also encapsulated by the larger associated NSR eigenvalues. The ordered eigentasks therefore represent the functions that are optimally-resolvable using this classical optical setup in the presence of the sampling noise that it is naturally, and unavoidably, subject to for finite $\NS$.




\hspace{0.5cm}

\section{Learning with Quantum Systems}
\label{sec:QSN}


\subsection{Sampling Noise in Quantum Systems}

Having developed our framework for REC in the most general context, in the remainder of this paper we will use it to analyze \textit{quantum} systems in greater depth. The same quantitative metrics -- REC, eigentasks, and NSR eigenvalues -- now carry the significance of being determined by a parameterized quantum state. To be more specific,
the classical data $\bm{u}$ is now encoded through a quantum channel parameterized by $\bm{\theta}$ acting on a known initial state, 
\begin{align}
    \hat{\rho}(\UI;\bm{\theta}) = \mathcal{U}(\UI;\bm{\theta} )\hat{\rho}_0, \label{eq:qchannel} 
\end{align}
whose data-dependence may be hard to model classically. The quantum channel $\mathcal{U}$ includes all quantum operations applied to the input data; to obtain the computational output or perform further classical processing, one must extract information from the quantum system via a set of measurements described most generally as a positive operator-valued measure (POVM). Specifically, we define a set of $K$ POVM elements \{$\hat{M}_k$\}, each associated with a distinct measurement outcome indexed $k$, and  constrained only by the normalization condition $\sum_{k=0}^{K-1}\hat{M}_k = \hat{\mathbf{I}}$ (and hence not necessarily commuting). 

Each shot then yields a discrete index $k^{(s)}(\UI)$ specifying the observed outcome: for input $\UI$, if outcome $k$ is observed in shot $s$ then $k^{(s)}(\UI) \gets k$. In this case, the single-shot random-valued feature $X^{(s)}_k(\UI)$ is exactly the indicator $\delta(k^{(s)}(\UI), k)$ of index $k$, so that the {\it measured features} are given by:
\begin{align}
    \bar{X}_k(\UI) = \frac{1}{\NS} \sum_s \delta(k^{(s)}(\UI), k). 
    \label{eq:Xsum0}
\end{align}
Hence $\bar{X}_k(\UI)$ in this case is the empirical frequency of occurrence of the outcome $k$ in $S$ repetitions of the experiment with the same input $\UI$. These measured features are formally random variables that are unbiased estimators of the expected value of the corresponding element $\hat{M}_k$ as computed from $\hat{\rho}(\UI)$. Explicitly
\begin{align}
    x_k({\UI}) = {\rm Tr}\{\hat{M}_k \hat{\rho}({\UI};\bm{\theta})\}, \label{eq:mapping}
\end{align}
so that $x_k$ is the probability of occurrence of the $k$th outcome as specified by the quantum state.  These probability amplitudes encompass the accessible information in $\hat{\rho}({\UI};\bm{\theta})$: any observable under this set can be written as a linear combination of POVM elements $\hat{O}_{\bm{W}} = \sum_k W_k \hat{M}_k$, such that $\langle \hat{O}_{\bm{W}} \rangle = \bm{W}^T \bm{x}$.

In quantum machine learning (QML) theory, it is standard to consider the limit $\NS \to \infty$, and to thus use expected features $\{x_k(\UI)\}$ for learning. In any actual implementation however, measured features $\{\bar{X}_k(\UI)\}$ must be constructed under finite $\NS$, in which case their fundamentally quantum-stochastic nature can no longer be ignored. The decomposition Eq.\,(\ref{eq:xbar}) is still applicable $\bar{\bm X}(\UI) =  {\bm x}(\UI) +  \boldsymbol{\zeta}(\UI)/\sqrt{S}$, where now $\bm{x}$ are the quantum-mechanical event
probabilities, and $\boldsymbol{\zeta}$ encodes the multinomial statistics of quantum sampling noise, whose covariance is explicitly
\begin{align}
    \cu_{jk}(\UI) = \delta_{jk} x_k(\UI) - x_j(\UI) x_{k}(\UI). 
    \label{eq:cov}
\end{align}
This is simply the expression for the covariance of multinomial distribution with  $\NS$ trials and $K$ mutually exclusive outcomes  with probabilities $p_k=x_k$. For arbitrary orders of cumulants of multinomial statistics, we refer to Ref.\,\cite{Wishart1949}. For the quadratic loss function considered here, only the cumulants up to second order turn out to be sufficient.

One may wonder what specifically distinguishes a quantum system from a classical stochastic system that can generate a multinomial distribution in its output. 
Firstly, certain $\{ p_k(u) \}$ combinations can be generated efficiently by only a quantum system. That is to say, given equal resources, a quantum system can access some $\{ p_k \}$ that may be inaccessible to any classical stochastic system and hence, as will be discussed later, the accessible space of functions is far richer. 
However, we will also find that resolvability of that function space in $S$ measurements is the key determinant in learning.  
Note that all statistical properties of stochastic readout features $\bar{\bm X}(\UI)$ -- namely first-order cumulants $\bm{x}(\UI)$, second-order cumulants $\bm{\Sigma}(\UI)$, and all higher-order cumulants -- are determined fully by the quantum state $\hat{\rho}({\UI})$, which itself may be hard to generate classically.

To proceed with our REC analysis in quantum systems, we write down the generalized eigenproblem Eq.\,(\ref{eq:eigenprob}) by computing $\gr = \EUI{\bm{x}\bm{x}^T}$ and $\ci = \EUI{\boldsymbol\Sigma}$. Eq.\,(\ref{eq:cov}) enables us to simplify the exact form of $\ci$, namely $\ci = \mathbf{D}-\gr$, where $\mathbf{D} \in \mathbb{R}^{K \times K}$ is a diagonal matrix with elements $\mathbf{D}_{kk} = \EUI{x_k}$.
Alternatively, for any encoding state ensemble $\{ p (\bm{u}) \dd \bm{u}, \hat{\rho} (\bm{u}) \}$, the matrices $\mathbf{D}$ and $\gr$ can be compactly expressed as (see Appendix\,\ref{sec:DefCap})
\begin{align}
    \mathbf{D}_{k k} & = \mathrm{Tr} \{\hat{M}_k \hat{\rho}^{(1)}\}, \\
    \gr_{j k} & = \mathrm{Tr} \{(\hat{M}_j \otimes \hat{M}_{k}) \hat{\rho}^{(2)}\}
\end{align}
by defining the $t$-th order \textit{quantum ensemble moment} $\hat{\rho}^{(t)} = \int \hat{\rho} (\bm{u})^{\otimes t} p (\bm{u}) \dd \bm{u}$ in the $t$-copy space of the quantum state \cite{Harrow2009}.

From Eq.\,\eqref{eq:EC}, we have $\lim_{S\to\infty} \EC = {\rm Rank}\{\gr\}$, where ${\rm Rank}\{\gr\} = K$, the number of measured features, provided no special symmetries exist. This important result reveals that in the absence of sampling noise all quantum systems -- independent of pararameterization -- have a capacity which is simply the number of independent accessible degrees of freedom \cite{dambre_information_2012, hermans_memory_2010}. 
The generic exponential scaling of measured degrees of freedom with the size of the quantum system (e.g.\,$K=2^L$ for $L$-qubit systems subject to a computational basis measurement) is often-cited as a motivator for studying ML with quantum systems~\cite{Kalfus_2022, wright_capacity_2019, martinez-pena_information_2020}. 
However, as will be demonstrated shortly, the REC of quantum systems can be significantly reduced from this limit for finite $S$ in a way that strongly depends on the encoding. By evaluating the ability of quantum systems to accurately express functions in the presence of quantum sampling noise, the capacity analysis above provides an important metric to assess the utility of quantum platforms for learning in practice.

\subsection{Resolvable Expressive Capacity of Quantum 2-designs}
\label{sec:twodesign}

We first consider the REC of quantum 2-designs: systems with fixed $\bm{\theta}$ that map inputs to a unitrary ensemble $\{p(\UI) \dd \UI, \hat{U}(\UI;\bm{\theta})\}$ whose first and second moments agree with those from a uniform (Haar) distribution of unitaries. Quantum 2-designs are important to recent QML studies \cite{cerezo_variational_2021, Holmes2022} due to their role in defining and studying ``expressibility'' \cite{sim_expressibility_2019, wu_expressivity_2021}: a metric quantifying how close a parameterized quantum system is to such a 2-design. The capacity eigenproblem Eq.\,(\ref{eq:eigenprob}) for any quantum 2-design over $K$-dimensions can be solved analytically (see Appendix \ref{app:2designsol}), yielding a flat spectrum of NSR eigenvalues $\beta^2_k = K (1-\delta_{k0})$.  This results in an REC
\begin{align}
    C_T = K \cdot \frac{S + 1}{S + K},
\end{align}
which at finite $\NS$ can be significantly lower than $K$.  For quantum systems with $K=2^L$, all $k\neq 0$ eigentasks have a noise strength $2^L/\NS$, requiring $S$ to grow exponentially with qubit-number $L$ in order to extract useful features.

A quantum 2-design is thought of as having maximal ``expressibility'', however we see that its REC always vanishes exponentially with system size for a fixed finite $S$. 
To emphasize the distinction with ``expressibility'', we note that REC reflects how much classical information can be extracted from  the entire ``quantum computational stack'' in practice: from an abstract algorithm, to the quantum hardware on which its implemented, and the classical electronics used for control and readout.  REC requires only noisy computational outputs $\{\bar{X}_k (\UI) \}$ and is thus efficiently-computable in experiment -- unlike more abstract metrics \cite{sim_expressibility_2019, wu_expressivity_2021, meyer_fisher_2021} -- yielding a directly relevant metric for learning with quantum hardware.

\section{Experimental Results in Quantum systems} 
\label{sec:ibmq}

In this section we discuss the implementation of the Eigentask construction in experiments we carried out on a 7-qubit IBMQ superconducting quantum processor \texttt{ibmq\_perth}. 

\subsection{The Quantum Circuit Ansatz implemented in Experiments}

To demonstrate the practical utility of our framework, we now show how the spectrum $\{\beta_k^2\}$, the REC, and eigentasks can all be computed for real quantum devices in the presence of parameter fluctuations and device noise. 
{We {reiterate} at the outset that our approach for quantifying the REC of a quantum system is very general, and can be applied to a variety of quantum system models.} 
For practical reasons, we perform experiments on $L$-qubit IBM Quantum (IBMQ) processors, whose dynamics is described by a parameterized quantum circuit containing single and two-qubit gates. However, as an example of the broad applicability of our approach, in Appendix \ref{app:H-ansztz} we compute the REC for $L$-qubit quantum annealers via numerical simulations, governed by the markedly different model of continuous-time Hamiltonian dynamics. 

On IBMQ devices, each input $u$ will generate a quantum circuit, hence the maximal number of distinct circuits places a resource constraint on input size. Specially, our experiment and computation is limited to $N=300$ 1D inputs $u$ that are also uniformly distributed, $p(u)=\mathrm{Unif}[-1,1]$, see Fig.\,\ref{fig:Genc1}(a). A 1D distribution then ensures features $\{\bar{X}_k (\UI) \}$ are sufficiently densely sampled to approach the continuum limit, and are also easy to visualize, as in the classical optical RC in Sec.\,\ref{sec:photonicRC}. We emphasize that this analysis can be straightforwardly extended to multi-dimensional and arbitrarily-distributed inputs given suitable hardware resources, without modifying the form of the Gram and covariance matrices.

We are only now required to specify the model of the quantum system, and choose an ansatz  tailored to be natively implementable on IBMQ processors (see Appendix \ref{sec:NISQRC_Architecture_Detail}). 
We fix $\hat{\rho}_0 = \ketbra{0}{0}^{\otimes L}$; note, however, that any other initial state may be implemented via an additional unitary and absorbed into the ``encoding'', i.e.\,the quantum channel $\mathcal{U}(u;\bm{\theta})$ of Eq.\,\eqref{eq:qchannel}. In this way, the dependence of REC on initial states  could be explored in future studies.

The circuit we choose consists of $\tau \in \mathbb{N}$ repetitions of the same input-dependent circuit block depicted in Fig.\,\ref{fig:Genc1}(a). The block itself is of the form $\mathcal{R}_{x}(\bm{\theta}^x/2) \mathcal{W}(J) \mathcal{R}_{z}(\bm{\theta}^z +\bm{\theta}^I u ) \mathcal{R}_{x}(\bm{\theta}^x/2) $, where $ \mathcal{R}_{x/z}$ are Pauli-rotations applied qubit-wise, e.g.\,$ \mathcal{R}_{z} = \bigotimes_{l} \hat{R}_{z}({\theta}^z_l +{\theta}^I _l u) $. A two-qubit coupling gate acts between physically connected qubits in the device and can be written as $\mathcal{W}(J) = \prod_{\langle l, l' \rangle} \mathcal{W}_{l, l'}(J) = \prod_{\langle l, l' \rangle} \mathrm{exp}\{- i \frac{J}{2} \hat{\sigma}^z_{l} \hat{\sigma}^z_{l'}\} $.
Within the structure of this ansatz, we will choose all single-qubit rotation parameters randomly: $\theta^{x/z}_l \sim \mathrm{Unif}[0,2\pi]$ and  $\theta^{I}_l\sim \mathrm{Unif}[0,10\pi]$, generally representing a circuit trained for a particular unspecified task. Each instance of random parameters, along with associated dissipative processes, specifies the quantum channel  $\mathcal{U}(u;\bm{\theta})$ which we refer to as an ``encoding''.  We will study the performance of an overall ansatz by looking at the behavior averaged across encodings as hyperparameters such as $J$ are varied.  In this work we also choose $\tau=3$, which limits circuit depth and associated prevalence of gate errors, while still generating a complex state with correlation generally distributed throughout all qubits.


Finally, we consider feature extraction via a computational basis measurement as is standard in quantum information processing: the POVM elements are the $K=2^L$ projectors $\hat{M}_k = \ketbra{\boldsymbol{b}_k }$, where $\boldsymbol{b}_k$ is the $L$-bit binary representation of the integer $k$. However, as with state preparation, measurements in any other basis can be (and in practice, are) realized using an additional unitary prior to computational basis readout, whose effect can similarly be analyzed as part of the general encoding $\mathcal{U}(u;\bm{\theta})$.


Note that for this ansatz, the choice $J=0~(\mbox{mod } \pi)$ yields either $\mathcal{W}_{l, l'}(J) = \hat{I}$ or $\hat{\sigma}^z \otimes \hat{\sigma}^z$, both of which ensure $\hat{\rho}(u)$ is a product state and measured features are simply products of uncorrelated individual qubit observables -- equivalent to a noisy classical system. Starting from this \textit{product system} (PS), tuning the coupling $J\neq 0~(\mbox{mod } \pi)$ provides a controllable parameter to realize a \textit{quantum correlated system} (CS), for which the $2^L$-dimensional multinomial distribution $\bm{x}(u)$ cannot be represented as a tensor product of $L$ marginal binomial distributions on each qubit. In general, such non-product systems are intuitively expected to result in $u$-dependent quantum states which exhibit entanglement and can potentially be more difficult to describe classically.
This control enables us to address a natural question regarding REC of quantum systems under finite $\NS$: what is the dependence of REC and realizable eigentasks on $J$, and hence on quantum correlations? 

\subsection{Resolvable Expressive Capacity of Quantum Circuits}
\label{subsec:RECqc}

\begin{figure}[t]
    \centering
    \includegraphics[width=\columnwidth]{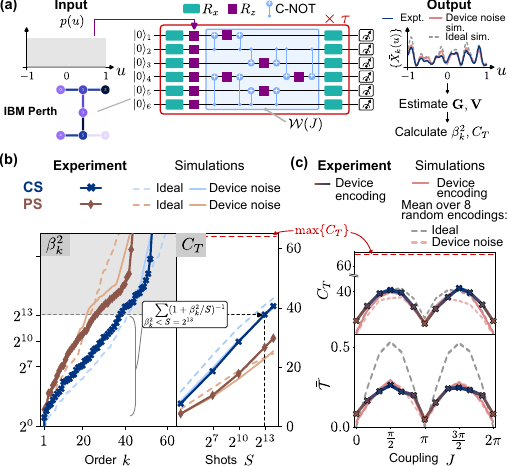}
    \caption{
    (a) A representation of the REC analysis, featuring the IBMQ Perth device and a schematic of the quantum circuit considered in this section.
    On the right, the specific feature plotted is $\bar{X}_1 (u)$ ($\bm{b}_1=000001$) with $S=2^{14}$ shots. 
    (b) Left panel: Device noise-to-signal spectrum $\beta^2_k$ for a specific encoding as a correlated system (CS), $J=\pi/2$ (blue crosses) and product system (PS), $J=0$ (brown diamonds). Ideal (solid) and device noise (dashed) simulations are also shown.  Note the agreement between device and simulation, along with distortion from more direct exponential growth in $\beta^2_k$ with $k$ in the ideal case, due to device errors. Right panel: $\EC$ vs. $\NS$ calculated from the left panel. At a given $\NS$, the $\EC$ can be approximated by performing the indicated sum over all $\bar{\beta}_k^2 < \NS$, where $\bar{\beta}_k^2$ denotes the estimate of $\beta_k^2$ computed from noisy data. (c) Resolvable expressive capacity $\EC$ (top panel) and expected total correlation $\bar{\mathcal{T}}$ (lower panel) for the chosen encoding under $S=2^{14}$ from the IBM device, and device noise simulations (dashed peach). Average metrics over 8 random encodings for device noise (solid peach) and ideal (solid gray) simulations are also shown. The $S\to\infty$ resolvable expressive capacity of these encodings always attains the ${\rm max}\{\EC\}=64$, indicated in dashed red. 
    }
    \label{fig:Genc1}
\end{figure}

To perform the capacity analysis, one must extract measured features from the quantum system as the input $u$ is varied, as exemplified in Fig.\,\ref{fig:Genc1}(a) for the IBMQ \texttt{ibmq\_perth} device.  For comparison, we also show ideal-device simulations (unitary evolution, no device noise), where slight deviations are observed. The agreement with experimental results is improved when the effects of gate errors, readout errors, and qubit relaxation are included, hereafter referred to as ``device noise'' simulations, highlighting both the non-negligible role of device nonidealities, and that our analysis incorporates them.

The measured features under finite $\NS$ are used to estimate the Gram and covariance matrices, and to therefore solve the eigenproblem Eq.\,(\ref{eq:eigenprob}) for NSR eigenvalues $\{\bar{\beta}_k^2\}$ and eigenvectors $\{\bar{\bm{r}}^{(k)}\}$, as estimators of $\{\beta_k^2\}$ and $\{\bm{r}^{(k)}\}$ (see Eqs.\,(\ref{eq:barbeta2k}-\ref{eq:barrk}) in Appendix \ref{app:tilde_correction} for detailed techniques). Typical NSR spectra computed for a random encoding (i.e.~set of rotation parameters) on the device are shown in Fig.\,\ref{fig:Genc1}(b), for $J=0$ (PS) and $J=\pi/2$ (CS), together with corresponding spectra from device noise simulations, with which they agree well. We note that at lower $k$, the device NSR eigenvalues are larger than those from ideal simulations, and at larger $k$ deviate from the direct exponential increase (with order) seen in ideal simulations. Both these effects are captured by device noise simulations as well and can therefore be attributed to device errors and dissipation. The NSR spectra therefore can serve as an effective diagnostic tool for quantum processors and encoding schemes. 


The NSR spectra can be used to directly compute the REC of the corresponding quantum device for finite $\NS$, via Eq.\,(\ref{eq:EC}). Practically, at a given $\NS$ only NSR eigenvalues $\bar{\beta}_k^2 \lesssim \NS$ contribute substantially to the REC. An NSR spectrum with a flatter slope therefore has more NSR eigenvalues below $\NS$, which gives rise to a higher capacity. Fig.\,\ref{fig:Genc1}(b) shows that the CS generally exhibits an NSR spectrum with a flatter slope than the PS, yielding a larger capacity for function approximation across all sampled $\NS$.

To more precisely quantify the role of quantum correlations in REC, we introduce the \textit{expected total correlation} (ETC) of the measured state over the input domain of $u$ \cite{Vedral2002, Modi2010},
\begin{align}
   \bar{\mathcal{T}} = \Eu{ \sum_{l = 1}^L \mathrm{S} ( \hat{\rho}_l^{M} (u) ) - \mathrm{S} (\hat{\rho}^{M} (u) ) }, 
\end{align}
where $\hat{\rho}^{M} (u) \equiv \sum_{k} \hat{\rho}_{kk}(u) \ket{\bm{b}_k}\!\bra{\bm{b}_k}$ is the post-measured state, $\mathrm{S}(\cdot)$ is the von Neumann entropy~(see Appendix \ref{app:QCM}),
and $\hat{\rho}_l = \mathrm{Tr}_{[L] \backslash \{ l \}} \{ \hat{\rho} \}$ is the reduced density matrix obtained by tracing over all qubits except qubit $l$. Therefore, non-zero ETC indicates the generation of quantum states over the input domain $u$ that on average have nontrivial correlations amongst their constituents, including for example pure many-body states that are entangled.

We now compute REC and ETC using $S=2^{14}$ in Fig.\,\ref{fig:Genc1}(c) as a function of $J$, for the same random encoding considered above on the device. We note that the experimental results show excellent agreement in both cases with the corresponding device noise simulation. We also show average REC at $S=2^{14}$ and ETC across 8 random encodings in both ideal and device noise simulations. We find that the influence of individual encodings, i.e.\,random rotation parameters, leads only to small deviations from the overall REC trend when global hyperparameters are held fixed. This implies that no crucial features of the REC are missed by us foregoing fine-tuning (e.g.\,via gradient descent) of individual rotation parameters \textit{in lieu} of sampling them from a given uniform probability distribution.

We note that product states by definition have $\bar{\mathcal{T}}=0$~\cite{nielsen2002quantum}; this is seen in ideal simulations for $J=0~(\mbox{mod}~ \pi)$. However, the actual device retains a small amount of correlation at this operating point, which is reproduced by device noise simulations. This can be attributed to gate or measurement errors as well as cross-talk, the latter being especially relevant for the transmon-based IBMQ platform with a parasitic always-on ZZ coupling \cite{sheldon_procedure_2016}. 
With increasing $J$, $\bar{\mathcal{T}}$ increases and peaks around $J \approx \pi/2~(\mbox{mod } \pi)$; interestingly, $\EC$ also peaks for the same coupling range. From the analogous plot of REC, we clearly see that at finite $S$, increased ETC appears directly correlated with higher REC. We have observed very similar behaviour using completely different quantum system models (see Appendix Fig.\,\ref{fig:app_Features_and_Capacity}~\cite{Giovannetti2006, martinez2021dynamical}). This indicates the utility of enhancing quantum correlations as a means of improving the general expressive capability of quantum systems.

We raise two notes of caution here. First, our analysis across different quantum system implementations has often (though not always) found that a certain threshold number of shots $\NS$ is required before the finite-$\NS$ capacity of a CS overtakes that of the corresponding PS (See Appendix~\ref{app:H-ansztz}). This higher resolvability of functions using a PS under restricted shots may be due to the comparative ease of estimating probabilities from an effectively product distribution, and merits further exploration. At a sufficiently large $\NS$, the increased complexity of the $u$-dependence imposed by the input-output map of a CS results in an REC that eventually surpasses that of the PS.

Secondly, we caution that the connection between measurement correlations and REC is an observed trend, rather than a law derived from first principles.  One can come up with contrived situations where increasing correlation has no effect on REC: for example, appending a layer of CNOT gates directly prior to measurement will generally increase the ideal ETC of any ansatz. For measured features however this amounts to a simple shuffling of labels $x_k(\UI)\leftrightarrow x_{k'}(\UI)$, thus yielding the same NSR spectrum and REC. The input, quantum-state, and feature mapping ultimately governs REC: only increases in correlation that also increase the complexity of the measured features' $u$-dependence (as achieved via the intermediate $\mathcal{W}$ gates here) are beneficial from the perspective of information processing.


As a final important point, note that at finite $\NS$, even with increased quantum correlations, the maximum REC is still substantially lower than the upper bound of $K=64$. This remains true even for ideal simulations, and over several random encodings, so the underperformance cannot be attributed to device noise or poor ansatz choice respectively.
It is worth emphasizing that the impact of device noise is captured in the small REC gap between the ideal and noisy simulation curves, with the remainder of the reduction from $K=64$ attributable to quantum sampling noise alone.
These results clearly indicate that the resulting sampling noise at finite $\NS$ is the fundamental limitation for QML applications on this particular IBM device, rather than other types of noise sources and errors. 

\subsection{A Robust Approach to Learning}
\label{sec:eigentasklearning}

While we have demonstrated the REC as an efficiently-computable metric of general expressive capability of a noisy quantum system, some important practical questions arise. First, does the general REC metric have implications for practical performance on \textit{specific} ML tasks? Secondly, given the limiting -- and unavoidable -- nature of correlated sampling noise, does the REC provide any insights on optimal learning using a particular noisy quantum system and the associated encoding?

\begin{figure}
    \centering
    \includegraphics[scale=1.0]{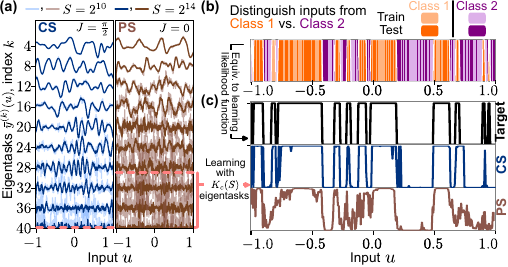}
    \caption{
    (a) Device eigentasks for correlated system (CS, left) and product system (PS, right), constructed from noisy features at $S=2^{10}$ and $S=2^{14}$.
    (b) Classification demonstration on IBMQ Perth. Binary distributions to be classified over the input domain are shown. (c) The classification task can be cast as learning the likelihood function separating the two distributions; this target function is shown in the upper panel. Lower panels show the learned estimate of this target based on the $N_{\rm train}=150$ points shown in (b), using only $K_c(\NS)$ eigentasks for $\NS = 2^{14}$; this cutoff is indicated by the dashed red lines. For the correlated system $K_c(S) = 40$, while for the product system $K_c(S) = 29$.
    }
    \label{fig:Genc2}
\end{figure}

Our formulation addresses both these important questions naturally, as we now discuss. {Recall that} beyond being a simple figure of merit, the REC is precisely the sum of capacities to approximate a particular set of orthogonal functions native to the given noisy quantum system: the eigentasks. {Furthermore}, these eigentasks 
$\bar{y}^{(k)}(u)$ can be directly estimated from a noisy quantum system via the generalized eigenvectors $\{\bar{\bm{r}}^{(k)}\}$, and are ordered by their associated NSR eigenvalues $\{\bar{\beta}_k^2\}$. In Fig.\,\ref{fig:Genc2}(a) show a selection of estimated eigentasks from the device for the CS $(J=\pi/2)$ and PS $(J=0)$ encodings of Fig.\,\ref{fig:Genc1}(b). For both systems, the increase in noise with eigentask order is apparent when comparing two sampling values, $\NS=2^{10}$ and $\NS=2^{14}$. Furthermore, for any order $k$, eigentasks for the PS are visibly noisier than the CS; this is consistent with NSR eigenvalues for PS being larger than those for CS (Fig.\,\ref{fig:Genc1}(b)). The higher resolvable expressive capacity of the CS can be interpreted the ability to accurately resolve more eigentasks at fixed $S$.


The resolvable eigentasks of a finitely-sampled quantum system are intimately related to its performance at specific QML applications. To demonstrate this result, we consider a concrete application: a binary classification task that is not linearly-separable. The domain $u \in [-1,1]$ over which REC was evaluated is separated into two classes, as depicted in Fig.\,\ref{fig:Genc2}(b). A selection of $N_{\rm train}=150$ total samples -- with equal numbers from each class -- are input to the IBMQ device, and 
eigentasks $\{\bar{y}^{(k)}(u^{(n)})\}$ are estimated using $S=2^{14}$ shots. A linear estimator applied to this set of eigentasks is then trained using logistic regression to learn the class label associated with each input. Finally, the trained IBMQ device is used to predict class labels of $N_{\rm test}=150$ distinct input samples for testing. Note that we use the \textit{random} circuits of the previous section to draw more direct comparisons between REC and task performance.  By training only external weights instead of internal parameters $\boldsymbol{\theta}$ we are employing the framework of quantum ELM \cite{mujal_opportunities_2021, wright_capacity_2019, wilson_quantum_2019, innocenti_potential_2022}, which allows one to avoid the computational overhead and difficulty associated with training quantum systems while still achieving comparable performance.

This task can equivalently be cast as one of learning the likelihood function that discriminates the two input distributions, shown in Fig.\,\ref{fig:Genc2}(c), with minimum error. The set of up to $K_{\rm L}$ eigentasks $\{\bar{y}^{(k)}(u)\}_{k \in [K_{\rm L}]}$, where $K_{\rm L} \leq K$, serves as the native orthonormal basis of readout features used to approximate \textit{any} target function using the quantum system. Importantly, the basis is \textit{ordered}, with eigentasks at higher $k$ contributing more noise, as dictated by the NSR eigenvalues $\bar{\beta}_{k}^2$. In particular, at any level of sampling $\NS$, there exists an eigentask order $K_c(\NS)$ after which the NSR eigenvalues $\bar{\beta}_k^2/\NS$ first drops below unity: $K_{c}(S) \equiv \max_k\{\bar{\beta}_{k}^2 < \NS\}$. Heuristically, including eigentasks $k > K_c(\NS)$ should contribute more `noise' to the function approximation task than `signal'. In Fig.\,\ref{fig:Genc2}(c), we plot the learned estimates of the likelihood function using $K_{\rm L} = K_c(\NS)$ eigentasks for both the CS and PS. First, we note that $K_c$ is lower for the PS than the CS; the former has fewer resolvable eigentasks at a given $\NS$. This limitation on resolvable features limits function approximation capacity: the learned estimate of the likelihood function using $K_c$ eigentasks is visibly worse for the PS than the CS. 




\begin{figure}[t]
    \centering
    \includegraphics[width=\columnwidth]{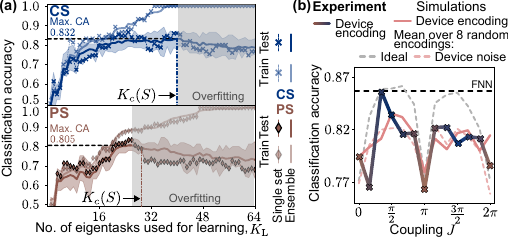}
    \caption{
    (a) Training (light and upwards trending curve) and testing (dark, peaking, then declining curve) accuracy for the 
    device encodings of Fig.\,\ref{fig:Genc2}(a), as a function of the number of eigentasks used to approximate the target function. Markers indicate performance on the dataset shown in Fig.\,\ref{fig:Genc2}(b), and solid lines are the average over $10$ random selections of training and test sets. 
    The shaded region denotes the maximum and minimum test accuracy observed. The optimal test set performance is found near the noise-to-signal cutoff $K_c(S=2^{14})$ (dash-dotted lines) informed by the quantum system's noise-to-signal spectra.
    (b) Testing set classification accuracy as a function of $J$ for our optimal learning method. In all cases, the average performance over the $10$ task permutations is reported, using $K_c(S=2^{14})$. Cross markers indicate device results for the chosen encoding, and the simulation result for this encoding is shown in solid peach. We further perform device noise simulations for a total of $8$ random encodings under finite $\NS$. Dashed peach shows the average testing accuracy over these 8 encodings. Dashed grey in contrast is the average testing accuracy for ideal simulations and in the $S\to\infty$ limit, where all $K=64$ features are used. The horizontal line denotes the performance of a software feed-forward neural network (FNN) with $K_{\rm L}=64$ nodes (and $1153 \gg K_c$ trained parameters) for comparison.
    }
    \label{fig:Genc3}
\end{figure}


In this way, higher REC allows noisy quantum systems to better approximate more functions, which translates to improved learning performance -- this result is explored systemically in Fig.\,\ref{fig:Genc3}(b). Of course, it is natural to ask whether using $K_c(\NS)\leq K$ eigentasks is optimal: exactly this question is investigated in Fig.\,\ref{fig:Genc3}(a), where we plot the training and test accuracy of both device encodings as a function of the number of measured eigentasks $K_{\rm L}$. The performance on the specific training and test set shown in  Fig.\,\ref{fig:Genc2}(b) is indicated with markers, and solid lines indicate the average performance over $10$ distinct divisions of the data into training and test sets. This permutation of the learning task is a standard technique to optimize hyperparameters in ML, and is done here to eliminate the sensitivity of these results to the choice of training set. First note that in all cases, using all eigentasks ($K_{\rm L} = K$) -- or equivalently all measured features $\{\bar{\bm{X}}\}$ -- leads to far lower test accuracy than is found in training.  The observed deviation is a distinct signature of overfitting: the optimized estimator learns noise in the training set (comprised of noisy eigentask estimates $\bar{y}^{(k)}(u^{(n)})$), and thus loses generalizability {to unseen samples} in testing.  

Improvements in model training performance with added features are only meaningful insofar as they also lead to better performance on new data: in both encodings we see test set classification accuracy peaks near $K_c(S)$. This is particularly clear for the averaged results, but even for individual datasets the test accuracy at $K_c(S)$ is within $\approx\!2\%$ of its maximum, thus confirming our heuristic reasoning that eigentasks beyond this order, with an NSR eigenvalues $<\!\!1$, hinder learning. The eigentask-learning approach naturally allows one to decompose the outputs from quantum measurements into a compressed basis with known noise properties, and then select the set of these which exactly captures the resolvable information at a given $S$.  This robust approach to learning enabled by the capacity analysis maximizes the ability of a noisy quantum system to approximate functions without overfitting to noise, in this case fundamental quantum sampling noise.


Finally, Fig.\,\ref{fig:Genc3}(b) shows the classification accuracy for this device encoding as $J$ is varied, where following the above approach, the optimal $K_c(S)$ set of eigentasks are used for each encoding.  We also show the performance of a similar-scale ($K_{\rm L}=64$ node) software neural network and ideal simulations in the $S\to\infty$ limit ($K_c(\infty)=64$) for comparison.  Note that only these infinite-shot results approach the classical neural network, with quantum sampling noise imposing a significant performance penalty even for $J \approx \pi/2~(\mbox{mod } \pi)$. We highlight the striking similarity with Fig.\,\ref{fig:Genc1}(c): encodings with larger quantum correlations and thus higher resolvable expressive capacity will perform generically better on learning tasks in the presence of noise, because they generate a larger set of eigentasks that can be resolved at a given sampling $\NS$. Resolvable Expressive Capacity is \textit{a priori} unaware of the specific problem considered here; this example thus emphasizes its power as a general metric predictive of performance on arbitrary tasks.

\section{Discussion}
\label{sec:discussion}

We have developed a straightforward approach to quantify the resolvable expressive capacity of any physical system in the presence of fundamental sampling noise. Crucially, this analysis extends to physical quantum systems where sampling noise is fundamentally imposed by quantum measurement theory. Our analysis is built upon an underlying framework that determines the native function set that can be most robustly realized by a finitely-sampled physical system: its eigentasks. We use this framework to introduce a methodology for optimal learning that we demonstrate using noisy quantum systems, which centers around identifying the minimal number of eigentasks required for a given learning task. The resulting learning methodology is resource-efficient, and the empirical evidence we provide indicates that it is also robust to overfitting. We demonstrate that eigentasks can be efficiently estimated from experiments on real devices using a limited number of training points and finite shots. We also demonstrate across two distinct qubit-based ans\"atze that the presence of measured quantum correlations enhances resolvable expressive capacity. 

We believe our work opens up several avenues of exploration in the field of learning with physical \textit{quantum} systems in particular. Firstly, our approach provides the tools to understand the limitations of sampling noise in noisy reservoir computing schemes (e.g.\,quantum reservoir computing~\cite{dambre_information_2012, fujii_harnessing_2017, chen_temporal_2020, wright_capacity_2019, garcia-beni_scalable_2022, Kalfus_2022}). In fact, during the final review of the present manuscript, work was posted to the arXiv~\cite{Polloreno2022} exploring limits to noisy reservoir computers using an approach closely aligned with our methods here. Secondly, our work has direct application to the design of circuits for learning with qubit-based systems. In particular, we propose the optimization of resolvable expressive capacity as a meaningful goal for the design of quantum circuits with finite measurement resources. This importantly includes the utilization of the eigentask formulation and eigentask learning as a useful tool for understanding the performance of physical quantum systems in practical learning tasks. Finally, the practical demonstration of our scheme under restrictions of finite input and output samples means that it can prove useful for studies on generalization and training. For example, any difference in REC and eigentasks computed with optimal weights estimated using only a finite number of input samples - as opposed to the ideal but impractical infinite input sampling limit - would constitute a generalization error over the input domain, which one can seek to minimize for optimal learning in future work.


\section*{Acknowledgement}
We express our sincere gratitude to the anonymous reviewers for their invaluable guidance, which significantly contributed to the refinement and enhancement of the final manuscript. We would like to thank Ronen Eldan, Fatih Dinç, Daniel Gauthier, Michael Hatridge, Benjamin Lienhard, Peter McMachon, Sridhar Prabhu, Shyam Shankar, Francesco Tacchino, Logan Wright, Xun Gao for stimulating discussions about the work that went into this manuscript. This research was developed with funding from the DARPA contract HR00112190072, AFOSR award FA9550-20-1-0177, and AFOSR MURI award FA9550-22-1-0203. The views, opinions, and findings expressed are solely the authors' and not the U.S. government's. 

\appendix

\bibliography{bibtex}


\begin{widetext}

\newpage





\startcontents[appendices]
\printcontents[appendices]{l}{1}{\section*{Appendices}\setcounter{tocdepth}{2}}

\makeatletter
\let\toc@pre\relax
\let\toc@post\relax
\makeatother

\newpage

\section{Table of main notations}
\label{app:table}


\begin{table}[htb]
\begin{tabular}{lp{0.72\textwidth}}

\toprule
    \multicolumn{2}{c}
    {\textbf{Abbreviations}}  \\
    \hline
    \hspace{8mm} REC & Resolvable Expressive Capacity, $C_T$ \\
    \hspace{8mm} (Q)ML & (Quantum) Machine Learning \\
    \hspace{8mm} PNN & Physical Neural Network \\
    \hspace{8mm} POVM & Positive Operator-Valued Measure \\
    \hspace{8mm} ELM & Extreme Learning Machine \\
    \hspace{8mm} RC  & Reservoir Computing \\
    \hspace{8mm} SLM & Spatial Light Modulator \\
    \hspace{8mm} NSR & Noise-to-Signal Ratio \\
    \hspace{8mm} PS  & Product System \\
    \hspace{8mm} CS  & Correlated System \\
    \hspace{8mm} ETC & Expected Total Correlation, $\bar{\mathcal{T}}$ \\
    \hline
    \multicolumn{2}{c}
    {\textbf{Symbols and Notation}} \\
    \hline    
    \hspace{8mm} $S$ & Number of shots \\
    \hspace{8mm} $N$ & Number of inputs; for each input we obtain $S$ output samples or shots \\
    \hspace{8mm} $L$ & Number of qubits \\
    \hspace{8mm} $K$ & Number of measured features; $K= 2^L$ for computational-basis projective measurement \\
    \hspace{8mm} $\UI$ & Input \\
    \hspace{8mm} $p$ & Input distribution \\
    \hspace{8mm} $X_k^{(s)}$ & Single-shot random-valued features in any physical system \\
    \hspace{8mm} $\Xs(\UI)$ & Collection of $k$ random-valued features for $S$ shots, $\equiv \{(X_{0}^{(s)}(\UI), X_{1}^{(s)}(\UI), \cdots, X_{K-1}^{(s)}(\UI))\}_{s \in [S]}$ \\
    \hspace{8mm} $\mathcal{D}$ & Complete dataset, $\equiv \{(\bm{u}^{(n)}, \mathcal{X}(\bm{u}^{(n)}))\}_{n \in [N]}$ \\
    \hspace{8mm} $\mathbb{E}_{\Xs}$ & Expectation over the output samples, conditioned on some fixed $\UI$\\
    \hspace{8mm} $\mathbb{E}_{\UI}$ & Expectation over the input, with underlying prior distribution $p(\UI)$, $\EUI{f} \equiv \int \dd \UI \, p(\UI) f(\UI)$\\
    \hspace{8mm} $\bar{X}_k$ & Empirical observed features, $(1/S) \sum_s X_k^{(s)}$ \\
    \hspace{8mm} $x_k$ & Expected features, $\Es{\bar{X}_k}$ \\
    \hspace{8mm} $\zeta_{k}$ & Noise component of $\bar{X}_{k}$ \\
    \hspace{8mm} $\bm{W}$ & General output weights \\
    \hspace{8mm} $\bm{w}$ & Learned optimal output weights for finite-$S$ features $\{\bar{X}_k\}$ \\
    \hspace{8mm} $\mathscr{L}$ & Loss function \\
    \hspace{8mm} $\gr$ & Gram matrix of expected features $\{x_k\}$ \\
    \hspace{8mm} $\ci$ & Expected covariance matrix of random variables $X^{(s)}_k(\UI)$ over input distribution \\
    \hspace{8mm} $\mathbf{D}$ & Expected second-order moment matrix of random variable $X^{(s)}_k(\UI)$ over input distribution, it is diagonal if $X^{(s)}_k(\UI)$ obeys multinomial distribution \\
    \hspace{8mm} $y^{(k)}$ & Eigentasks, $\sum_{k'} r_{k'}^{(k)} x_{k'}$ \\
    \hspace{8mm} $\beta_k^2$ & NSR eigenvalue associated with eigentask $y^{(k)}$ \\
    \hspace{8mm} $\bm{r}^{(k)}$ & Linear combination coefficients of expected features $\{x_{k'}\}$ forming  $y^{(k)}$ \\
    \hspace{8mm} $\bar{\beta}_k^2$ & Finite-$S$ estimate of $\beta_k^2$ \\
    \hspace{8mm} $\bar{\bm{r}}^{(k)}$ & Finite-$S$ estimate of $\bm{r}^{(k)}$ \\
    \hspace{8mm} $\bar{y}^{(k)}$ & Finite-$S$ estimate of eigentasks, $\sum_{k'} r_{k'}^{(k)} \bar{X}_{k'}$ \\
    \hspace{8mm} $\bm{\theta}$ & Quantum system parameters \\
    \hspace{8mm} $\hat{\rho}$ & Generated quantum state \\
    \hspace{8mm} $\mathcal{U}$ & Quantum channel \\
    \hspace{8mm} $\hat{M}_k$ & POVM elements, $\equiv \ketbra{\boldsymbol{b}_k }$ for computational-basis projective measurement\\
    \hspace{8mm} ${\bm b}_k$ & Computational basis eigenstate labels\\
    \hspace{8mm} $k^{(s)}$ & Measurement outcome for shot $s$ \\
    \hspace{8mm} $\hat{\rho}^{M}$ & Diagonal post-measurement state, $\sum_{k}\hat{\rho}_{kk}(\UI) \ket{\bm{b}_k}\!\bra{\bm{b}_k}$\\
    \hspace{8mm} $K_c(\NS)$ \hspace{8mm} & Cutoff index where $\beta_k^2$ approaches $\NS$, $\max_k\{\beta_{k}^2 < \NS\}$\\
    \hline
    \label{tab}
\end{tabular}
\caption{Table of abbreviations and symbols used in main text and appendices}
\end{table}


\section{Feature maps generated by quantum systems}
\label{sec:NISQRC_Architecture_Detail}

\label{DetailsEncodings}

In the main text, we introduce the idea of encoding inputs into the state of a quantum system via a parameterized quantum channel, reproduced below:
\begin{align}
    \hat{\rho}(\bm{u};\bm{\theta}) = \mathcal{U}(\bm{u};\bm{\theta})\hat{\rho}_0,
    \label{appeq:quantumChannel}
\end{align}
one then measures this state to approximate desired functions of the input. 
Fig.\,\ref{fig:Schematic_4Inputs} gives a simple example of this mapping from classical inputs $\UI$ (here in a 2D compact domain) to a quantum state generated by a $\bm{u}$-dependent encoding, and finally to the measured features in a $2$-qubit system undergoing commuting local measurements in the computational basis. The measurement outcomes are therefore bitstrings, of which there are $K= 2^L = 4$, namely: $\bm{b}_k \in \{00,01,10,11\}$. A given shot will yield one of these possible bitstrings.

On the right we plot samples of $S$-shot features $\{\bar{X}_k\}$ constructed for different numbers of shots $S=100, 1000, 10000$ (here we enumerate the feature $k$ with the associated bitstring $\bm{b}_k$). The feature-space for a 2-qubit system is four-dimensional. Owing to the normalization condition $\sum_k \bar{X}_{k} = 1$, only three of these dimensions are independent. For ease of visualization we only plot a two-dimensional projection in the  $\bar{X}_{00} - \bar{X}_{11}$ plane. Each dot in this plot is an average (cf. \Eq{eq:Xsum}) over the associated $S$ shots holding the input $\UI = (u_1, u_2)$ identical over those experiments.

As expressed in Eq.\,\eqref{eq:xbar}, the structure of the noise and thus the correlations in the distribution is determined by the associated quantum state, subject to an overall scaling with $S$. It is important to notice here that as $S\to \infty$ this distribution collapses to a single deterministic point, the corresponding quantum probability $\bm{x}(\UI)$.  It is also evident from this plot that the shape and orientation of these clusters depends on the underlying quantum state $\hat{\rho}(\bm{u};\bm{\theta})$ and associated probabilities $\bm{x}(\UI)$ via Eq.\,\eqref{eq:cov}.  In the remainder of this section, we will consider more complex quantum models, such that they generate mappings which can be useful for learning. A descriptive pseudo-algorithm for learning scheme based circuit-ansatz can be found in Algorithm\,\ref{alg:arc_prob}.

To describe these models, we begin by first limiting to 1-D inputs $u$ as analyzed in the main text; generalizations to multi-dimensional inputs $\bm{u}$ are straightforward. Then, we write Eq.\,(\ref{appeq:quantumChannel}) in the form
\begin{align}
    \hat{\rho}(u;\bm{\theta}) = \hat{U}(u; \boldsymbol{\theta}) \hat{\rho}_0 \hat{U}^{\dagger}(u; \boldsymbol{\theta}).
\end{align}
In the main text, we have considered a model for dynamics of an $L$-qubit quantum system that is natively implementable on modern quantum computing platforms: namely an ansatz of quantum circuits with single and two-qubit gates. We refer to this encoding as the \textit{circuit ansatz} (or \textit{C-ansatz} for short) for which the operator $\hat{U}(u; \boldsymbol{\theta})$ takes the precise form
\begin{align}
    \hat{U}(u; \boldsymbol{\theta}) = \left[ \mathcal{R}_{x} \!\left(\frac{\bm{\theta}^x}{2}\right) \mathcal{W}(J) \mathcal{R}_{z} \!\left(\bm{\theta}^z +\bm{\theta}^I u \right) \mathcal{R}_{x} \!\left(\frac{\bm{\theta}^x}{2} \right)
    \right]^{\tau}~~~~~~~\textit{(C-ansatz)}
    \label{appeq:cansatz}
\end{align}
\begin{figure}
    \centering
    \includegraphics[width = 0.8\columnwidth]{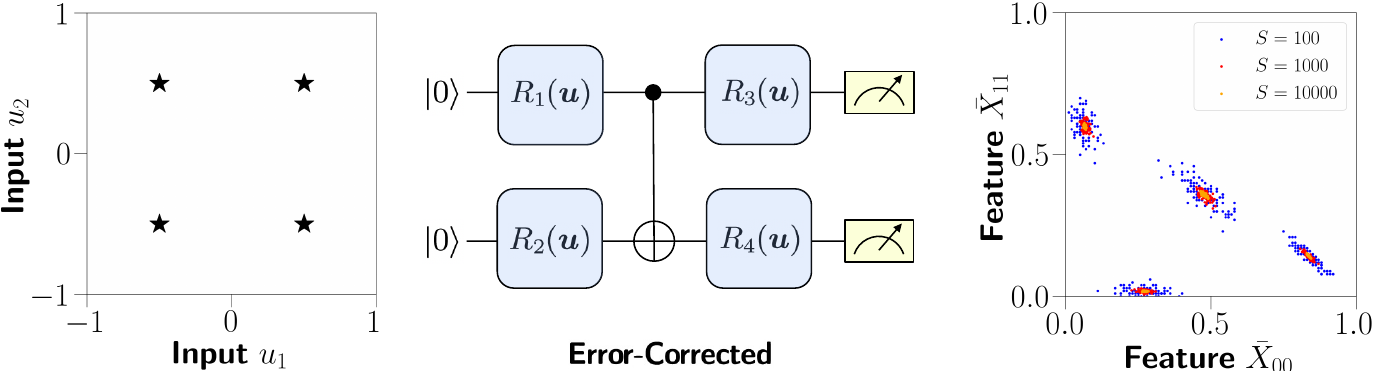}
    \caption{Schematic of a simple $L=2$ qubit circuit, comprised of a CNOT gate sandwiched by input-dependent local $x$-rotation gates $\{R_i(\UI)\}$.  Different $2$D inputs shown on the left are mapped to the finite-$S$ feature space on the right via this circuit. Specifically, a $2$D slice ($\bar{X}_{00}$ and $\bar{X}_{11}$) of the $4$D feature space is shown. Each point represents an individual sample or experiment, i.e. an output constructed with $S<\infty$ shots via Eq.\,\protect\eqref{eq:Xsum}. Distinct values of $S=10^2,10^3,10^4$ are shown in different colors (blue, red, green). For each input $\UI$ and shots $S$, the simulation is repeated $100$ times, resulting in the distribution shown. 
    }
    \label{fig:Schematic_4Inputs}
\end{figure}

For completeness, we recall that $ \mathcal{R}_{x/z}$ are Pauli-rotations applied qubit-wise, e.g.\,$ \mathcal{R}_{z} = \bigotimes_{l} \hat{R}_{z}({\theta}^z_l +{\theta}^I _l u) $, while the coupling gate acts between physically connected qubits in the device and can be written as $\mathcal{W}(J) = \prod_{\langle l, l' \rangle} \mathrm{exp}\{- i \frac{J}{2} \hat{\sigma}^z_{l} \hat{\sigma}^z_{l'}\} $. We emphasize here again that $\tau \in \mathbb{N}^+$ is an integer, representing the number of repeated blocks in the C-ansatz encoding. We note that the actual operations implemented on IBMQ processors also include dynamics due to noise, gate, and measurement errors, and thus must be represented as a general quantum channel as in Eq.\,\eqref{appeq:quantumChannel}. As discussed in the main text, the REC of a quantum system can be computed in the presence of these more general dynamics, and is sensitive to the limitations introduced by them.

An alternative ansatz analyzed here is the \textit{Hamiltonian ansatz} (or \textit{H-ansatz} for short) where the operator $\hat{U}(u; \boldsymbol{\theta})$ describes continuous Hamiltonian dynamics. This ansatz is relevant to computation with general quantum devices, such as quantum annealers and more generally quantum simulators: 
\begin{align}
    \hat{U}(u; \boldsymbol{\theta}) = {\rm exp}\{-i\hat{H}(u)t \},~\hat{H}(u) = \hat{H}_0 + u \cdot \hat{H}_1~~~~~~~\textit{(H-ansatz)}
    \label{eq:Hu=H0+uH1}
\end{align}
Here $t$ is a continuous parameter defining the evolution time; and $\hat{H}_0 = \sum^L_{\langle l,l' \rangle} J_{l, l'} \hat{\sigma}^z_l \hat{\sigma}^z_{l'} + \sum^L_{l=1} h^x_{l} \hat{\sigma}^x_l  + \sum^L_{l=1} h^z_{l} \hat{\sigma}^z_l$ and $\hat{H}_1 = \sum^L_{l=1} h^I_{l} \hat{\sigma}^z_l$. The transverse $x$-field strength $h^x_{l} = \bar{h}^x + \varepsilon^x_{l}$ and longitudinal $z$-drive strength $h^{z,I}_{l} = \bar{h}^{z,I} + \varepsilon^{z,I}_{l}$ are all randomly chosen and held fixed for a given realization of the quantum system,
\begin{align}
    \varepsilon^{x,z,I}_{l} \sim h^{x,z,I}_{\mathrm{rms}}~\mathcal{N}(0, 1),
\end{align}
where $\mathcal{N}(0,1)$ defines the standard normal distribution with zero mean and unit variance. We consider nearest-neighbor interactions $J_{l,l'}$, which can be constant $J_{l,l'} \equiv J$, or drawn from $J_{l,l'} \sim \mathrm{Unif}[0, J_{\rm max}]$, where $\mathrm{Unif}[a,b]$ is a uniform distribution with non-zero density within $[a,b]$. 

As an aside, we note that the C-ansatz quantum channel described by Eq.\,(\ref{appeq:cansatz}) can be considered a Trotterization-inspired implementation of the H-ansatz in Eq.\,(\ref{eq:Hu=H0+uH1}). In particular, if we set $\theta^{x/z/I} = h^{x/z/I} \Delta\cdot\tau$, where $t=\Delta \cdot \tau$, and consider the limit $\Delta \to 0$ while keeping $t$ fixed, Eq.\,(\ref{appeq:cansatz}) corresponds to a Trotterized implementation of Eq.\,(\ref{eq:Hu=H0+uH1}). This correspondence is chosen for practical reasons, but is not necessary in our analysis.


\RestyleAlgo{ruled}
\SetKwComment{Comment}{/* }{ */}
\SetKwInOut{input}{Input}
\SetKwInOut{output}{Output}
\SetKwFor{For}{For}{}{EndFor}
\SetKwFor{If}{If}{}{EndIf}
\begin{algorithm}[t]
    \caption{Measured features under multinomial sampling in quantum system}
    \label{alg:arc_prob}
    \input{$u \in [-1, +1]$}
    \output{$\bar{\boldsymbol{X}}(u)$, which approximates $x_k(u) := \mathrm{Tr}\left\{\hat{\rho}(u) \ket{\boldsymbol{b}_k} \! \bra{\boldsymbol{b}_k}\right\}$}
    \For{$s \gets 1$ to $\NS$}{
        Initialize overall state $\hat{\rho}_{0} \gets \ket{0}\bra{0}^{\otimes \NQ}$\;
        Evolve under quantum channel $\mathcal{U}(u)$: $\hat{\rho}(u) \gets \mathcal{U}(u) \hat{\rho}_0$\;
        Measure all $\NQ$ qubits: $\boldsymbol{b}^{(s)}(u) \gets \boldsymbol{b}_k =  \left(b_{k,1}, b_{k,2} \cdots, b_{k,L}\right) \in \{0, 1\}^{\NQ}$\;
    }
    \For{$k \gets 0$ to $K-1$}{
        Take the ensemble averages as readout features:\\
        \hspace{5mm} $\bar{X}_k(u) \gets \frac{1}{\NS}\sum_{s=1}^{\NS} \delta (\boldsymbol{b}_k, \boldsymbol{b}^{(s)}(u))$ \Comment*[r]{Notice $x_k(u) := \mathrm{Tr}\left\{\hat{\rho}(u) \ket{\boldsymbol{b}_k} \! \bra{\boldsymbol{b}_k}\right\} = \lim_{\NS \to \infty} \bar{X}_k(u)$}
    }
\end{algorithm}

\begin{algorithm}[t]
    \caption{Training of output weights}\label{alg:training}
    \input{$\{ u^{(1)}, \cdots, u^{(N)} \} \in [-1, +1]^N$}
    \output{$\widetilde{\boldsymbol{w}}_N$, such that $y = \widetilde{\boldsymbol{w}}_N \cdot \bar{\boldsymbol{X}}(u)$ can approximate $f(u)$}
    \For{$n \gets 1$ to $N$}{
        Generate features $\bar{\boldsymbol{X}}(u^{(n)})$  through Algorithm 1
    }
    Collect the features into a regression matrix $\regmat \in \mathbb{R}^{N \times K}$\;
    Compute empirical Gram matrix $\bar{\gr} \gets \frac{1}{N} \regmat^T \regmat$ \Comment*[r]{For finite $S$, $\lim_{N \to \infty} \bar{\gr} = \tilde{\gr} := \gr + \frac{1}{S} \ci$}
    Compute target vector $\boldsymbol{Y} \gets \left(f(u^{(1)}), \cdots, f(u^{(N)}) \right)^T$\;
    $\boldsymbol{w}^{\ast} \gets (\regmat^T \regmat)^{-1} \regmat^T \boldsymbol{Y} $ \Comment*[r]{For finite $S$, $\lim_{N \to \infty} \boldsymbol{w}^{\ast} = \boldsymbol{w} :=$ Eq.\,(\ref{eq:wopt})}
\end{algorithm}

\section{Information capacity with sampling noise}
\label{sec:Information_capacity_saturation}

In this Appendix \ref{sec:Information_capacity_saturation}, we provide a detailed \textit{theoretical} construction of the resolvable expessive capacity (REC) analysis. Important related \textit{numerical} techniques required for the \textit{practical} calculation of REC are addressed in Appendix \ref{app:Spectral_finite_statistics}.

We start building the theory by rewriting the functional capacity Eq.\,(\ref{eq:fcap}) into a generalized Rayleigh quotient Eq.(\ref{eq:Rayleigh_quotient}) in Appendix \ref{sec:DefCap}. By solving for the critical point of this generalized Rayleigh quotient, we prove in Appendix \ref{app:eigentasks} that it naturally defines a set of orthonormal functions -- the \textit{eigentasks} $\{y^{(k)}\}$. Each eigentask of the Rayleigh quotient is associated with a particular eigenvalue $\{\beta^2_k\}$, called eigen-noise-to-signal ratio (NSR), whose meaning is interpreted in Appendix \ref{sec:noisy_ET}. In Appendix \ref{app:EC}, we derive a formula $\EC = \sum_{k} 1/(1 + \beta_k^2/S)$ to compute the resolvable expressive capacity, which reproduces the main result Eq.\,(\ref{eq:EC}) in the main text. 
In many practical scenarios in machine learning, like classification problems, there is usually one more nonlinear postprocessing function $\sigma_{\mathrm{NL}}$ acting on $\bm{W}\cdot\bar{\bm{X}}(u)$. We generalize our methodology to those cases in Appendix \ref{app:ComplxNonLin}, via an approximation truncating the third and higher order derivatives of $\sigma_{\mathrm{NL}}$. 
Finally, we give a simplified, equivalent form of eigenproblem in Appendix \ref{app:IntroD} specifically for systems obeying multinomial statistics, which allows solving for both the eigentasks and NSR eigenvalue more simply.

\subsection{Definition of capacity for physical systems with sampling noise}
\label{sec:DefCap}
Suppose an arbitrary probability distribution $p(u)$ for a random (scalar) variable $u$ defined in $D \subseteq \mathbb{R}$. This naturally defines a function space $L^2_p (D)$ containing all functions $f:D \to \mathbb{R}$ with $\int f^2(u) p(u) \dd u < \infty$. The space is equipped with the inner product structure $\langle f_1, f_2 \rangle_{p} = \int f_1(u) f_2(u) p(u) \dd u$. We first review the definition of the function approximation capacity known as the {\it Information Processing Capacity} (IPC) introduced in Ref.~\cite{dambre_information_2012}. This deterministic quantity is based on a metric quantifying the accuracy of a physical system to approximate one of the functions $f_\ell(u)$ of its input through a linear estimator based on its accessible (measurable) degrees of freedom $x_k (u)$:
\begin{equation}
    C [f_\ell] = 1 - \min_{\boldsymbol{W}_\ell \in \mathbb{R}^K}  \frac{\int \left( \sum_{k = 0}^{K-1} W_{\ell k} x_k (u) - f_{\ell} (u) \right)^2 p(u) \dd u}{\int f_{\ell} (u)^2 p(u)  \dd u},
\end{equation}
where functions $f_{\ell}(u)$ are orthogonal target functions $\langle f_{\ell}, f_{\ell'} \rangle_{p} = \int f_{\ell}(u) f_{\ell'}(u) p(u) \dd u = 0$ for $\ell \neq \ell'$. The IPC is defined as $\EC \equiv \sum_{\ell=0}^{\infty} C[f_{\ell}]$, capturing the ability of what type of function the linear combination of physical system readout features can produce. Ref.~\cite{dambre_information_2012} provides an upper bound for the IPC; the IPC of any generic dynamical system is bounded by the accessible degrees of freedom, $\EC \leq K$. 

While this result is quite general, it neglects the limitations due to noise in readout features, which is unavoidable when using physical systems in the presence of finite computational and measurement resources. It is generally accepted that the capacity is reduced in the presence of additive noise, but there are no general results on how to quantify that reduction in the presence of {\it general physical noise}. This is our goal here, to arrive at an exact result for capacity reduction under well-defined but sufficiently general conditions given the physical system. Additional desideratum on this metric is that it provides a practical, calculable metric that can be calculated (1) either from numerical solution of the dynamics of the physical system incorporating a sufficiently accurate noise model, or (2) if desired, from experimental data extracted from said physical system. In the main text, we provide a comparative analysis of these two modalities for a 7-qubit superconducting quantum processor.

We start from considering the noisy readout features $\bar{\bm{X}}(u)$ whose expectations and covariances are
\begin{align}
    \Es{\bar{\boldsymbol{X}} (u)} & \equiv \boldsymbol{x}(u), \\
    \Covs [\bar{\boldsymbol{X}} (u)] & \equiv \frac{1}{\NS} \cu(u),
\end{align}
where the expectation and covariance are evaluated under the product distribution of $S$-shot i.i.d.\,variable $\{X^{(s)}_k(u)\}$. To determine the optimal capacity to compute an arbitrary normalized function $f (u)$ using the noisy readout features $\bar{\bm{X}}(u)$ extracted from the physical system, we need to find an optimal $\boldsymbol{W}$ such that
\begin{equation}
    C [f] = 1 - \frac{\min_{\boldsymbol{W}}  \int \Es{\left( \sum_{k = 0}^{K-1} W_k \bar{X}_k (u) - f (u) \right)^2} p(u) \dd u}{\int f^2 (u) p(u) \dd u}. \label{eq:defIPC}
\end{equation}

By expanding the numerator of the right-hand side for a given, finite number of shots $\NS$, we find
\begin{align}
     & \int f^2 (u) p(u) \dd u - \int \Es{\left( \sum_{k = 0}^{K-1} W_k \bar{X}_k (u) - f (u) \right)^2} p(u) \dd u \nonumber\\
     =~&\! - \sum_{k_1 = 0}^{K-1} \sum_{k_2 = 0}^{K-1} W_{k_1} W_{k_2}  \int \Es{\bar{X}_{k_1} (u) \bar{X}_{k_2} (u)} p(u) \dd u + 2 \sum_{k = 0}^{K-1} W_k  \int \Es{\bar{X}_k (u)} f (u) p(u) \dd u \nonumber\\
    =~& - \sum_{k_1 = 0}^{K-1} \sum_{k_2 = 0}^{K-1} W_{k_1} W_{k_2}  \int \left( x_{k_1} (u) x_{k_2} (u) + \frac{1}{\NS} \cu(u)_{k_1 k_2} \right) p(u) \dd u + 2 \sum_{k = 0}^{K-1} W_k  \int x_k (u) f (u) p(u) \dd u. \label{eq:Lambdatilde}
\end{align}
where the final lines comes from the property of covariance matrix $\Es{\bar{X}_{j} (u) \bar{X}_{k} (u)} - \Es{\bar{X}_{j} (u)} \Es{\bar{X}_{k} (u)} = \frac{1}{\NS} \cu(u)_{j k}$.

The goal of the remaining part of this section is deducing a more compact generalized Rayleigh quotient form of functional capacity. The dependence of readout features $x_k(u)$ on the input $u$ can always be written in the form of a Taylor expansion,
\begin{align}
    x_k (u) = \sum_{j = 0}^{\infty} (\mathbf{T})_{k j} u^j
    \label{eq:TMat}
\end{align}
where we define the \textit{transfer matrix} $\mathbf{T}(\bm{\theta})\equiv \mathbf{T} \in \mathbb{R}^{K \times \infty}$ that depends on the density matrix $\hat{\rho}(u)$, and in particular on parameters $\bm{\theta}$ characterizing the physical system. 
The first term in Eq.\,(\ref{eq:Lambdatilde}) does not depend explicitly on the function $f(u)$ being constructed, and introduces quantities that are determined entirely by the response of the physical system of interest to inputs over the entire domain of $u$. In particular, we introduce the \textit{Gram matrix} $\gr \in \mathbb{R}^{K \times K}$ as
\begin{align}
    (\gr)_{k_1k_2} &= \int x_{k_1}(u) x_{k_2} (u) p(u) \dd u  = \sum_{j_1 = 0}^{\infty} \sum_{j_2 = 0}^{\infty} (\mathbf{T})_{k_1 j_1} \left( \int u^{j_1 + j_2} p(u) \dd u \right) (\mathbf{T})_{k_2 j_2} \equiv (\mathbf{T}\gh\mathbf{T}^T)_{k_1k_2}
\end{align}
where we have also introduced the \textit{generalized Hilbert matrix} $\gh \in \mathbb{R}^{\infty \times \infty}$ as
\begin{equation}
    (\gh)_{j_1 j_2} = \int u^{j_1 + j_2} p(u) \dd u .
\end{equation}
Secondly, we introduce the noise matrix $\ci \in \mathbb{R}^{K \times K}$,
\begin{align}
    (\ci)_{k_1 k_2} & = \int \cu(u)_{k _1k_2}~p(u) \dd u
    \label{eq:V}
\end{align}
The second term in Eq.\,(\ref{eq:Lambdatilde}) depends on $f(u)$ and can be simplified using the $\gh$ matrix as well. Introducing the Taylor series expansion $f (u) = \sum_{j = 0}^{\infty} (\mathbf{Y})_j u^j$
\begin{align} 
    \Eu{x_k f} = \int x_{k} (u) f (u) p(u) \dd u & = \sum_{j_1 = 0}^{\infty} \sum_{j_2 = 0}^{\infty} (\mathbf{T})_{k j_1} \left( \int u^{j_1 + j_2} p(u) \dd u \right) (\mathbf{Y})_{j_2} = (\mathbf{T}\gh\mathbf{Y})_{k}.
\end{align}
With these definitions, Eq.\,(\ref{eq:defIPC}) can be compactly written in matrix form as a Tikhonov regularization problem: 
\begin{equation}
    C[f] =
    1 - \min_{\boldsymbol{W}} \left( \frac{ \left\| \gh^{\frac{1}{2}}\mathbf{T}^T \boldsymbol{W} - \gh^{\frac{1}{2}}\mathbf{Y} \right\|^2 + \frac{1}{\NS} \boldsymbol{W}^T \ci \boldsymbol{W}} {\mathbf{Y}^T\gh\mathbf{Y}} \right). \label{eq:Cf}
\end{equation}
The least-squares form ensures that the optimal value (argmin) $\boldsymbol{w}$ of $\boldsymbol{W}$ has closed form 
\begin{equation}
    \boldsymbol{w} = \left( \gr + \frac{1}{\NS} \ci \right)^{- 1} \Eu{\bm{x}f} = \left( \mathbf{T}\gh\mathbf{T}^T + \frac{1}{\NS} \ci \right)^{- 1} \mathbf{T}\gh\mathbf{Y}. \label{eq:wopt}
\end{equation}
Substituting $\bm{w}$ into the expression for $C$, we obtain the optimal capacity with which a function $f$ can be constructed, which takes the form of a \textit{generalized Rayleigh quotient}
\begin{equation}
    C[f] = \frac{\Eu{\bm{x}^Tf} \left( \gr + \frac{1}{\NS} \ci \right)^{- 1} \Eu{\bm{x}f}}{\Eu{f^2}} = \frac{\mathbf{Y}^T \gh\mathbf{T}^T \left( \gr + \frac{1}{\NS} \ci \right)^{- 1} \mathbf{T}\gh\mathbf{Y}}{\mathbf{Y}^T\gh\mathbf{Y}}. \label{eq:Rayleigh_quotient}
\end{equation}

In scenario of fundamental quantum measurement noise, or quantum sampling noise, given $u$ and $\NS$, the quantum readout features $\bar{X}_k (u) = \frac{1}{\NS} \sum_{s=1}^{\NS} \delta(k^{(s)} (u), k)$ are stochastic variables. The expectation vector and covariance matrix of $\bar{\bm{X}} (u)$ can be expressed in terms of $\hat{\rho} (u)$
\begin{align}
    \Es{\bar{\boldsymbol{X}} (u)} & \equiv \boldsymbol{x}(u) = \mathrm{Tr}\{\hat{M}_k \hat{\rho}(u)\}, \label{eq:Ux}\\
    \Covs [\bar{\boldsymbol{X}} (u)] & \equiv \frac{1}{\NS} \cu(u) = \frac{1}{\NS} \left(\mathrm{diag}(\bm{x})-\bm{x}\bm{x}^T \right). \label{eq:Sigma}
\end{align}
To understand Eq.\,(\ref{eq:Sigma}), we provide brief proofs of some important identities involving the second order statistics of multinomial distribution. In the case of quantum measurement noise, the single-shot random-valued feature is the indicator $X^{(s)}_k (u) = \delta (k^{(s)} (u), k)$ for shot $s$ and input $u$. By definition, for any $s$, the expectation of indicator is always the probability of obtaining index $k$, given the input $u$: $\Es{X^{(s)}_k (u)} = x_k (u)$. If $s \neq s'$, the experiment for different shots must be independent, so we have $\Es{X^{(s)}_k (u) X^{(s')}_{k'} (u)} = x_k (u) x_{k'} (u)$; while if $s = s'$, the mutual exclusion for getting different indices in one shot implies that $\Es{X^{(s)}_k (u) X^{(s)}_{k'} (u)} = \delta_{k k'} x_k (u)$. Thus, we can unify them into one equation
\begin{equation}
    \Es{X^{(s)}_k (u) X^{(s')}_{k'} (u)} = \delta_{s s'} \delta_{k k'} x_k (u) + (1 - \delta_{s s'}) x_k (u) x_{k'} (u) = x_k (u) x_{k'} (u) + \delta_{s s'} (\delta_{k k'} x_k (u) - x_k (u) x_{k'} (u)) .
\end{equation}
Then, the expectation of the product of $\bar{X}_k$ and $\bar{X}_{k'}$ is 
\begin{align}
    \Es{\bar{X}_k (u) \bar{X}_{k'} (u)} & = \frac{1}{S^2} \sum_{s, s'} (x_k (u) x_{k'} (u) + \delta_{s s'} (\delta_{k k'} x_k (u) -  x_k (u) x_{k'} (u))) \nonumber\\
    & = x_k (u) x_{k'} (u) + \frac{1}{S} (\delta_{k k'} x_k (u) - x_k (u) x_{k'} (u)), \label{eq:2ndMat}
\end{align}
which can be directly used to derive Eq.\,(\ref{eq:Sigma}), namely the element $\cu_{k k'}(\UI)$ is:
\begin{align}
    \Es{ [(\bar{X}_k (u) - x_k (u)) (\bar{X}_{k'} (u) - x_{k'} (u))} & = \Es{\bar{X}_k (u) \bar{X}_{k'} (u)} - \Es{\bar{X}_k (u) x_{k'} (u)} -\Es{x_k (u) \bar{X}_{k'} (u)} + \Es{x_k (u) x_{k'} (u)} \nonumber\\
    & = x_k (u) x_{k'} (u) + \frac{1}{S} (\delta_{k k'} x_k (u) - x_k (u) x_{k'} (u)) - x_k (u) x_{k'} (u) \nonumber\\
    & = \frac{1}{S} (\delta_{k k'} x_k (u) - x_k (u) x_{k'} (u)) . 
\end{align}
Therefore for quantum sampling noise, we can express the noise matrix $\ci$ more explicitly,
\begin{align}
    (\ci)_{k_1 k_2} & = \int \cu(u)_{k _1k_2}~p(u) \dd u = \int (\delta_{k_1 k_2} x_{k_1} (u) - x_{k_1}\!(u) x_{k_2}\!(u)) p(u) \dd u \equiv (\mathbf{D})_{k_1k_2}-(\gr)_{k_1k_2}
    \label{eq:defV}
\end{align}
Here we have also introduced the \textit{second-order-moment} matrix $\mathbf{D} \in \mathbb{R}^{K \times K}$ such that
\begin{align}
    (\mathbf{D})_{k_1 k_2} = \delta_{k_1 k_2} \sum_{k} \gr_{k k_1} = \delta_{k_1 k_2} \int x_{k_1} (u) p(u) \dd u. \label{eq:defD}
\end{align}
Then, the noise matrix simply defines the covariance of readout features, and is therefore given by $\ci = \mathbf{D} - \gr$.

In the study of quantum machine learning, it is convenient to define the $t$-th order \textit{quantum ensemble moment} of ensemble $\mathcal{E} = \{ p (\bm{u}) \dd \bm{u}, \hat{\rho} (\bm{u}) \}$ in a $t$-copy space \cite{Harrow2009}:
\begin{equation}
  \hat{\rho}^{(t)} = \int \hat{\rho} (\bm{u})^{\otimes t} p (\bm{u}) \dd \bm{u} .
\end{equation}
The Gram matrix and second-order moment matrix can then be compactly expressed as
\begin{align}
  \mathbf{D}_{k k} & = \int x_k (\bm{u}) p (\bm{u}) \dd \bm{u} = \int \mathrm{Tr} \{\hat{M}_k \hat{\rho} (\bm{u})\} p (\bm{u}) \dd \bm{u} = \mathrm{Tr} \! \left\{ \int \hat{M}_k \hat{\rho} (\bm{u}) p (\bm{u}) \dd \bm{u} \right\} = \mathrm{Tr} \{\hat{M}_k \hat{\rho}^{(1)}\}, \\
  \gr_{k k'} & = \int x_k (\bm{u}) x_{k'} (\bm{u}) p (\bm{u}) \dd \bm{u} = \int \mathrm{Tr} \{\hat{M}_k \hat{\rho} (\bm{u})\} \mathrm{Tr} \{\hat{M}_k \hat{\rho} (\bm{u})\} p (\bm{u}) \dd \bm{u} \nonumber\\
  & = \mathrm{Tr} \! \left\{ (\hat{M}_k \otimes \hat{M}_{k'}) \left( \int \hat{\rho} (\bm{u}) \otimes \hat{\rho} (\bm{u}) p (\bm{u}) \dd \bm{u} \right) \right\} = \mathrm{Tr} \{(\hat{M}_k \otimes \hat{M}_{k'}) \hat{\rho}^{(2)}\}. 
\end{align}

While most of the results in our paper do not utilize this representation, we do find that it provides a compact and natural representation and we use this representation to derive the analytical results in \Sec{sec:twodesign}, see Appendix~\ref{app:2designsol}.

\subsection{Eigentasks}
\label{app:eigentasks}
Eq.\,(\ref{eq:Rayleigh_quotient}) defines the optimal capacity of approximating an arbitrary function $f (u) = \sum_{j = 0}^{\infty} (\mathbf{Y})_j u^j$. We can therefore naturally ask which functions $f$ maximize this optimal capacity. To this end, we first note that the denominator of Eq.\,(\ref{eq:Rayleigh_quotient}) is simply a normalization factor that can be absorbed into the definition of the function $f(u)$ being approximated, without loss of generality. More precisely, we consider:
\begin{align}
    \langle f, f \rangle_p = 1 = \left( \gh^{\frac{1}{2}} \mathbf{Y} \right)^T \left( \gh^{\frac{1}{2}} \mathbf{Y} \right) = \mathbf{Y}^T\gh\mathbf{Y}.
\end{align}
Then, we can rewrite the optimal capacity from Eq.\,(\ref{eq:Rayleigh_quotient}) as
\begin{align}
    C[f] = \mathbf{Y}^T\gh^{\frac{1}{2}} \mathbf{Q} \gh^{\frac{1}{2}}\mathbf{Y}.
    \label{eq:Cp}
\end{align}
Here we have defined the matrix $\mathbf{Q} \in \mathbb{R}^{\infty \times \infty}$ as
\begin{align}
    \mathbf{Q} &= \mathbf{B}\left(\mathbf{I} + \frac{1}{\NS} \mathbf{R} \right)^{- 1} \!\!\! \mathbf{B}^{T}, \label{eq:Q} \\
    \mathbf{B} &= \gh^{\frac{1}{2}} \mathbf{T}^T\gr^{-\frac{1}{2}},
    \label{eq:B} \\
    \mathbf{R} & = \gr^{-\frac{1}{2}}\mathbf{V}\gr^{-\frac{1}{2}}
\end{align}
by introducing the matrix square root of $\gr^{\frac{1}{2}} \in \mathbb{R}^{K \times K}$, and the \textit{noise-to-signal ratio} (NSR) matrix $\mathbf{R}$. The decomposition in Eq.\,(\ref{eq:Q}) may be verified by direct substitution into Eq.\,(\ref{eq:Cp}). The ability to calculate matrix powers and in particular the inverse of $\gr$ requires constraints on its rank. 

Before we analytically find the eigenvectors of $\mathbf{Q}$, we need to show in general that the number of linearly independent features always equals to the rank of the Gram matrix $\gr$, no matter what symmetries the system is subject to. Let us consider any vector $\bm{c}\in \mathbb{R}^K$, the quadratic form
\begin{align}
    \sum_{k_1,k_2=0}^{K-1} c_{k_1} c_{k_2} (\gr)_{k_1, k_2} =  \int \left( \sum_{k_1=1}^{K} c_{k_1} x_{k_1} (u) \right)\!\! \left( \sum_{k_2=1}^{K} c_{k_2} x_{k_2} (u) \right) p(u) \dd u = \left\langle \sum_{k=0}^{K-1} c_{k} x_{k}, \sum_{k=0}^{K-1} c_{k} x_{k} \right\rangle_p. 
\end{align}
gives the norm of function $\sum_{k=0}^{K-1} c_{k} x_{k} (u)$ in RHS. The summation $\sum_{k_1,k_2=1}^{K} c_{k_1} c_{k_2} (\gr)_{k_1, k_2}=0$ vanishes if and only if function $\sum_{k=0}^{K-1} c_{k} x_{k} (u)$ is a zero function, namely $\bm{c}$ is in the null space of $\gr$. We conclude that the rank Gram matrix $\gr$ is equal to the number of linearly independent features. One way for the rank of $\gr$ to be deficient is through the use of the input encoding that leads to identical dependence of the features in a way that is not broken by the rest of the system's interactions (dictated by $\bm{\theta}$). We do not consider such symmetries. All simulations utilize encodings that result in a full rank $\gr$. In case of a rank-deficient $\gr$, one should replace all appearances of  $\gr^{-1}$ above with the pseudoinverse $\gr^{+}$.


We now consider the measure-independent part of the eigenvectors of $\mathbf{Q}$, indexed $\mathbf{Y}^{(k)}$, satisfying the standard eigenvalue problem:
\begin{align}
    \mathbf{Q} \gh^{\frac{1}{2}} \mathbf{Y}^{(k)} = C_k\gh^{\frac{1}{2}}\mathbf{Y}^{(k)}. \label{eq:eigQ}
\end{align}
where $k=0,\cdots,K-1$. From Eq.\,(\ref{eq:Cp}), it is clear that these eigenvectors have a particular meaning. Consider the function $y^{(k)}(u)$ defined by the eigenvector $\mathbf{Y}^{(k)}$, namely
\begin{align}
    y^{(k)}(u) = \sum_{j = 0}^{\infty} \mathbf{Y}^{(k)}_{j} u^j, \label{eq:fkdef}
\end{align}
which we will refer to from now on as \textit{eigentasks}. Suppose we wish to construct the function $y^{(k)}(u)$ using outputs obtained from the physical system defined by $\mathbf{Q}$ in the $\NS\to\infty$ limit (namely, with \textit{deterministic} outputs). At a first glance, before we dive into solving the eigenproblem Eq.(\ref{eq:eigQ}), we do not know any relationship between $y^{(k)}$ and $\bm{x}(u)$.The rest part of this subsection is aiming to  prove that $y^{(k)}$ must be a specific linear combination of features $\bm{x}(u)$. Then, the physical system's capacity for this construction is simply given by the corresponding eigenvalue $C_{k}$, as may be seen by substituting Eq.\,(\ref{eq:eigQ}) into Eq.\,(\ref{eq:Cp}). Formally, the $y^{(k)}(u)$ serves as the \textit{critical point} (or \textit{stationary point}) of the generalized Rayleigh quotient in Eq.\,(\ref{eq:Rayleigh_quotient}). Consequently, the function that is constructed with largest capacity then corresponds to the nontrivial eigenvector with largest eigenvalue. 

To obtain these eigentasks, we must solve the eigenproblem defined by Eq.\,(\ref{eq:eigQ}). Here, the representation of $\mathbf{Q}$ in Eq.\,(\ref{eq:Q}) becomes useful, as we will see that the eigensystem of $\mathbf{Q}$ is related closely to that of the NSR matrix $\mathbf{R}$. In particular, we first define the eigenproblem of $\mathbf{R}$,
\begin{align}
    \nsr \gr^{\frac{1}{2}}\bm{r}^{(k)} & = \beta_k^2 \gr^{\frac{1}{2}}\bm{r}^{(k)} \label{eq:Reigen}
\end{align}
with NSR eigenvalues $\beta_k^2$ and corresponding eigenvectors $\bm{r}^{(k)}$, which satisfy the orthogonality relation $\bm{r}^{(k')T} \gr \bm{r}^{(k)} = \delta_{k, k'}$. Here the $\bm{r}^{(k)}$ can also be computed from the solution to a simpler generalized eigen-problem, where matrix square root operation $\gr^{\frac{1}{2}}$ is not needed:
\begin{align}
    \ci \bm{r}^{(k)} = \beta^2_k \gr \bm{r}^{(k)}. \label{eq:eigenproblem}
\end{align}
This is because $ \ci \bm{r}^{(k)} = \gr^{\frac{1}{2}} \nsr \gr^{\frac{1}{2}} \bm{r}^{(k)} = \beta_k^2 \gr^{\frac{1}{2}} \gr^{\frac{1}{2}} \bm{r}^{(k)} = \beta^2_k \gr \bm{r}^{(k)}$. The prefactor $\gr^{\frac{1}{2}}$ is introduced for later convenience. Eq.\,(\ref{eq:Reigen}) then allows us to define the related eigenproblem
\begin{align}
    \left(\mathbf{I} + \frac{1}{\NS} \mathbf{R} \right)^{- 1}\!\!\! \gr^{\frac{1}{2}}\bm{r}^{(k)}  = \left( 1 + \frac{\beta_k^2}{\NS}  \right)^{-1} \gr^{\frac{1}{2}}\bm{r}^{(k)}
\label{eq:eigRinvSol}
\end{align}
Next, we note that $\mathbf{Q}$ is related to the matrix in brackets above via a \textit{generalized} similarity transformation defined by $\mathbf{B}$, Eq.\,(\ref{eq:Q}). In particular, $\mathbf{B}^T \mathbf{B} = \gr^{-\frac{1}{2}} \gr \gr^{-\frac{1}{2}} = \mathbf{I} \in \mathbb{R}^{K \times K}$, while we remark that $\mathbf{B}\mathbf{B}^T \neq \mathbf{I}$ since it is in $\mathbb{R}^{\infty \times \infty}$. This connection allow us to show that
\begin{align}
    \mathbf{Q} \mathbf{B} \gr^{\frac{1}{2}} \bm{r}^{(k)} = \mathbf{B} \left( \mathbf{I} + \frac{1}{\NS} \mathbf{R} \right)^{- 1}\!\!\! \mathbf{B}^T \mathbf{B} \gr^{\frac{1}{2}} \bm{r}^{(k)} = \frac{1}{1+\beta_k^2/S} \mathbf{B} \gr^{\frac{1}{2}} \bm{r}^{(k)}. 
    \label{eq:eiqQSol}
\end{align}
Comparing with Eq.\,(\ref{eq:eigQ}), we can now simply read off both the eigenvalues and eigenvectors of $\mathbf{Q}$,
\begin{align}
    \left. \begin{array}{rl}
        C_k & = \frac{1}{1+\beta_k^2/\NS}  \\
        \gh^{\frac{1}{2}}\mathbf{Y}^{(k)} & = \mathbf{B} \gr^{\frac{1}{2}} \bm{r}^{(k)} 
    \end{array} \right\}
    \implies \mathbf{Y}^{(k)} = \mathbf{T}^T \bm{r}^{(k)}
\end{align}
where we have used the definition of $\mathbf{B}$ from Eq.\,(\ref{eq:B}). The functions defined by the eigenvectors $\mathbf{Y}^{(k)}$ are automatically orthonormal:
\begin{equation}
    \left\langle y^{(k_1)}, y^{(k_2)} \right\rangle_{p} = \left( \gh^{\frac{1}{2}} \mathbf{Y}^{(k_1)} \right)^T\!\! \left( \gh^{\frac{1}{2}} \mathbf{Y}^{(k_2)} \right) = \boldsymbol{r}^{(k_1)T} \gr^{\frac{1}{2}} \mathbf{B}^T \mathbf{B} \gr^{\frac{1}{2}} \boldsymbol{r}^{(k_2)} = \boldsymbol{r}^{(k_1)T} \gr \boldsymbol{r}^{(k_2)} = \delta_{k_1 k_2}. 
\end{equation}

\subsection{Noisy eigentasks from readout features}
\label{sec:noisy_ET}

We can now also discuss the interpretation of $\{\beta_k^2\}$ for a physical system for which $\{\bm{r}^{(k)}\}$ are known. Consider the evaluation by the physical system (for a given $u$) under finite shots $S$, which yields a single instance of the readout features $\bar{\bm{X}}(u)$. We can simply construct a \textit{noisy} estimator of the $k$th eigentask, $\bar{y}^{(k)}(u)$
\begin{align}
    \bar{y}^{(k)} (u) = \sum_{k' = 0}^{K-1} r_{k'}^{(k)} \bar{X}_{k'} (u)
\end{align}
which is equivalent to requiring the output weights $\bm{W} = \bm{r}^{(k)}$.The corresponding set of noisy function is also orthogonal, this is because $\ci \bm{r}^{(k)} = \beta^2_k \gr \bm{r}^{(k)}$ implies $\bm{r}^{(k)T} \ci \bm{r}^{(k')} = \beta^2_k \delta_{k, k'}$ and hence
\begin{align}
    \Eu{\Es{\bar{y}^{(k_1)} \bar{y}^{(k_2)} }} = \boldsymbol{r}^{(k_1)T} \left( \gr +\frac{1}{S} \ci \right) \boldsymbol{r}^{(k_2)} = \left(1+\frac{\beta^2_k}{S}\right) \delta_{k_1 k_2} \label{eq:baryk1baryk2}
\end{align}
Let us define $\bar{y}^{(k)}(u) = y^{(k)}(u) + \xi^{(k)}(u)$. It means: for each $k$, the noisy eigentask $\bar{y}^{(k)}(u)$ contains a signal part $y^{(k)}(u)$ and a noise part in $\xi^{(k)}(u)$, where the latter one is computed from the linear combination $\xi^{(k)}(u) = \frac{1}{\sqrt{S}} \sum_{k=0}^{K-1} r^{(k)}_{k'} \zeta_{k'}(u)$. One can check $\Eu{\Es{y^{(k_1)} \xi^{(k_2)} }} = \Eu{\Es{\xi^{(k_1)} y^{(k_2)} }} =0$, and 
\begin{align}
    \Eu{ y^{(k_1)} y^{(k_2)} } & =
    \boldsymbol{r}^{(k_1)T} \gr \boldsymbol{r}^{(k_2)} = \delta_{k_1 k_2}, \\
    \Eu{\Es{\xi^{(k_1)} \xi^{(k_2)} }} & = 
    \frac{1}{S} \boldsymbol{r}^{(k_1)T} \ci \boldsymbol{r}^{(k_2)} = \frac{\beta^2_{k_1}}{S} \delta_{k_1 k_2}. \label{eq:ortnoise}
\end{align}
It means that taking linear combinations of $\{x_k(u)\}$ and $\{\zeta_k(u)/\sqrt{S}\}$ with coefficients $\{\bm{r}^{(k)}\in\mathbb{R}^K\}_{k \in [K]}$, not only produces orthonormal eigentasks $\{ y^{(k)}(u) \}$ for signal, but also induces a set of orthogonal noise functions $\{ \xi^{(k)} (u) \}$.

If the physical system can be run multiple times for a given $S$, multiple instances of $\bar{\bm{X}}(u)$ can be obtained, from each of which an estimate of the $k$th eigentask $\bar{y}^{(k)}(u)$ can be constructed. The expectation value of these estimates then simply yields
\begin{align}
    \Es{\bar{y}^{(k)} (u)} = \sum_{k' = 0}^{K-1} r_{k'}^{(k)} \Es{\bar{X}_{k'} (u)} = \sum_{k' = 0}^{K-1} r_{k'}^{(k)} {x}_{k'} (u) = {y}^{(k)}(u)
\end{align}

If we have access to only a single instance of $\bar{\bm{X}}(u)$, however, and thus only one estimate $\bar{y}^{(k)}(u)$ (as $y^{(k)}(u)$ and $\bar{y}^{(k)}(u)$ depicted in Fig.\,\ref{fig:app_Features_and_Capacity}), it is useful to know the expected error in this estimate. This error can be extracted from Eq.\,(\ref{eq:Cf}). In particular, requiring $\mathbf{Y}^{(k)} = \mathbf{T}^T \bm{r}^{(k)}$, we have
\begin{align}
    &\frac{ \left\| \gh^{\frac{1}{2}} \mathbf{T}^T \bm{r}^{(k)} - \gh^{\frac{1}{2}} \mathbf{Y}^{(k)} \right\|^2 + \frac{1}{\NS} \bm{r}^{(k)T} \ci \bm{r}^{(k)} }{\mathbf{Y}^{(k)T} \gh \mathbf{Y}^{(k)}} = \frac{1}{S} \bm{r}^{(k)T} \ci \bm{r}^{(k)} = \frac{\beta_k^2}{S}. 
\end{align}
This mean squared error in using $\bar{y}^{(k)}(u)$ to estimate ${y}^{(k)}(u)$ over the domain of $u$ decreases to zero for $S\to\infty$ as expected, since the noise in $\bar{\bm{X}}$ decreases with $S$. However, $\beta_k^2$ defines the $S$-independent contribution to the error. In particular, this indicates that at a given $S$, certain functions with lower NSR eigenvalues $\beta_k^2$ may be better approximated using this physical system than others. We present in Fig.\,\ref{fig:app_Features_and_Capacity} the measured features $\bar{\bm{X}}$, the eigentasks $\bm{y}$ and their $S$-finite version $\bar{\bm{y}}$ in a 6-qubit Hamiltonian based system. The associated NSR spectrum, resolvable expressive capacity, and total correlations are also depicted for both CS ($J \neq 0$) and PS ($J = 0$) encodings.

\subsection{$S$-shot resolvable expressive apacity: derivation of the bound}
\label{app:EC}
Given an arbitrary set of complete orthonormal basis functions $f_{\ell} (u) = \sum_{j = 0}^{\infty} (\mathbf{Y}_{\ell})_j u^j$,
\begin{equation}
     \langle f_{\ell}, f_{\ell'} \rangle_p = \left( \gh^{\frac{1}{2}} \mathbf{Y}_{\ell} \right)^T \left( \gh^{\frac{1}{2}} \mathbf{Y}_{\ell'} \right) = \delta_{\ell \ell'} . \label{eq:S-shot-eigentask}
\end{equation}
The total capacity is independent of the basis choice
\begin{align}
    \EC(\NS) & = \sum_{\ell = 0}^{\infty} C [f_{\ell}] = \sum_{\ell = 0}^{\infty} \mathbf{Y}_{\ell}^T \gh^{\frac{1}{2}} \left( \gh^{\frac{1}{2}} \mathbf{T}^T \left( \mathbf{T}\gh\mathbf{T}^T + \frac{1}{\NS} \ci \right)^{- 1} \!\!\! \mathbf{T}\gh^{\frac{1}{2}} \right) \gh^{\frac{1}{2}} \mathbf{Y}_{\ell} \nonumber\\
    & = \mathrm{Tr} \left( \gh^{\frac{1}{2}} \mathbf{T}^T \left( \mathbf{T}\gh\mathbf{T}^T + \frac{1}{\NS} \ci \right)^{- 1} \!\!\! \mathbf{T}\gh^{\frac{1}{2}} \right) = \mathrm{Tr} \left( \left( \gr + \frac{1}{\NS} \ci \right)^{- 1} \!\!\! \gr \right) = \sum_{k=0}^{K-1} \frac{1}{1 + \frac{\beta_k^2}{\NS}}. 
\end{align}

\subsection{Eigentask learning training procedure for a nonlinear postprocessing layer}
\label{app:ComplxNonLin}

The definition of REC metric and the training scheme underlying Eigentask Learning considered in the main text is based on a linear estimator $\bm{W} \cdot \bar{\bm{X}}$ fed into a quadratic loss function. Note that in an experimental context, their calculations are performed on a classical processor after the measurement results are collected. This is what is done in the experiments on the superconducting quantum processor in \Sec{sec:ibmq}. This choice of a linear estimator and a quadratic loss function may seem arbitrary, but the rationale behind it, as explained in the main text, is the desire to quantify the function expression capacity of solely the physical system itself, rather than the classical post-processing layer.   

The Eigentask Learning training methodology introduced in the present work is however sufficiently general to be adapted to non-linear post-processing scenarios as well, which is the subject of this Appendix. The central finding is that a cumulant expansion of the non-linear loss function produces extra regularization terms whose magnitudes can be characterized by the NSR spectra $\{\beta^2_k\}$. The non-linear training loss can then be well-approximated using a truncated set of eigentasks. To be more specific, we are going to demonstrate that those eigentasks $y^{(k)}$ whose corresponding $\beta^2_k/S$ is larger should \textit{qualitatively} contribute a larger penalty to the loss function. 

The most general case of the output layer can involve a nonlinear activation function or kernel which may subsequently be fed into a nonlinear loss function. A unified description of these two aspects can be achieved through the usage of a differentiable non-linear function $\sigma_{\mathrm{NL}}$ 
\begin{align}
    \mathscr{L} = \Eu{\Es{\sigma_{\mathrm{NL}}(\bar{\bm{X}})}}. \label{eq:L=EusigmaX}
\end{align} 
For our proof, we do not consider the most general form of $\sigma_{\mathrm{NL}}$, but employ some reasonable assumptions on it that allow us to qualitatively demonstrate the role of eigentasks in the presence of nonlinear post-processing. First, we assume that it is legitimate to truncate a series expansion of $\sigma_{\mathrm{NL}}$ to second order; that is, all terms of the third and higher order derivatives of $\sigma_{\mathrm{NL}}$ are assumed to be much smaller in comparison. Secondly, we assume that the second order derivative (namely the \textit{Hessian matrix}) $\nabla_{\bm{x}}\nabla_{\bm{x}}^T \sigma_{\mathrm{NL}}(u)$ does not vary too strongly with respect to $u$. For mean-square loss in Eq.\,(\ref{eq:fcap}), the Hessian matrix of $\nabla_{\bm{x}}\nabla_{\bm{x}}^T \sigma_{\mathrm{NL}}(u) = 2 \bm{W}\bm{W}^T$ is a constant matrix.
Another typical example of $\sigma_{\mathrm{NL}}$ is the cross-entropy loss function of logistic regression used in the toy binary classification problem considered in \Sec{sec:eigentasklearning}. Here the target function is the conditional probability distribution $f(u) := \mathrm{Prob}[u \in C_1 | \, u]$, where $C_1$ represents the class labeled by $1$. The eventual loss function contains a softmax layer and a cross-entropy function $\mathscr{L} = \Eu{\Es{\mathrm{H} (f(u), \sigma(\bm{W} \cdot \bar{\bm{X}}(u)))}}$ where $\sigma$ is sigmoid function (\textit{e.g.} softmax function $\sigma(z) = 1/(1+\mathrm{exp}(-z))$), and $\mathrm{H}(p,q) = - p \ln q - (1-p) \ln (1-q)$ is the cross-entropy. One can check that $\nabla_{\bm{x}}\nabla_{\bm{x}}^T \sigma_{\mathrm{NL}} = \sigma(\bm{W}\cdot \bm{x}) (1-\sigma(\bm{W}\cdot \bm{x})) \bm{W} \bm{W}^T$, where $\sigma(\bm{W}\cdot \bm{x}) (1-\sigma(\bm{W}\cdot \bm{x})) \in [0, 1/4]$ is a bounded function. 

Suppose the eigentasks $\bar{\bm{y}}$ have been determined by solving the generalized eigenvalue problem \Eq{eq:eigenprob}. We proceed by expressing the non-linear loss function \Eq{eq:L=EusigmaX} in terms of $\bar{\bm{y}}$. This can be done by expressing $\bar{\bm{X}}$ in terms of $\bar{\bm{y}}$, $\bar{\bm{X}} = \bm{\Gamma} \bar{\bm{y}}$, with $\bm{\Gamma}^T = (\bm{r}^{(0)}, \cdots, \bm{r}^{(K-1)})^{-1}$. This is possible by virtue of $\{\bm{r}^{(k)}\}$ being the eigenvectors of the problem Eq.\,(\ref{eq:eigenproblem}). All noisy measured features $\{\bar{X}_k\}$ can now be expressed in terms of the orthogonal signal-basis $\{y^{(k)}\}$ and the noise-basis $\{\xi^{(k)}\}$
\begin{align}
    \bar{X}_{k'}(u) \equiv \sum_{k=0}^{K-1} \Gamma_{k'k} (y^{(k)}(u) + \xi^{(k)} (u)).
\end{align}
Using a cumulant expansion for the non-linear loss function and recalling that $\Es{\xi^{(k)}(u)} = 0$ and $\Es{\xi^{(k)} \xi^{(k')} }=\bm{r}^{(k)T} \mathbf{\Sigma} \bm{r}^{(k')}$ where $\mathbf{\Sigma}$ is the covariance of original sampling noise $\bm{\zeta}$,
\begin{align}
    \mathscr{L} & = \Eu{\Es{\sigma_{\mathrm{NL}} (\bar{\bm{X}})}} = \Eu{\Es{\sigma_{\mathrm{NL}} (\mathbf{\Gamma} \bar{\bm{y}})}} = \Eu{\Es{\sigma_{\mathrm{NL}} \! \left( \sum_k \Gamma_{0, k} (y^{(k)} + \xi^{(k)}), \cdots, \sum_k \Gamma_{K - 1, k} (y^{(k)} + \xi^{(k)}) \right) }} \nonumber\\
    & = \Eu{\sigma_{\mathrm{NL}} (\mathbf{\Gamma} \bm{y})} + \sum_{k = 0}^{K - 1} \Eu{\Es{\frac{\partial \sigma_{\mathrm{NL}}}{\partial y^{(k)}} \xi^{(k)} }} + \frac{1}{2} \sum_{k_1 = 0}^{K - 1} \sum_{k_2 = 0}^{K - 1} \Eu{\Es{\frac{\partial^2 \sigma_{\mathrm{NL}}}{\partial y^{(k_1)} \partial y^{(k_2)}} \xi^{(k_1)} \xi^{(k_2)} }} + \mathcal{O}\!\left(\frac{1}{S^2}\right) \nonumber\\
    & = \Eu{\sigma_{\mathrm{NL}} (\mathbf{\Gamma} \bm{y})} + \sum_{k = 0}^{K - 1} \Eu{\frac{\partial \sigma_{\mathrm{NL}}}{\partial y^{(k)}} \Es{\xi^{(k)}}} + \frac{1}{2} \sum_{k_1 = 0}^{K - 1} \sum_{k_2 = 0}^{K - 1} \Eu{\frac{\partial^2 \sigma_{\mathrm{NL}}}{\partial y^{(k_1)} \partial y^{(k_2)}} \Es{\xi^{(k_1)} \xi^{(k_2)}} } + \mathcal{O}\!\left(\frac{1}{S^2}\right) \nonumber\\
    & = \Eu{\sigma_{\mathrm{NL}} (\mathbf{\Gamma} \bm{y})} + \frac{1}{2} \sum_{k_1 = 0}^{K - 1} \sum_{k_2 = 0}^{K - 1} \Eu{ \frac{\partial^2 \sigma_{\mathrm{NL}}}{\partial y^{(k_1)} \partial y^{(k_2)}} \bm{r}^{(k_1)T} \mathbf{\Sigma} \bm{r}^{(k_2)} } + \mathcal{O}\!\left(\frac{1}{S^2}\right), \label{eq:Lapprox}
\end{align}
We see here that all terms including third and higher order derivatives of $\sigma_{\mathrm{NL}}$ are of $\mathcal{O}\!\left(\frac{1}{S^2}\right)$, which can be neglected in comparison to lower order terms for large enough $\NS$, in accordance with the discussion following Eq.\,(\ref{eq:L=EusigmaX}).
Secondly, the slow variation of the second order derivative $\nabla_{\bm{y}}\nabla_{\bm{y}}^T \sigma_{\mathrm{NL}}(u) = \mathbf{\Gamma}^T \nabla_{\bm{x}}\nabla_{\bm{x}}^T \sigma_{\mathrm{NL}}(u) \mathbf{\Gamma}$ with respect to $u$ allow us to make a further simplification of Eq.\,(\ref{eq:Lapprox}) by taking the mean-value approximation 
\begin{align}
    \Eu{ \frac{\partial^2 \sigma_{\mathrm{NL}}}{\partial y^{(k_1)} \partial y^{(k_2)}} \bm{r}^{(k_1)T} \mathbf{\Sigma} \bm{r}^{(k_2)} } \approx \Eu{ \frac{\partial^2 \sigma_{\mathrm{NL}}}{\partial y^{(k_1)} \partial y^{(k_2)}} } \Eu{\bm{r}^{(k_1)T} \mathbf{\Sigma} \bm{r}^{(k_2)} }
\end{align}
and using the equality $\Eu{\bm{r}^{(k_1)T} \mathbf{\Sigma} \bm{r}^{(k_2)}} = \delta_{k_1 k_2} \beta_{k_1}^2 / S$:
\begin{equation}
   \mathscr{L} \approx 
   \Eu{\sigma_{\mathrm{NL}} (\mathbf{\Gamma} \bm{y})} + \sum_{k = 0}^{K - 1} \frac{\beta_k^2}{S} \cdot \Eu{\frac{\partial^2 \sigma_{\mathrm{NL}}}{(\partial y^{(k)})^2} } =
   \Eu{\sigma_{\mathrm{NL}} (\mathbf{\Gamma} \bm{y})} + \sum_{k} \frac{\beta_k^2}{S} \cdot (\mathbf{\Gamma}^T \Eu{ \nabla_{\bm{x}}\nabla_{\bm{x}}^T \sigma_{\mathrm{NL}}} \mathbf{\Gamma})_{k k}. \label{eq:Lapprox-1}
\end{equation} 

In typical scenarios, such as for instance the case of logistic regression, the loss function depends on the trainable parameters $W_k$ through a linear combination $\bm{W} \cdot \bar{\bm{X}}(u)$. For such scenarios, it proves convenient to introduce $\bm{\Omega}$ such that $\bm{\Omega} = \bm{\Gamma}^T \bm{W}$
\begin{align}
    \bm{W} \cdot \bar{\bm{X}}(u) \equiv \sum_{k=0}^{K-1} \Omega_{k} \cdot (y^{(k)}(u) + \xi^{(k)} (u)) ,
\end{align}
For any loss function of the form $\sigma_{\mathrm{NL}}(\bm{x}) = g(\bm{W} \cdot \bm{x})$, the second term of Eq.\,(\ref{eq:Lapprox-1}) can then be expressed as $\mathbf{\Gamma}^T \nabla_{\bm{x}}\nabla_{\bm{x}}^T \sigma_{\mathrm{NL}} \mathbf{\Gamma} = g''(\bm{\Omega}\cdot \bm{y}) \mathbf{\Omega} \mathbf{\Omega}^T$. In the case of logistic regression $\sigma_{\mathrm{NL}}(\bm{x}) = g(\bm{W} \cdot \bm{x}) = \operatorname{H} (f, \sigma(\bm{W} \cdot \bm{x}))$. The final loss function can be further simplified to 
\begin{align}
    \mathscr{L} & \approx 
    \Eu{\operatorname{H} \! \left(f, \sigma\!\left(\bm{\Omega} \cdot \bm{y}\right)\right)} + \left(\sum_{k=0}^{K-1} \frac{\beta_k^2}{\NS} \Omega^2_{k}\right) \cdot \Eu{ \sigma(\bm{\Omega} \cdot \bm{y}) (1-\sigma(\bm{\Omega} \cdot \bm{y})) }. \label{eq:LapproxCls}
\end{align}
Here the first term is the $S$-infinity value of the loss function, $\lim_{S \to \infty} \mathscr{L} = \Eu{\operatorname{H} \! \left(f, \sigma\!\left(\bm{\Omega} \cdot \bm{y}\right)\right) }$. For the softmax function, we have $\sigma(\bm{\Omega} \cdot \bm{y}(u)) (1-\sigma(\bm{\Omega} \cdot \bm{y}(u)))\in[0,1/4]$, which is assumed not to fluctuate strongly as a function of $u$, compared to the fluctuation of $\bm{r}^{(k_1)T} \mathbf{\Sigma}(u) \bm{r}^{(k_2)}$.

While the final result \Eq{eq:LapproxCls} is obtained under certain assumptions on $\sigma_{\mathrm{NL}}$ as detailed at the outset, the form of the last term suggests the interpretation of $\beta_k^2 / \NS$ as a natural regularization, and the use of $\beta_k^2 / \NS$ as a metric for truncation. This truncation is successfully employed in calculations behind the generation of plots in \Fig{fig:Genc2} and \Fig{fig:Genc3} for the binary classification task, using the cross-entropy loss function of logistic regression. \Eq{eq:LapproxCls} therefore provides some theoretical justification for the use of this truncation scheme when a nonlinear post-processing layer is employed.

\subsection{Simplifying the noise-to-signal matrix and its eigenproblem for quantum systems}
\label{app:IntroD}

We have shown that the problem of obtaining the eigentasks for a generic quantum system, and deducing its resolvable expressive capacity under finite measurement resources, can be reduced simply to solving the eigenproblem of its NSR matrix $\mathbf{R}$, Eq.\,(\ref{eq:Reigen}). Note that constructing $\mathbf{R}=\gr^{-\frac{1}{2}}\mathbf{V}\gr^{-\frac{1}{2}}$ requires computing the inverse of $\gr$. However, $\gr$ can have small (although always nonzero) eigenvalues, especially for larger systems, rendering it ill-conditioned and making the computation of $\mathbf{R}$ numerically unstable. Fortunately, certain simplifications can be made to derive an equivalent eigenproblem that is much easier to solve. We begin by employing the fact that the second-order moment matrix $\mathbf{D}$ of multinomial sampling is diagonal.
In particular,
\begin{align}
    (\mathbf{D})_{k_1k_2} = \left\{
    \begin{array}{cc}
        \sum_{k = 0}^{K-1} (\gr)_{k k_1}, &\text{if } k_1=k_2 \\
        0, & \text{if } k_1 \neq k_2
    \end{array} \right.
\end{align}

Using $\mathbf{V} =\mathbf{D}-\gr$, we can rewrite the eigenproblem for $\mathbf{R}$,
\begin{align}
    \nsr \left( \gr^{\frac{1}{2}}\bm{r}^{(k)} \right) & = \beta_k^2 \gr^{\frac{1}{2}}\bm{r}^{(k)} \nonumber \\
    \implies \gr^{-\frac{1}{2}}(\mathbf{D}-\gr)\gr^{-\frac{1}{2}} \left( \gr^{\frac{1}{2}}\bm{r}^{(k)} \right) &= \beta_k^2 \gr^{\frac{1}{2}}\bm{r}^{(k)} \nonumber \\
    \implies \gr^{-1}\mathbf{D} \bm{r}^{(k)} &= (1+\beta_k^2) \bm{r}^{(k)}
\end{align}
Finally, considering the inverse of the matrix on the left hand side, we obtain the simplified eigenproblem for the matrix $\mathbf{D}^{-1}\gr$,
\begin{align}
    \mathbf{D}^{-1}\gr \bm{r}^{(k)} = (1+\beta_k^2)^{-1} \bm{r}^{(k)} \equiv \alpha_k \bm{r}^{(k)},
    \label{eq:DinvGeigen}
\end{align}
which shares eigenvectors with $\nsr$, and whose eigenvalues are a simple transformation of the NSR eigenvalues $\beta_k^2$. Importantly, constructing $\mathbf{D}^{-1}\gr$ no longer requires calculating any powers of $\gr$, and it relies only on the straightforward inversion of a diagonal matrix $\mathbf{D}$.

\section{Spectral analysis based on finite statistics in quantum systems}
\label{app:Spectral_finite_statistics}

While Eq.\,(\ref{eq:DinvGeigen}) is a numerically simpler eigenproblem to solve than Eq.\,(\ref{eq:Reigen}), it still requires the approximation of $\gr$ (recall that $\mathbf{D}$ can be obtained from $\gr$) from readout features $\bar{\bm{X}}(u)$ under finite sampling of the input ($N$) and finite shots ($\NS$). To be more precise, in experiment one only has access to measured features sampled at finite-$S$ $\bar{\bm{X}}$ (indeed, this distinction is the underlying premise of this article).  However, in Eq.\,\eqref{eq:eigenprob} $\gr$ and $\ci$ are defined with respect to the ideal $\bm{x}$. Let $\widetilde{\gr} \equiv \EUI{\Es{\bar{\bm{X}}\bar{\bm{X}}^T}}$ and $\widetilde{\ci} \equiv \EUI{\Es{\mathrm{diag}(\bar{\bm{X}})-\bar{\bm{X}}\bar{\bm{X}}^T}}$. The objective of Appendix \ref{app:tilde_correction} is showing that the eigen-analysis $\{\beta^2_k, \bm{r}^{(k)}\}$ can be accurately expressed with
\begin{align}
    \beta_k^2 = \frac{S \cdot \tilde{\beta}_k^2}{(S-1)-\tilde{\beta}_k^2}, \label{eq:btilde}
\end{align}
and $\bm{r}^{(k)} = \tilde{\bm{r}}^{(k)}$ from solving generalized eigenvalue problem $\widetilde{\ci} \tilde{\bm{r}}^{(k)} = \tilde{\beta}_k^2 \widetilde{\gr} \tilde{\bm{r}}^{(k)}$. In what follows, we show how an approximation $\widetilde{\gr}_N$ of $\gr$ can be constructed from finitely-sampled readout features, as relevant for practical quantum devices. Secondly, we also describe an approach in Appendix \ref{sec:GramMatrix-free} to obtain the eigentasks $y^{(k)}(u)$ and corresponding NSR eigenvalues $\beta_k^2$ in a singlular-value decomposition (SVD) problem that avoids explicit construction of the Gram matrix, and is thus numerically even more robust.

\subsection{Approximating eigentasks and NSR eigenvalues under finite $S$ and $N$}
\label{app:tilde_correction}

For practical computations, readout features $\bar{\bm{X}}(u)$ from the quantum system for finite $\NS$ can be computed for a discrete set of $u^{(n)} \in [-1,1]$ for $n = 1,\ldots,N$. Labelling the corresponding readout features $\bar{\bm{X}}(u^{(n)})$, we can define the \textit{regression matrix} constructed from these readout features,
\begin{equation}
    \regmat \equiv (\bar{\bm{X}} (u^{(1)}), \bar{\bm{X}} (u^{(2)}), \cdots, \bar{\bm{X}} (u^{(N)}))^T = \left(\begin{array}{ccc}
        \bar{X}_0 (u^{(1)}) & \cdots & \bar{X}_{K-1} (u^{(1)})\\
        \vdots &  & \vdots\\
        \bar{X}_0 (u^{(N)}) & \cdots & \bar{X}_{K-1} (u^{(N)})
    \end{array}\right) . \label{eq:regmat}
\end{equation}
Here, $\regmat \in \mathbb{R}^{N\times K}$, with subscript $N$ indicating its construction from a finite set of $N$ inputs, is a random matrix due to the stochasticity of readout features; in particular it can be written as:
\begin{align}
    \regmat = \RM  + \frac{1}{\sqrt{\NS}}\mathbf{Z}(\RM)
\end{align}
where $(\RM)_{nk} = \Es{\bar{X}_k(u^{(n)})} = x_k(u^{(n)})$, and $\mathbf{Z}$ is the centered multinomial stochastic process, so that $\Es{\widetilde{\mathbf{F}}_N} = \RM$.

Using this regression matrix $\regmat$, we can obtain an estimation of the Gram matrix and second order moment matrix, which we denote $\widetilde{\gr}_N$ and $\widetilde{\mathbf{D}}_N$, and whose matrix elements are defined via
\begin{align}
     (\widetilde{\gr}_N)_{k_1 k_2} & \equiv \frac{1}{N}\sum_{n=1}^N \bar{X}_{k_1}(u^{(n)})\bar{X}_{k_2}(u^{(n)}) = \frac{1}{N} (\regmat^T \regmat)_{k_1k_2} \approx \int \bar{X}_{k_1}(u)\bar{X}_{k_2}(u) p(u) \dd u, \\
     (\widetilde{\mathbf{D}}_N)_{k_1 k_2} & \equiv \delta_{k_1, k_2} \frac{1}{N}\sum_{n=1}^N \bar{X}_{k_1}(u^{(n)}) \approx \delta_{k_1, k_2} \int \bar{X}_{k_1}(u) p(u) \dd u.
\end{align}
While the quantities $\widetilde{\gr}_N$ and $\widetilde{\mathbf{D}}_N$ are computed from stochastic readout features, their stochastic contributions are suppressed in the large $N$ limit by the Hoeffding inequality for sums of bounded stochastic variables. 

In what follows, our goal is to prove that by solving the eigenproblem $\widetilde{\mathbf{D}}_N^{-1} \widetilde{\gr}_N \tilde{\bm{r}}_{N}^{(k)} = (1 + \tilde{\beta}_{N, k}^2)^{-1} \tilde{\bm{r}}_{N}^{(k)}$, the true NSR eigenvalue $\beta_k^2$ and eigentask coefficients $\bm{r}^{(k)}$ can be well approximated by $\NS \tilde{\beta}_{N,k}^2/(\NS - \tilde{\beta}_{N,k}^2 - 1)$ and $\tilde{\bm{r}}_{N}^{(k)}$, respectively. In particular, to achieve the goal stated above, the first step is still taking the $N \to \infty$ limit and defining the deterministic limit of $\widetilde{\gr}_N$ and $\widetilde{\mathbf{D}}_N$, according to Eq.\,(\ref{eq:2ndMat}), as
\begin{align}
    \widetilde{\gr} &\equiv \lim_{N \to \infty} \frac{1}{N} (\regmat^T \regmat) = \gr + \frac{1}{\NS} \ci = \gr + \frac{1}{\NS} (\mathbf{D}-\gr), \label{eq:Gtilde1}
    \\ \widetilde{\mathbf{D}} &\equiv \lim_{N \to \infty} \widetilde{\mathbf{D}}_N = \mathbf{D}. \label{eq:Dtilde1}
\end{align}
In a generic physical system, the covariance matrix $\ci$ can be reconstructed by computing the empirical covariance of measured features and employing the well-known rule of \textit{Bessel's correction}, $\Es{ \frac{1}{S} \sum_s (X^{(s)}_k (u) - \bar{X}_k (u)) (X^{(s)}_{k'} (u) - \bar{X}_{k'} (u)) } = \frac{S-1}{S} \mathbf{\Sigma} (u)$. In this way, $\mathbf{V}$ can be approximated from the whole record of single-shot random-valued features $\Xs(u) = \{ X_k^{(s)} (u) \}_{k \in [K], s \in [S]}$
\begin{equation}
    \lim_{N \rightarrow \infty} \frac{1}{N (S - 1)} \sum_{n = 1}^N \sum_{s = 1}^S (X^{(s)}_k (u^{(n)}) - \bar{X}_k (u^{(n)})) (X^{(s)}_{k'} (u^{(n)}) - \bar{X}_{k'} (u^{(n)})) = \mathbf{V}_{k k'}. \label{eq:Bessel}
\end{equation}
However, in the scenario of quantum sampling noise, the covariance matrix has a special structure. In particular, $\cu$ is not independent of the first order moments $\bm{x}(u)$. This allows us to invert Eq.\,(\ref{eq:Gtilde1}) and Eq.\,(\ref{eq:Dtilde1}) to express the Gram matrix $\gr$ and second-order moment matrix $\mathbf{D}$ in terms of the estimators $\widetilde{\mathbf{G}}$ and $\widetilde{\mathbf{D}}$ computed using a finite number of shots $\NS$, in a numerically cheap way without using full readout record of $\Xs(u) = \{ X_k^{(s)} (u) \}_{k \in [K], s \in [S]}$,
\begin{align}
    \gr &= \frac{\NS}{\NS-1}\widetilde{\mathbf{G}} - \frac{1}{\NS-1}\widetilde{\mathbf{D}},  \\
    \mathbf{D} &= \widetilde{\mathbf{D}}.
\end{align}
We see that to lowest order in $\frac{1}{\NS}$, $\gr \approx \widetilde{\gr}$ and $\mathbf{D} \approx \widetilde{\mathbf{D}}$, which is what one might expect naively. However, we clearly see that the estimation of $\gr$ can be improved by including a higher-order correction in $\frac{1}{\NS}$. This contribution arises due to the highly-correlated nature of noise and signal for quantum systems: we are able to estimate the noise matrix $\widetilde{\mathbf{G}}$ and $\widetilde{\mathbf{D}}$ using knowledge of the readout features, and correct for the contribution to $\widetilde{\gr}$ and $\widetilde{\mathbf{D}}$ that arises from this noise matrix. We will see that this contribution will be important in more accurately approximating quantities of interest derived from $\gr$, $\mathbf{D}$.

\begin{figure}[t]
    \centering
    \includegraphics[width=0.5\textwidth]{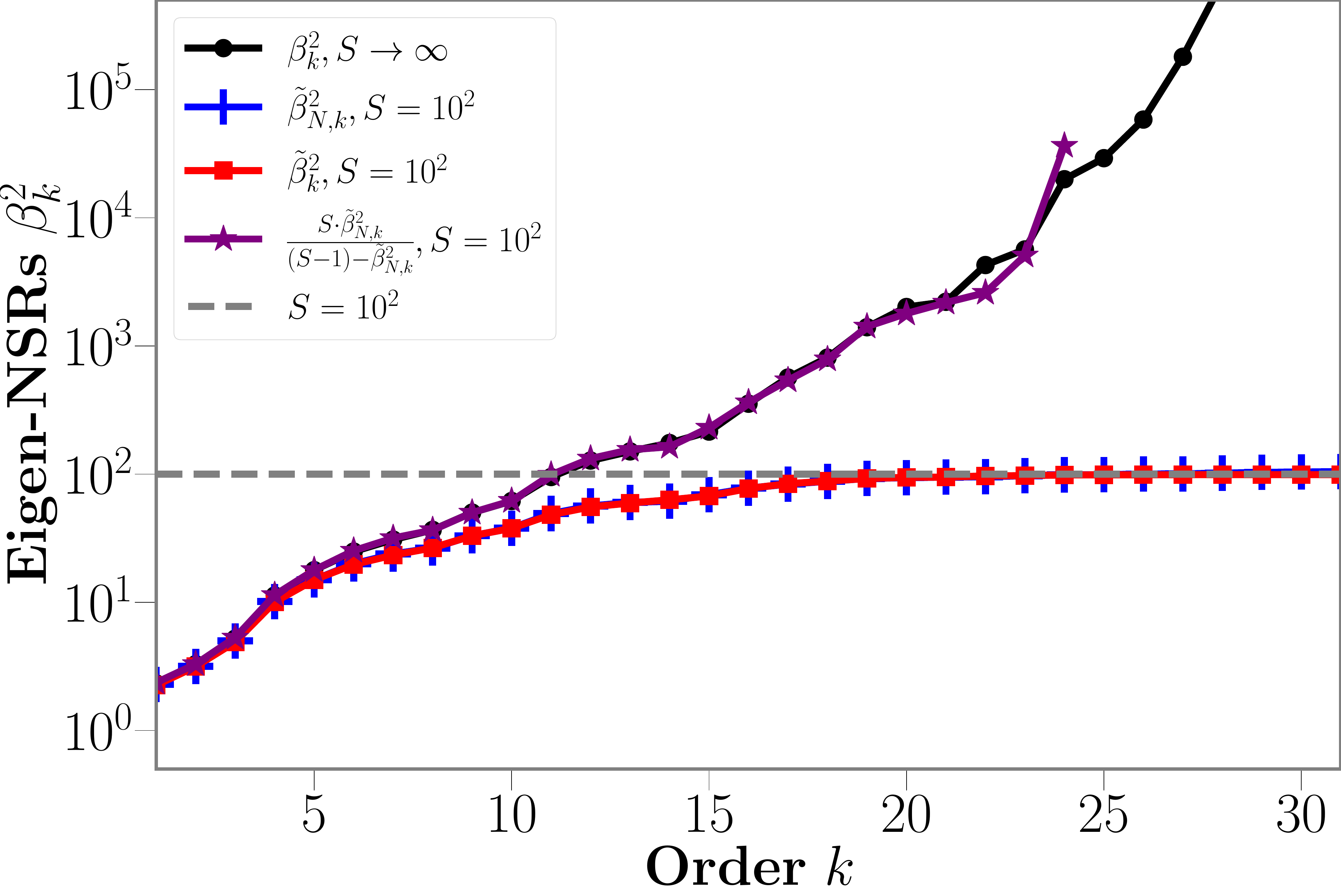}
    \caption{Eigen-analysis in $L=5$ H-ansatz system by taking $\NS=10^2$ shots on each of $N = 10^4$ input samples, with true NSR eigenvalues  $\beta_k^2$ (black), $\NS$-finite sampled $\tilde{\beta}_{N,k}^2$ (blue) and corrected NSR $(\NS \cdot \tilde{\beta}_{N,k}^2)/((\NS - 1) - \tilde{\beta}_{N,k}^2)$ (purple). $\tilde{\beta}_k^2$, the large $N$ limit of $\tilde{\beta}_{N,k}^2$ is also plotted in red for comparison. The data correction is necessary since all $\tilde{\beta}_{N,k}^2$ are below the gray dashed line (representing $S=10^2$), and the corrected data (in purple) show much better performance even if $\beta_k^2 \gg \NS$. The corrected data (in purple) has a cutoff at $k=25$, because in this example all sampled $\tilde{\beta}_{N,k}^2$ with $k > 25$ are larger than $S-1$, and hence are not correctable. 
    }
    \label{fig:NSR_Tilde_Renormalization}
\end{figure}

To this end, we recall that our ultimate aim is not just to estimate $\gr$ and $\mathbf{D}$, but to solve the eigenproblem of Eq.\,(\ref{eq:DinvGeigen}). Using the above relation, we can then establish $\widetilde{\mathbf{D}}^{-1}\widetilde{\mathbf{G}} = \frac{\NS-1}{\NS}\mathbf{D}^{-1}\gr +\frac{1}{\NS} \mathbf{I}$, and write Eq.\,(\ref{eq:DinvGeigen}) in a form entirely in terms of $\widetilde{\mathbf{G}}$ and $\widetilde{\mathbf{D}}$,
\begin{align}
    \mathbf{D}^{-1}\gr \bm{r}^{(k)} &= (1+\beta_k^2)^{-1} \bm{r}^{(k)}, \nonumber \\
    \implies \widetilde{\mathbf{D}}^{-1}\widetilde{\mathbf{G}}\bm{r}^{(k)} &= \left[\frac{\NS-1}{\NS}(1+\beta_k^2)^{-1} +\frac{1}{\NS} \right]\bm{r}^{(k)}. 
\end{align}
Note that the final form is conveniently another eigenproblem, now for the finite-$\NS$ matrix $\widetilde{\mathbf{D}}^{-1}\widetilde{\mathbf{G}}$:
\begin{align}
    \widetilde{\mathbf{D}}^{-1}\widetilde{\mathbf{G}} \tilde{\bm{r}}^{(k)} = (1+\tilde{\beta}_k^2)^{-1} \tilde{\bm{r}}^{(k)} \equiv \tilde{\alpha}_k \tilde{\bm{r}}^{(k)},
    \label{eq:DinvGeigenFinS}
\end{align}
whose eigenvalues and eigenvectors can be easily related to the desired eigenvalues $\beta_k^2$ and eigenvectors $\bm{r}^{(k)}$ of Eq.\,(\ref{eq:DinvGeigen}). Following some algebra, we find:
\begin{align}
    \beta_k^2 &= \frac{S}{(\NS-1) -\tilde{\beta}_k^2}\cdot \tilde{\beta}_k^2 = \tilde{\beta}_k^2 +  
    \sum_{j=1}^{\infty} \tilde{\beta}_k^2 \left( 1+\tilde{\beta}_k^2 \right)^{j} \left( \frac{1}{S} \right)^j, \label{eq:betaEst} \\
    \bm{r}^{(k)} &= \tilde{\bm{r}}^{(k)}. 
    \label{eq:rEst}
\end{align}
From Eq.\,(\ref{eq:betaEst}), we see that to lowest order in $\frac{1}{\NS}$, $\beta_k^2 \approx \tilde{\beta}_k^2$. We note that $\tilde{\beta}_k^2$ are always smaller than $S-1$, and the zero-th order truncation of Taylor series expression for $\beta_k^2$ above is valid only for those $k$ satisfying $\tilde{\beta}_k^2 \ll \NS-1$ (see Fig.\,\ref{fig:NSR_Tilde_Renormalization}). However, this expression also supplies corrections to higher orders in $\frac{1}{S}$, which are non-negligible for $\tilde{\beta}_k^2 \lesssim S-1$, as we see in example of Fig.\,\ref{fig:NSR_Tilde_Renormalization}. In contrast, the estimated eigenvectors $\tilde{\bm{r}}^{(k)}$ to \textit{any} order in $\frac{1}{S}$ equal the desired eigenvectors ${\bm{r}}^{(k)}$ without any corrections.

\begin{figure}
    \centering
    \includegraphics[width=1.0\textwidth]{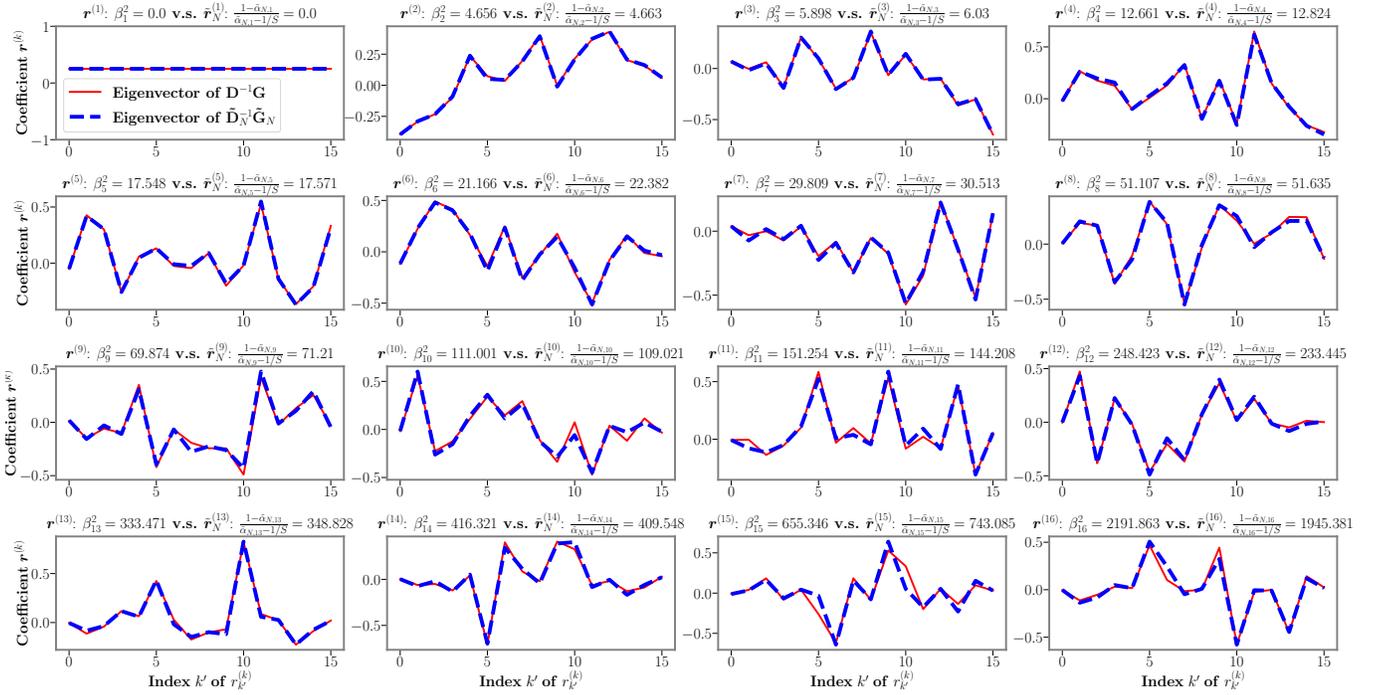}
    \caption{Estimating noise-to-signal ratio eigenvalues and corresponding eigentask coefficients under finite statistics ($N=300, \NS=1000$) in a 4-qubit H-encoding system, and comparison with theoretical value for $N\to\infty,\NS\to\infty$. }
    \label{fig:Tilde_Renormalization}
\end{figure}

Of course, in practice we do not have access to the matrices $\widetilde{\gr}$ and $\widetilde{\mathbf{D}}$, as these are only defined precisely in the limit where $N\to\infty$. However, for sufficiently large $N$, we can approximate these matrices to lowest order by their finite $N$ values, $\widetilde{\gr} = \widetilde{\gr}_N + \mathcal{O}\left(\frac{1}{N}\right)$ and $\widetilde{\mathbf{D}} = \widetilde{\mathbf{D}}_N + \mathcal{O}\left(\frac{1}{N}\right)$. 
Then, the eigenproblem in Eq.\,(\ref{eq:DinvGeigenFinS}) can be expressed in the final form,
\begin{align}
    \widetilde{\mathbf{D}}_N^{-1} \widetilde{\gr}_N \tilde{\bm{r}}_{N}^{(k)} = (1 + \tilde{\beta}_{N, k}^2)^{-1} \tilde{\bm{r}}_{N}^{(k)} \equiv \tilde{\alpha}_{N,k} \tilde{\bm{r}}_{N}^{(k)}, 
    \label{eq:DinvGeigenApprox}
\end{align}
where the eigenvalues $\tilde{\beta}_{N,k}^2, \tilde{\alpha}_{N,k}$ and eigenvectors $\tilde{\bm{r}}_{N}^{(k)}$ in the large $N$ limit must satisfy
\begin{align}
    \lim_{N \to \infty} \tilde{\beta}_{N,k}^2 = \tilde{\beta}_{k}^2, \quad \lim_{N \to \infty} \tilde{\alpha}_{N,k} = \tilde{\alpha}_k, \quad 
    \lim_{N \to \infty} \tilde{\bm{r}}_{N}^{(k)} = \tilde{\bm{r}}^{(k)} \equiv \bm{r}^{(k)}. 
\end{align}
Here the invertibility of the empirically-computed matrix $\widetilde{\mathbf{D}}_N$ required for Eq.\,(\ref{eq:DinvGeigenApprox}) is numerically checked, based on which we can establish a better numerical method in Appendix \ref{sec:GramMatrix-free}. 

Eq.\,(\ref{eq:DinvGeigenApprox}) represents the eigenproblem whose eigenvalues $\tilde{\beta}_{N,k}^2$ and eigenvectors $\tilde{\bm{r}}_{N}^{(k)}$ we actually calculate. For large enough $N$ and under finite $\NS$, we can use these as valid approximations to the eigenvalues and eigenvectors of Eq.\,(\ref{eq:DinvGeigenFinS}). According to using Eqs.\,(\ref{eq:betaEst}), (\ref{eq:rEst}), we are finally able to directly estimate the $N,\NS \to \infty$ quantities $\beta_k^2$ and $\bm{r}^{(k)}$ by the following two quantities:
\begin{align}
    \bar{\beta}_k^2 &\equiv \frac{\NS \cdot \tilde{\beta}_{N,k}^2}{(\NS - 1) - \tilde{\beta}_{N,k}^2} = \frac{1 - \tilde{\alpha}_{N,k}}{\tilde{\alpha}_{N,k} - \frac{1}{\NS}}, \label{eq:barbeta2k} \\
    \bar{\bm{r}}^{(k)} &\equiv \tilde{\bm{r}}_N^{(k)}. \label{eq:barrk} 
\end{align}
It is clear that the approximation of $\beta_k^2$ to lowest order will be an underestimate, as the contribution of order $\frac{1}{\NS}$ is positive. 
In Fig.\,\ref{fig:NSR_Tilde_Renormalization}, we demonstrate a match between $\bar{\beta}_k^2 = (\NS \cdot \tilde{\beta}_{N,k}^2)/((\NS - 1) - \tilde{\beta}_{N,k}^2)$ and $\beta_k^2$ for a wide range of eigentasks with lower order $k$.
In Fig.\,\ref{fig:Tilde_Renormalization}, we plot the estimated eigenvectors $\tilde{\bm{r}}_N^{(k)}$ computed under finite statistics ($N=300,S=1000$, where these two numbers are relevant for IBM quantum processors) in H-encoding, together with the $N,\NS \to \infty$ eigenvectors ${\bm{r}}^{(k)}$, and the estimated eigenvalues.

\subsection{Gram matrix-free construction to approximate eigentasks and NSR eigenvalues}
\label{sec:GramMatrix-free}


If we consider Eq.\,(\ref{eq:DinvGeigenApprox}) and multiply through by $\mathbf{D}_N^{-\frac{1}{2}}$, the resulting equation can be written as an equivalent eigenproblem, 
\begin{align}
    \frac{1}{N} \widetilde{\mathbf{D}}_N^{-\frac{1}{2}} \regmat^T \regmat \widetilde{\mathbf{D}}_N^{-\frac{1}{2}} \left(\widetilde{\mathbf{D}}_N^{\frac{1}{2}} \tilde{\bm{r}}_N^{(k)}\right) = \tilde{\alpha}_{N,k}\left(\widetilde{\mathbf{D}}_N^{\frac{1}{2}} \tilde{\bm{r}}_{N}^{(k)}\right) 
\end{align}
where we have also written $\widetilde{\gr}_N = \frac{1}{N}\widetilde{\mathbf{F}}_N^T\widetilde{\mathbf{F}}_N$ as in the previous section. Note that as written above, the eigenproblem is entirely equivalent to obtaining the singular value decomposition of the matrix $\frac{1}{\sqrt{N}} \widetilde{\mathbf{D}}_N^{-\frac{1}{2}} \regmat^T = U \Sigma V^T$, where $U\in \mathbb{R}^{K\times K}$ and $V\in \mathbb{R}^{N\times N}$ are unitary matrix, and $\Sigma$ is a non-negative diagonal matrix with non-increasing diagonal entries: 
\begin{align}
    \Sigma = \mathrm{diag}(\tilde{\alpha}_{N,0}^{\frac{1}{2}}, \cdots, \tilde{\alpha}_{N, K-1}^{\frac{1}{2}}) \approx \mathrm{diag}(\tilde{\alpha}_0^{\frac{1}{2}}, \cdots, \tilde{\alpha}_{K-1}^{\frac{1}{2}}).
\end{align}
To obtain the estimation of combination coefficients $\bm{r}^{(k)}$, let $\bar{\bm{t}}^{(k)} \in \mathbb{R}^K$ be the normalized left singular vector of $\frac{1}{\sqrt{N}} \widetilde{\mathbf{D}}_N^{-\frac{1}{2}} \regmat^T$ (which is also the eigenvector of $\frac{1}{N} \widetilde{\mathbf{D}}_N^{-\frac{1}{2}} \regmat^T \regmat \widetilde{\mathbf{D}}_N^{-\frac{1}{2}} \approx \mathbf{D}^{-\frac{1}{2}} \widetilde{\gr} \mathbf{D}^{-\frac{1}{2}}$ in the large $N$ limit). Then $\bm{r}^{(k)} \approx \tilde{\alpha}^{-\frac{1}{2}}_{N,k} \widetilde{\mathbf{D}}_N^{-\frac{1}{2}} \bar{\bm{t}}^{(k)} = \bar{\bm{r}}^{(k)} \in \mathbb{R}^K$, and 
\begin{align}
    U = (\bar{\bm{t}}^{(0)}, \cdots, \bar{\bm{t}}^{(K-1)}) = \widetilde{\mathbf{D}}_N^{\frac{1}{2}} ( \tilde{\alpha}^{\frac{1}{2}}_{N,0} \bar{\bm{r}}^{(0)}, \cdots, \tilde{\alpha}^{\frac{1}{2}}_{N,K-1} \bar{\bm{r}}^{(K-1)}). 
\end{align}
Here $\bar{\bm{r}}^{(k)} = \tilde{\alpha}^{-\frac{1}{2}}_{N,k} \widetilde{\mathbf{D}}_N^{-\frac{1}{2}} \bar{\bm{t}}^{(k)}$ can be treated as the combination prefactor of $\hat{M}_k$, to obtain the observables which correspond to the eigentasks. The merit of an SVD analysis of $\frac{1}{\sqrt{N}} \widetilde{\mathbf{D}}_N^{-\frac{1}{2}} \regmat^T$ is that we only need to work with a $K$-by-$N$ matrix of features $\regmat$, which is numerically cheaper than further constructing a Gram matrix $\frac{1}{N} \regmat^T \regmat$. Therefore,
\begin{align}
    \Sigma V^T = U^T \frac{1}{\sqrt{N}} \widetilde{\mathbf{D}}_N^{-\frac{1}{2}} \regmat^T = \frac{1}{\sqrt{N}} \left(
    \begin{array}{c}
        \tilde{\alpha}^{\frac{1}{2}}_{N,0} \bar{\bm{r}}^{(0)T}\\
        \vdots\\
        \tilde{\alpha}^{\frac{1}{2}}_{N,K-1} \bar{\bm{r}}^{(K-1)T}
    \end{array}\right) \widetilde{\mathbf{D}}_N^{\frac{1}{2}} \widetilde{\mathbf{D}}_N^{-\frac{1}{2}} \regmat^T = \frac{1}{\sqrt{N}} \left(
    \begin{array}{c}
        \tilde{\alpha}^{\frac{1}{2}}_{N,0} \bar{\bm{r}}^{(0)T}\\
        \vdots\\
        \tilde{\alpha}^{\frac{1}{2}}_{N,K-1} \bar{\bm{r}}^{(K-1)T}
    \end{array}\right) \regmat^T \label{eq:SigmaVT}
\end{align}
The entries of Eq.\,(\ref{eq:SigmaVT}) are 
\begin{align}
    (\Sigma V^T)_{k,n} = \frac{1}{\sqrt{N}} \tilde{\alpha}^{\frac{1}{2}}_{N,k} \sum_{n=1}^{N} \bar{r}^{(k)}_{k'} (\regmat)_{n,k'} = \frac{1}{\sqrt{N}} \tilde{\alpha}^{\frac{1}{2}}_{N,k} \sum_{n=1}^{N} \bar{r}^{(k)}_{k'} \bar{X}_{k'}(u^{(n)}) \propto \bar{y}^{(k)}(u^{(n)}).
\end{align}
This means that for each data sample $u^{(n)}$, the value of the $k$th order eigentask $\bar{y}^{(k)}(u^{(n)})$ is exactly the principal component coordinate of $\frac{1}{\sqrt{N}} \widetilde{\mathbf{D}}_N^{-\frac{1}{2}} \regmat^T$, up to a constant factor.

The appearance of the SVD above brings comparisons to a popular, powerful data-compressing tool: \textit{principal component analysis}, or PCA, which is used to project a relatively high-dimensional data set into a smaller space, without losing much information. In standard PCA, the original data set is cast into a feature matrix $F \in \mathbb{R}^{N \times K}$, representing $N$ data samples and $K$ features. Let $\mu_k = \frac{1}{N} \sum_n F_{n,k}$ and $\sigma^2_k = \frac{1}{N} \sum_n (F_{n,k}-\mu_k)^2$. Then the standard-scored (or the z-scored) matrix $F'$ is defined by 
\begin{align}
    F'_{n,k} = \frac{F_{n,k} - \mu_{k}}{\sigma_k}. \label{z-score}
\end{align}
The SVD of $F'^{T} = U \Sigma V^T$ gives the well-known principal component analysis. Each row of $\Sigma V^T \in \mathbb{R}^{K \times N}$ is called a \textit{principal component} of the standard score data set $F'$. To be more specific, for each data sample labeled by $n$, the coordinate of its $k$-th principal component is $(\Sigma V^T)_{k,n}$.

While classical PCA focuses on how to \textit{reconstruct} a data set with minimal representative features, REC analysis focuses on \textit{minimizing} the effect from quantum sampling noise. 
Since the aim of REC analysis is about the affect from quantum sampling noise, the normalization factor in  REC is taking the reciprocal of $\sqrt{N}(\widetilde{\mathbf{D}}_N^{\frac{1}{2}})_{kk} = (\sum_n \bar{X}_{k}(u^{(n)}))^{1/2}$, which is quite different from usual PCA, where one uses the reciprocal of the standard deviation $\sigma_k$ of each feature over the whole data set samples for normalization, as is described by the form of Eq.\,(\ref{z-score}). 



\section{H-ansatz quantum systems: NSR spectra, resolvable expressive capacity, and eigentasks}
\label{app:H-ansztz}

In this section, we evaluate the REC for quantum systems described by the H-ansatz introduced in Appendix \ref{DetailsEncodings}, as an example of how REC can be efficiently computed for a variety of general quantum systems, and is not just restricted to parameterized quantum circuits.  The results of the analysis are compiled in Fig.\,\ref{fig:app_Features_and_Capacity}, and discussed below.


\begin{figure}
    \centering
    \includegraphics[width = 0.9\columnwidth]{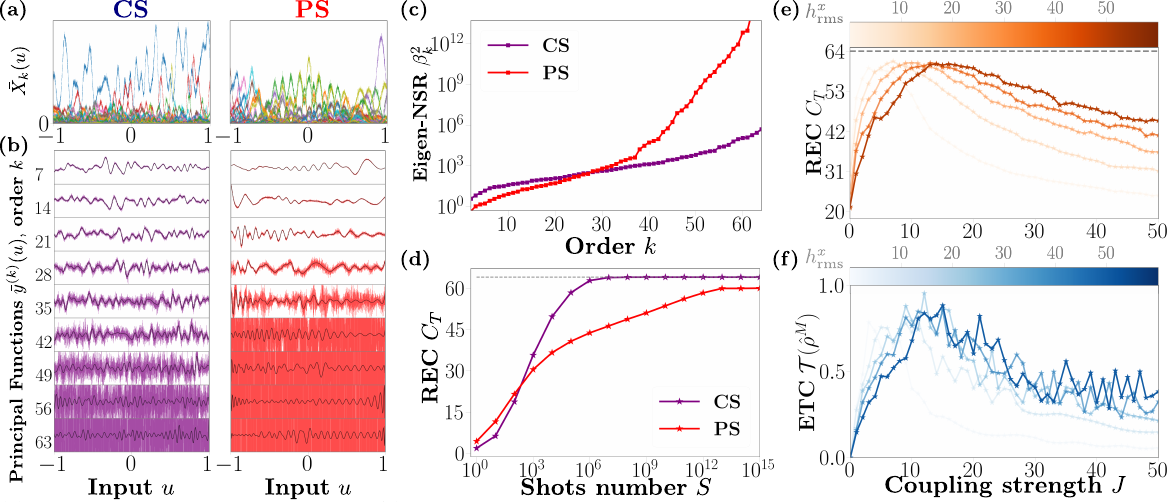}
    \caption{Eigen analysis in a $6$-qubit H-ansatz system (with $N = 5000$ and $\NS = 1000$) forming a 1D ring. The Hamiltonian parameters are chosen randomly with zero-mean and variance $(h^x_{\mathrm{rms}},  h^z_{\mathrm{rms}}, h^I_{\mathrm{rms}}) = (20, 5, 5)$, and $t = 5$ (See Appendix \ref{DetailsEncodings} for details). Coupling strength is uniformly $J \neq 0$ (correlated system) or $J=0$ (product system).
    (a) All $2^L=64$ noisy features $\bar{X}_k(u)$ and (b) noisy eigentasks $\bar{y}^{(k)}(u) = \bm{r}^{(k)}\cdot\bar{\bm{X}}(u)$ for selected $k$ from the features in (a), as well as their expected values $y^{(k)}(u) = \lim_{\NS\to\infty}\bar{y}^{(k)}(u)= \bm{r}^{(k)} \cdot \bm{x}(u)$ (black). 
    (c) Noise-to-signal ratio spectrum $\beta_k^2$ and (d) $\EC$ vs shots $S$ for both correlated system and product system encodings. 
    (e) $\EC$ at $\NS=10^5$ and (f) ETC $\bar{\mathcal{T}}(\hat{\rho}^{M})$ in representative random $6$-qubit H-ansatz, as a function of coupling strength $J$. The peaks of capacity and correlation coincide, around $J \sim h_{\mathrm{rms}}^x$.
    }
    \label{fig:app_Features_and_Capacity}
\end{figure}


Fig.\,\ref{fig:app_Features_and_Capacity}(a) presents the set of features $\{\bar{X}_k(u)\}$ for typical $L=6$ qubit CS and PS at $\NS=1000$ with randomly chosen parameters (referred to as encodings, see caption). The resultant noisy eigentasks $\{\bar{y}^{(k)}(u)\}$ and NSR spectra $\{\beta_k^2\}$ extracted via the eigenvalue analysis are shown in  Figs.\,\ref{fig:app_Features_and_Capacity}(b) and \ref{fig:app_Features_and_Capacity}(c) respectively. In the side-by-side comparison in Fig.\,\ref{fig:app_Features_and_Capacity}(b), we clearly see the $J=0$ ansatz transitioning to a regime with more noise at much lower $k$ than the $J \neq 0$ ansatz. This is reflected in Fig.\,\ref{fig:app_Features_and_Capacity}(c), the $\beta_k^2$ spectrum, having a much flatter slope for larger $k$ (note the plot is semilog). Finally, Fig.\,\ref{fig:app_Features_and_Capacity}(d) shows the REC of both systems as a function of $\NS$. REC rapidly rises for small $S$ for both systems, but the rise of the $J=0$ system is steeper. After a certain threshold in $S$, however, the CS grows more rapidly, approaching the upper bound $2^6=64$ with $\NS \sim 10^8$; in contrast, the PS has a significantly lower $\EC$.


Just like the case of parameterized quantum circuits considered in the main text, we also explore how the REC $\EC$ changes with $J$ for the H-ansatz, and compare it to the total correlation ETC $\bar{\mathcal{T}}$,  as shown in Fig.\,\ref{fig:app_Features_and_Capacity}(f). For $J \to 0$ we have a PS with $\bar{\mathcal{T}}=0$, whereas in the $J \to \infty$ we also have $\bar{\mathcal{T}}=0$ because $\hat{\rho}_0 = \ket{0}\!\bra{0}^{\otimes L}$ is an eigenstate of the encoding ($\hat{\rho}(u)=\hat{\rho}_0$).  This implies there must be a peak at some intermediate $J$, which for both REC and ETC occurs when the coupling is proportional to the transverse field $J \sim h^x$.  At finite $S$, increased ETC is directly related to a higher REC.

Another interesting aspect is the clear trend seen in the maximization of REC around $J \sim h^x_{\mathrm{rms}}$ for various $h^x_{\mathrm{rms}}$, possibly hinting at the role of increased correlation around the MBL phase transition in random spin systems~\cite{martinez2021dynamical}. 
This trend is consistent with results in quantum metrology -- in general, the SNR obtained from averaging $L$ uncorrelated probes scales as $1/\sqrt{L}$. This scaling can become favorable in the presence of quantum correlation and other non-classical correlations, in which case the scaling of the SNR can show up as a quadratic improvement $1/L$~\cite{Giovannetti2006}. 
For even larger $J$, we find that $\hat{\rho}(u)\to\hat{\rho}_0 = \ket{0}\!\bra{0}^{\otimes L}$, which clearly reduces $\ETC$, but also $\EC$ as the quantum system state becomes $u$-independent.

\section{Analytic solution to the quantum 2-design resolvable expressive capacity}
\label{app:2designsol}


There are many system-specific factors that can contribute to the scaling of resolvable expressive capacity with system size, making it challenging to create a general model that describes all systems.
However, we can analytically solve for the REC of a class of quantum models for a specific system: \textit{$2$-design parametric quantum circuits} $\{ p(\bm{u}) \dd \bm{u}, \hat{U} (\bm{\theta} ; \bm{u}) \}$.  We clarify that we are referring here to systems with specific parameters $\bm{\theta}$ which result in 2-designs with respect to the input distribution $p(\bm{u})$; the ensemble average is taken with respect to  inputs $u$.  Quantum literature \cite{Holmes2022}  often refers to general ans\"atze which form $2$-designs with respect to parameters $\bm{\theta}$ instead, which is not what we are considering here.

To be more specific, an ensemble $\{ p(\bm{u}) \dd \bm{u}, \hat{U} (\bm{\theta} ; \bm{u}) \}$ is a $2$-design if the following two quantum channels, defined on \textit{any} $2L$-qubit state $\hat{\tau}$ are equal
\begin{equation}
    \mathcal{C} (\hat{\tau}) = \int \hat{U} (\bm{\theta} ; \bm{u})^{\otimes 2} \hat{\tau} (\hat{U} (\bm{\theta} ; \bm{u})^{\dagger})^{\otimes 2} p (\bm{u}) \dd \bm{u} = \int \hat{U}^{\otimes 2} \hat{\tau} (\hat{U}^{\dagger})^{\otimes 2} \dd \mu_H (\hat{U}) .
\end{equation}
where $\mu_{H}$ is the uniform (Haar) measure. We can verify that all information in the Gram matrix is explicitly contained in the elements of $\mathcal{C} (\hat{\rho}_0 \otimes \hat{\rho}_0)$. To be more specific,
\begin{align}
    & \bra{\bm{b}_{k_1}, \bm{b}_{k_2}} \mathcal{C} (\hat{\rho}_0 \otimes \hat{\rho}_0) \ket{\bm{b}_{k_1}, \bm{b}_{k_2}} \nonumber\\
    =~& \bra{\bm{b}_{k_1}, \bm{b}_{k_2}} \left( \int (\hat{U} (\bm{\theta} ; \bm{u}) \otimes \hat{U} (\bm{\theta} ; \bm{u})) \ket{\bm{b}_0, \bm{b}_0} \bra{\bm{b}_0, \bm{b}_0} (\hat{U} (\bm{\theta} ; \bm{u})^{\dagger} \otimes \hat{U} (\bm{\theta} ; \bm{u})^{\dagger}) p (\bm{u}) \dd \bm{u} \right) \ket{\bm{b}_{k_1}, \bm{b}_{k_2}} \nonumber\\
    =~& \int \left| \bra{\bm{b}_{k_1}} \hat{U} (\bm{\theta} ; \bm{u}) \ket{\bm{b}_0} \right|^2 \cdot \left| \bra{\bm{b}_{k_2}} \hat{U} (\bm{\theta} ; \bm{u}) \ket{\bm{b}_0} \right|^2 p (\bm{u}) \dd \bm{u} \nonumber\\
    =~& \int x_{k_1} (\bm{u}) x_{k_2} (\bm{u}) p (\bm{u}) \dd \bm{u} = (\mathbf{G})_{k_1 k_2} . 
\end{align}
However, $\mathcal{C} (\hat{\rho}_0 \otimes \hat{\rho}_0) = \int U^{\otimes 2} (\hat{\rho}_0 \otimes \hat{\rho}_0) (U^{\dagger})^{\otimes 2} \dd \mu_H (U)$ implies that we can compute the Gram matrix by instead integrating over the Haar measure \cite{Puchaa2017}:
\begin{align}
  (\mathbf{G})_{k_1 k_2} = \int | U_{0, k_1} |^2 | U_{0, k_2} |^2 \dd \mu_H (U) = \left\{\begin{array}{ll}
    \frac{2}{K (K + 1)}, & \text{if } k_1 = k_2,\\
    \frac{1}{K (K + 1)}, & \text{if } k_1 \neq k_2 .
  \end{array}\right. 
\end{align}
Then the corresponding second-order moment matrix $\mathbf{D}$ is given by
\begin{equation}
    (\mathbf{D})_{k k} = \frac{2}{K (K + 1)} + (K - 1) \times \frac{1}{K (K + 1)} = \frac{1}{K} .
\end{equation}
It is self-consistent that the matrix $\mathbf{D}= \mathrm{diag} \left( \frac{1}{K}, \frac{1}{K}, \cdots, \frac{1}{K} \right)$ obeys the normalization condition $\mathrm{Tr} (\mathbf{D}) = K \cdot \frac{1}{K} = 1$. Then we can solve the eigenvalues $\{ \alpha_k \}_{k \in [K]}$ of random walk matrix $(\mathbf{D}^{- 1} \mathbf{G})_{k_1 k_2} = \frac{1}{K + 1} (1+\delta_{k_1 k_2})$. It gives
\begin{equation}
     \alpha_k = \frac{1}{K + 1}(1 + K \delta_{k0}).
\end{equation}
Furthermore, we use $\alpha_k = \frac{1}{1 + \beta_k^2}$ or $\beta^2_k = \frac{1}{\alpha_k} - 1$ to compute the NSR eigenvalue:
\begin{equation}
    (\beta^2_0, \beta^2_1, \beta^2_2, \cdots, \beta^2_{K - 2}, \beta^2_{K - 1}) = (0, K, K, \cdots, K, K) .
\end{equation}
Then the resolvable expressive capacity of any 2-design system is given by
\begin{equation}
    C_T = 1 + \frac{K - 1}{1 + \frac{K}{S}} = K \times \frac{1 + \frac{1}{S}}{1 + \frac{K}{S}} = 2^L \times \frac{S + 1}{S + 2^L} .
\end{equation}


\section{Quantum correlation metrics}
\label{app:QCM}

There is no one standard metric to quantify correlation in a many-body state. The metric we would like to utilize here, the \textit{quantum total correlation}, is a quantity inspired by the classical total correlation of $L$ random variables $(b_1, \cdots, b_L)$, that is $\sum_{l = 1}^L \mathrm{H} (b_l) - \mathrm{H} (b_1 , \cdots, b_L )$. Using the chain rule of Shannon entropy $\mathrm{H} (b_1, b_2, \cdots, b_L) = \mathrm{H} (b_1) + \mathrm{H} (b_2 | b_1 ) + \cdots + \mathrm{H} (b_L | b_1, b_2, \cdots, b_{L - 1} )$
\begin{align}
    & \sum_{l = 2}^L \mathrm{H} (b_l) - \mathrm{H} (b_1, b_2, \cdots, b_L) = \sum_{l = 1}^L \mathrm{H} (b_l) - \sum_{l = 1}^L \mathrm{H} (b_l | b_1, b_2, \cdots, b_{l - 1} ) = \sum_{l = 2}^L \mathrm{I}(b_1, \cdots, b_{l - 1} ; b_l) \in [0, L-1], 
\end{align}
we can see that the classical total correlation tells us how a set of random variables reveals information about each other. Similarly, the quantum total correlation can be defined as \cite{Vedral2002, Modi2010}
\begin{align}
    \mathcal{T} ( \hat{\rho} ) = \sum_{l = 1}^L \mathrm{S} ( \hat{\rho}_l ) - \mathrm{S} ( \hat{\rho} )
\end{align}
where $\mathrm{S}$ is von Neumann entropy and $\hat{\rho}_l := \mathrm{Tr}_{[L] \backslash \{ l \}} \left\{ \hat{\rho} \right\} $ is the subsystem state at qubit $l$. Due to the subadditivity of von-Neumann entropy $\sum_{l = 1}^L \mathrm{S} ( \hat{\rho}_l )  \geq \mathrm{S} ( \hat{\rho} )$, the quantum total correlation is non-negative, and is zero iff the state $\hat{\rho} = \bigotimes_{l=1}^{L} \hat{\rho}_l$ is a product state. 

In this paper's measurement scheme, the specific readout POVMs are the projectors onto the computational states $\{ \ket{\boldsymbol{b}_k} \bra{\boldsymbol{b}_k} \}_{k \in [K]}$. Thus, we are in particular interested in analyzing the post-measurement state $\hat{\rho}^M (u) = \sum_k \rho_{k k} (u)  \ket{\boldsymbol{b}_k} \bra{\boldsymbol{b}_k}$ whose subsystems are correspondingly in states $\hat{\rho}_l^M (u) = \mathrm{Tr}_{[L] \backslash \{ l \}} \left\{ \hat{\rho}^M (u) \right\}$. We compute the average or \textit{expected} quantum total correlation over the input domain $u$ with respect to the input probability distribution $p(u)$: 
\begin{align}
    \bar{\mathcal{T}}\! \left( \hat{\rho}^M \right) = \Eu{ \sum_{l = 1}^L \mathrm{S} ( \hat{\rho}_l^M (u) ) - \mathrm{S} (\hat{\rho}^M (u) ) } = \Eu{ \sum_{l = 1}^L \mathrm{H} (b_l (u)) - \mathrm{H} (b_1 (u), \cdots, b_L (u)) }
\end{align}
where the second equality comes from the diagonal nature of the post-measurement state which reduces the quantum total correlation to a normal classical total correlation.

The post-measurement quantum total correlation always reaches its maximum $L-1$ when the post-measurement state (which just constitutes the diagonal entries of the pre-measurement state) is a GHZ-typed state. As an additional example, for a $W$-state $\ket{W} = \frac{1}{\sqrt{L}} \left( \ket{10 \cdots 0} + \ket{01 \cdots 0} + \cdots + \ket{00 \cdots 1} \right)$, the post-measurement quantum total correlation $\mathcal{T} ( \ket{W} \! \bra{W} )$ is
\begin{align}
     L \left( - \left( \frac{1}{L} \right) \log_2 \left( \frac{1}{L} \right) - \left( \frac{L - 1}{L} \right) \log_2 \left( \frac{L - 1}{L} \right) \right) - L \left( - \left( \frac{1}{L} \right) \log_2 \left( \frac{1}{L} \right) \right) = (L - 1) \log_2 \left( \frac{L}{L - 1} \right) ,
\end{align}
which is upper bounded by $\lim_{L \rightarrow \infty} \mathcal{T} ( \ket{W}  \! \bra{W} ) = \frac{1}{\ln (2)} \approx 1.443$.

\section{REC and eigentasks for classical systems: a basic optical PNN example}
\label{app:photonicRC}


In this appendix section, we present some additional details of the REC analysis of the optical setup considered in Sec.~\ref{sec:photonicRC} of the main text. Our focus here is (1) on the details of the sampling noise statistics, which are different when compared to quantum systems, and (2) the form of the REC eigenproblem for this special case.

\subsection{Sampling noise statistics for a classical optical system}
\label{app:photonicRC1}

For convenience, we recall the form of the electric field of propagating radiation presented in Sec.~\ref{sec:photonicRC}. The electric field after the SLM can be written generally in the form~\cite{zhu_arbitrary_2014}:
\begin{align}
    E_0(u;\vec{d}) = A_0 \cos \left(\frac{\varphi_1(u;\vec{d})}{2} \right) \exp\left\{i\left(\frac{\varphi_1(u;\vec{d})+2\varphi_2(u;\vec{d})}{2}\right)\right\}
\end{align}
where $\varphi_l(u;\vec{d})$ are input encoding functions, and $\vec{d}$ is the position vector describing the coordinates where the electric field is evaluated in the plane orthogonal to the propagation direction; in particular $\vec{d} = (q^1,q^2)$. The specific form of the encoding functions is given by:
\begin{subequations}
    \begin{align}
    \varphi_1(u;\vec{d}) &= B\left( \cos u \left[A_1(\vec{d}) \cos q^1 + A_2(\vec{d}) \sin q^2 \right] + \sin u\left[A_1(\vec{d}) \sin q^1 + A_2(\vec{d}) \cos q^2 \right] \right) \label{eq:inputenc1} \\
    \varphi_2(u;\vec{d}) &= B u( A_1(\vec{d}) q^1 + A_2(\vec{d}) q^2) \label{eq:inputenc2}
\end{align}
\end{subequations}
where we set $B=3.75$, and $A_{1,2}(\vec{d})$ are fixed \textit{input-independent} spatial mask functions whose values are sampled from a normal distribution with zero mean and unit variance; more precisely, $A_{1,2}(\vec{d}) \sim \mathcal{N}(0,1)$ for every $\vec{d} = (q^1,q^2)$.

Following the input encoding, the light propagates through a lens and is measured in the lens' focal plane. The electric field in the focal plane $E(u;\vec{d})$ can be shown to be related to the initial field $E_0(u;\vec{d})$ via a Fourier transform~\cite{saleh_fundamentals_1991, yariv_photonics_2007}:
\begin{align}
    E(u;\vec{d}) = \int\int \dd^2 \vec{d}'~E_0(u;\vec{d}')~\exp\left\{\frac{i2\pi}{\lambda f}\left( \vec{d}\cdot\vec{d}' \right) \right\} 
\end{align}
where $\lambda = \frac{2\pi}{k}$ is the wavelength of the propagating field with wavevector $k$, and $f$ is the focal length of the lens being used.

Finally, information must be extracted from this optical system via measurement for its use as a PNN, which also requires us to address the associated measurement noise in a classical setting. We consider photodetection using a camera in the focal plane of the lens. Furthermore, we consider the camera plane as being comprised of a discrete set of $K=P^2$ photodetectors, arranged in an $P$-by-$P$ square spatial grid, such that the $k$th photodetector is identified with coordinates $\vec{d}_k = (q^1_k,q^2_k)$. This spatial grid ultimately defines the coarse-graining level at which the propagating fields can be probed, and is set by the spatial resolution of the photodetection apparatus, as expected. Then, it is known~\cite{wiseman_quantum_2009} that the differential, stochastic photocurrent generated in a given photodetector in a single measurement, which we name for reference $\dd I(\vec{d}_k,t)$, can be written as a Poisson point process,
\begin{align}
    \dd I(\vec{d}_k,t) = \dd N(\vec{d}_k,t)
\end{align}
where $\dd N(\vec{d}_k,t)$ describes the increment in photodetector counts in a time $\dd t$. The stochastic increments, which are independent at different time, has the specific statistical properties:
\begin{subequations}
\begin{align}
    \mathrm{Prob}[\dd N(\vec{d}_k,t) = 1] &= \eta~P(\vec{d}_k)~\dd t + o({\dd t}), \label{appeq:meandN} \\
    \mathrm{Prob}[\dd N(\vec{d}_k,t) \geq 2] & = o({\dd t}), \\
    \mathbb{E}[\dd N(\vec{d}_k,t)\dd N(\vec{d}_{k'},t')] &= \mathbb{E}[\dd N(\vec{d}_k,t)]\mathbb{E}[\dd N(\vec{d}_{k'},t')], \quad {\rm if }~\vec{d}_k\neq \vec{d}_{k'} \label{appeq:vardN}
\end{align} 
\end{subequations}
where expectation values are formally computed over the distribution of Poisson point processes $\dd N(\vec{d}_k,t)$. Importantly, $P(\vec{d}_k)$ is the power incident on the photodetector with spatial coordinate $\vec{d}_k$. Then, the expectation value of the increment in counts is directly proportional to the intensity of the incident radiation and a measurement efficiency factor $\eta$. The second line defines the fact that the probability of more than a single increment in counts in the time interval $\dd t$ is $o(\dd t)$, and hence higher-order. Finally, the third line indicates that counts on spatially distinct photodetectors and at distinct times are uncorrelated. 


The single-shot measured features of this photonic learning scheme become the integrated photocurrent values over an integration time $T_{\rm int}$ (the input $u$ is not written explicitly for notational simplicity): 
\begin{align}
    X^{(s)}_k \equiv I(\vec{d}_k) = \int_0^{T_{\rm int}}~\dd I(\vec{d}_k,t) \in \mathbb{N}
\end{align}
which are once again stochastic quantities, as they vary from one measurement to the next. This also allows us to easily write down the statistical properties of the integrated photocurrent using Eqs.~(\ref{appeq:meandN})-(\ref{appeq:vardN}):
\begin{subequations}
\begin{align}
    \mathbb{E}[I(\vec{d}_k)] &= \eta~P(\vec{d}_k)~T_{\rm int}  \\
    \mathbb{E}[I(\vec{d}_k)I(\vec{d}_{k'})] &= \mathbb{E}[I(\vec{d}_k)] \mathbb{E}[I(\vec{d}_{k'})], \quad {\rm if }~\vec{d}_k\neq\vec{d}_{k'} \\
    \mathbb{E}[I^2(\vec{d}_k)] &= \eta~P(\vec{d}_k)~T_{\rm int} + (\eta~P(\vec{d}_k)~T_{\rm int})^2 = \eta~P(\vec{d}_k)~T_{\rm int} + (\mathbb{E}[I(\vec{d}_k)])^2
\end{align} 
\end{subequations}
The final expression then provides the variance of the integrated photocurrent for a given photodetector indexed by $k$:
\begin{align}
    \sigma_{\rm I}^2(\vec{d}_k) = \mathbb{E}[I^2(\vec{d}_k)]-(\mathbb{E} [I(\vec{d}_k)])^2 = \eta~P(\vec{d}_k)~T_{\rm int}
\end{align}
As an aside, we calculate the measured power signal-to-noise ratio of the integrated photocurrent, ${\rm SNR}_I$, which takes the form
\begin{align}
    {\rm SNR}_I = \frac{(\mathbb{E}[I(\vec{d}_k)])^2}{\sigma_{\rm I}^2(\vec{d}_k)} = \eta~P(\vec{d}_k)~T_{\rm int},
\end{align}
and therefore grows with incident power $P(\vec{d}_k)$ and integration time $T_{\rm int}$ (assuming an unsaturated photodetector). 

We now make the connection between the measured photocurrents and the propagating fields reaching the photodetector, described by $E(u;\vec{d}_k)$. More precisely, the power incident on the photodetector is simply set by the Poynting flux of the propagating fields, and can be related to the electric field intensity, $P(\vec{d}_k) = \alpha|E(u;\vec{d}_k)|^2$, where $\alpha$ is as introduced in Sec.~\ref{sec:photonicRC}.

The complete input-output map defined above fits within our very general framework. To emphasize this, we now define the measured features $\bar{X}_k(u)$ extracted from this classical machine analogously to the case of measured features extracted from quantum systems, namely Eq.~(\ref{eq:Xsum}). Precisely, we define $\bar{X}_k(u)$ as averages over individual shots $s$ of integrated photocurrents for each photodetector,
\begin{align}
    \bar{X}_k(u) = \frac{1}{\NS}\sum_s X_k^{(s)}(u),
\end{align}
We have here restored the $u$-dependence of $X_k$, which arises via the encoded amplitudes and phases in the electric field $E(u;\vec{d}_k)$. 

$\bar{X}_k(u)$ are therefore sums over i.i.d.\,random variables $X_k(u)$. Hence we can directly write for the mean and covariance of these measured features calculated from infinitely-many samples:
\begin{subequations}
\begin{align}
    \Es{\bar{X}_k(u)} &= \eta\alpha|E(u;\vec{d}_k)|^2T_{\rm int}  \\
    \Covs[\bar{X}_j, \bar{X}_k](u) & = 
    \frac{1}{\NS} \delta_{jk} \eta\alpha|E(u;\vec{d}_k)|^2T_{\rm int}
\end{align}  
\end{subequations}
To connect with our prior notation, we further write:
\begin{align}
    \bar{X}_k(u) = x_k(u) + \frac{1}{\sqrt{\NS}}\zeta_k(u)
    \label{appeq:xbarCML}
\end{align}
which is simply Eq.~(\ref{eq:xbar}) of the main text. Note that for most classical machine learning schemes, $\NS=1$; however, in our analysis we allow that the shot number $S$ can be any integer. Here $x_k(u)$ are deterministic quantities defined as:
\begin{align}
    x_k(u) \equiv \Es{\bar{X}_k(u)} = \eta\alpha|E(u;\vec{d}_k)|^2T_{\rm int}.
    \label{appeq:xCML}
\end{align}
Then, it follows that the remaining term $\frac{1}{\sqrt{\NS}}\zeta_k(u)$ is a stochastic process with zero mean (as taking expectation values on both sides of Eq.~(\ref{appeq:xbarCML}) will demonstrate) and whose second-order moment (equivalent to the variance as it has zero mean) encodes the variance of the Poisson point process in one shot of experiment,
\begin{align}
    \Covs [\bar{X}_j, \bar{X}_k](u)  = \frac{1}{\NS} \Covs [{\zeta}_j, {\zeta}_k](u) = \frac{1}{\NS}\eta\alpha|E(u;\vec{d}_k)|^2 T_{\rm int} \delta_{jk} = \frac{1}{\NS} \delta_{jk}x_k(u)
\end{align}
where we have used Eq.~(\ref{appeq:xCML}). This finally yields
\begin{align}
    \Covs[{\zeta}_j, {\zeta}_k](u) 
    \equiv
    \cu_{jk}(u)
    = \delta_{jk} x_k(u)
    \label{appeq:sigmaCML}
\end{align}
as presented in Sec.~\ref{sec:photonicRC}.

\subsection{REC analysis for a classical optical system}

For the Poisson noise process of photodetection, the noise matrix $\ci$ is different compared to the case of the multinomial noise process. Fortunately, for the specific case we have considered with spatially uncorrelated detectors, $\ci$ is in fact simpler. We note that:
\begin{align}
    \ci_{jk} = \int du~p(u) \cu_{jk}(u) = \int du~p(u) \delta_{jk}x_k(u) .
\end{align}
Hence $\ci$ is itself now diagonal. The eigenproblem in question, Eq.~(\ref{eq:eigenprob}), can therefore be simplified to:
\begin{align}
    \ci \bm{r}^{(k)} = \beta_k^2 \gr \bm{r}^{(k)} \implies \ci^{-1}\gr \bm{r}^{(k)} = \frac{1}{\beta_k^2}\bm{r}^{(k)}
    \label{appeq:eigenprobCML}
\end{align}
where we have computed the inverse of $\ci$ since it is a diagonal matrix; we also assume none of its diagonal entries vanish. This is not a strong constraint, since these entries are simply equal to the measured features, which are the sum of intensities incident on the photodetector, and will typically be nonzero (assuming integrated photocurrents from any `dead' photodetectors or pixels are excluded from the measured features). 

Solving Eq.~(\ref{appeq:eigenprobCML}) allows us to calculate the infinite-shot eigentasks and the $\EC$ as a function of $\NS$ for the photonic learning system. We can also construct the Gram and covariance matrices using finitely-sampled features $\bar{X}_k(u)$ over the input domain, as would be done in a real experiment. One do not need to employ the general Eq.\,(\ref{eq:Bessel}) to estimate the eigentasks. In fact, there is simpler procedure which is similar to Eq.\,(\ref{eq:Gtilde1}) and Eq.\,(\ref{eq:Dtilde1}). The following two quantities computed from the $S$-finite statistics 
\begin{align}
     (\widetilde{\gr}_N)_{k_1 k_2} & \equiv \frac{1}{N}\sum_{n=1}^N \bar{X}_{k_1}(u^{(n)})\bar{X}_{k_2}(u^{(n)}) = \frac{1}{N} (\regmat^T \regmat)_{k_1k_2} \approx \int \bar{X}_{k_1}(u)\bar{X}_{k_2}(u) p(u) \dd u, \\
     (\widetilde{\mathbf{V}}_N)_{k_1 k_2} & \equiv \delta_{k_1, k_2} \frac{1}{N}\sum_{n=1}^N \bar{X}_{k_1}(u^{(n)}) \approx \delta_{k_1, k_2} \int \bar{X}_{k_1}(u) p(u) \dd u.
\end{align}
Similar to the scenario in Appendix \ref{app:tilde_correction}. We can show that by solving the eigenproblem $\widetilde{\ci}_N^{-1} \widetilde{\gr}_N \tilde{\bm{r}}_N^{(k)} = \tilde{\alpha}_{N,k} \tilde{\bm{r}}_N^{(k)}$, the true eige-NSR $\beta_k^2$ and eigentasks coefficients $\bm{r}^{(k)}$ can be well approximated by $1/(\tilde{\alpha}_{N,k} - 1/S)$ and $\tilde{\bm{r}}_{N}^{(k)}$, respectively. The derivation still starts from taking their limits for $N \to \infty$:
\begin{align}
    \widetilde{\gr} &= \lim_{N\to \infty}\widetilde{\gr}_N = \gr + \frac{1}{\NS} \mathbf{V}, \label{eq:Gtilde2}
    \\ \widetilde{\mathbf{V}} &= \lim_{N\to \infty}\widetilde{\mathbf{V}}_N = \mathbf{V}. \label{eq:Dtilde2}
\end{align}
Therefore, $\widetilde{\gr}$ and $\widetilde{\mathbf{V}}$ will provide an eigen-problem:
\begin{align}
    \widetilde{\ci}^{-1} \widetilde{\gr} \bm{r}^{(k)} = \tilde{\alpha}_k \bm{r}^{(k)}
\end{align}
where $\tilde{\alpha}_k = \frac{1}{\beta_k^2} + \frac{1}{S}$, or equivalently $\beta^2_k = \frac{1}{\tilde{\alpha}_k - 1/S}$. It means that by numerically solving $\widetilde{\ci}_N^{-1} \widetilde{\gr}_N \tilde{\bm{r}}_N^{(k)} = \tilde{\alpha}_{N,k} \tilde{\bm{r}}_N^{(k)}$, one can use
\begin{align}
    \bar{\beta}_k^2 &\equiv \frac{1}{\tilde{\alpha}_{N,k} - 1/S}, \\
    \bar{\bm{r}}^{(k)} &\equiv \tilde{\bm{r}}_N^{(k)}.  
\end{align}
to accurately approxiamte to $\beta_k^2$ and $\bm{r}^{(k)}$, which finishes the proof of the desired statement.

\begin{center}
    \rule{30mm}{1pt}
\end{center}

\stopcontents[appendices]


\end{widetext}

\end{document}